\documentclass[aps,prb,twocolumn,amsmath,amssymb,superscriptaddress,floatfix,eqsecnum]{revtex4-1}

\usepackage{amsfonts}
\usepackage{amsthm}
\usepackage{cases}
\usepackage{url}
\usepackage{framed}
\usepackage{cancel,soul}
\usepackage{graphicx}
\usepackage{multirow}
\usepackage[pdftex]{color}
\usepackage{bm}
\usepackage[normalem]{ulem}
\usepackage[colorlinks=true,linktoc=page,linkcolor=blue,citecolor=blue]{hyperref}

\newcommand{\bea}{\begin{eqnarray}}
\newcommand{\eea}{\end{eqnarray}}
\def\:={\,\raisebox{0.85pt}{.}\hspace{-2.78pt}\raisebox{2.85pt}{.}\!\!=\,}
\def\=:{\,=\!\!\raisebox{0.85pt}{.}\hspace{-2.78pt}\raisebox{2.85pt}{.}\,}

\begin{document}

\title{A model of chiral spin liquids with Abelian and non-Abelian
       topological phases}

\author{Jyong-Hao Chen}
\affiliation{Condensed Matter Theory Group, Paul Scherrer Institute, 
CH-5232 Villigen PSI, Switzerland}
\author{Christopher Mudry}
\affiliation{Condensed Matter Theory Group, Paul Scherrer Institute, 
CH-5232 Villigen PSI, Switzerland}
\author{Claudio Chamon}
\affiliation{Department of Physics, Boston University, 
Boston, MA, 02215, USA}
\author{A. M. Tsvelik}
\affiliation{
Condensed Matter Physics and Materials Science Division,
Brookhaven National Laboratory, Upton, NY 11973-5000, USA
            }

\date{\today}

\begin{abstract} 
We present a two-dimensional lattice model for quantum spin-1/2 for
which the low-energy limit is governed by four flavors of strongly
interacting Majorana fermions.
We study this low-energy effective theory using two alternative approaches.
The first consists of a mean-field approximation.
The second consists of a Random Phase approximation (RPA)
for the single-particle Green's functions of the Majorana fermions
built from their exact forms in a certain one-dimensional limit.
The resulting phase diagram consists of two competing chiral phases,
one with Abelian and the other with non-Abelian topological order,
separated by a continuous phase transition. 
Remarkably, the Majorana fermions propagate in the two-dimensional bulk,
as in the Kitaev model for a spin liquid on the honeycomb lattice.
We identify the vison fields, which are mobile 
(they are static in the Kitaev model)
domain walls propagating along only one of the two space directions.
\end{abstract}

\maketitle
\tableofcontents

\section{Introduction and results}

\begin{figure*}[t]
\begin{center}
\includegraphics[width=0.75\textwidth]{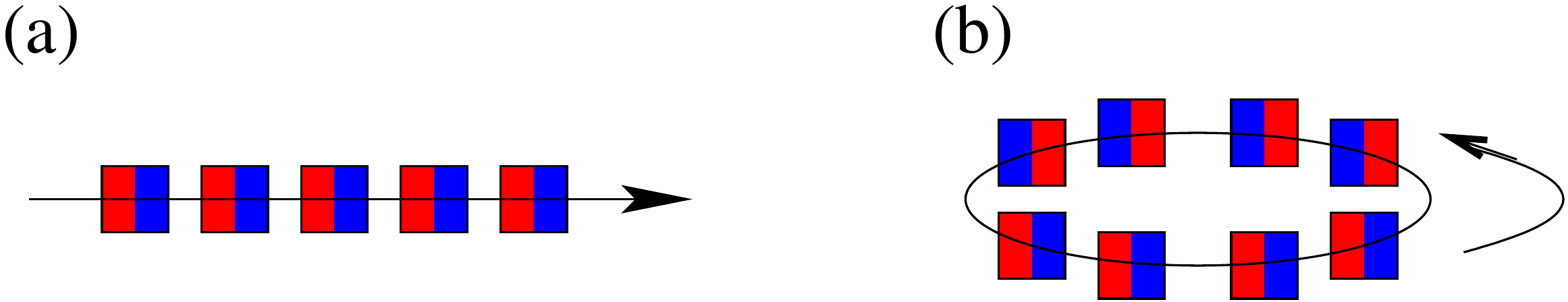}
\caption{(Color online)
(a) Alignment of blocks along an open line.
Each block represents a non-chiral conformal field theory in $(1+1)$-dimensional
spacetime. One half of the gapless modes are left movers, the other half
are right movers. This partition into movers with opposite chirality is
represented by the coloring blue and red, respectively.
(b) Alignment of blocks along a circle.
\label{Fig: open versus closed BCS}
         }
\end{center}
\end{figure*}

\subsection{Motivation}

Most of the observed low-temperature phases
in condensed matter physics are characterized by 
spontaneous symmetry breaking (SSB) through the
onset of a local order parameter acquiring a non-vanishing
expectation value. Antiferromagnetism is the paradigmatic example of
SSB with the staggered magnetization as the local order parameter.
On the other hand, it has been  found that such states as exist in the
fractional quantum Hall effect (FQHE) possess a hidden (topological)
order not associated with any  local order parameter.
This type of order may exist only if the bulk is incompressible,
in which case it reveals itself in several ways. 
In particular, a sufficient condition for the topological order is
the existence of robust gapless boundary excitations.
If the system is situated on a manifold without boundaries,
the ground state is degenerate and the degeneracy depends on
the genus of the manifold. These characteristics of topological order 
were formulated by Wen in Ref.\ \onlinecite{Wen91a}, but the notion 
has been later refined by relating it to the presence  of
long-range quantum entanglement in Refs.\
\onlinecite{Kitaev06b,Levin06,Chen10}.
The other feature in $(2+1)$-dimensional spacetime is the
presence of gapped   point-like excitations
obeying braiding statistics that is neither fermionic nor bosonic.
This sharpening of what constitutes the essence of topological order
in $(2+1)$-dimensional spacetime 
has opened the possibility of its classification.
\cite{Wen16,Lan16a}
However, these discussions of topological order
have been conducted with little reference to microscopic models.
There are very few of them which can be treated by controlled approximations;
most notably the quantum dimer model on the triangular lattice
\cite{Moessner01}
and the Kitaev model on the honeycomb lattice \cite{Kitaev06a}.
The latter is a model of interacting quantum spins, whose
excitations are Majorana fermions.
Their propagation is facilitated by the presence of the so-called visons which 
in this model are immobile $\mathbb{Z}^{\,}_{2}$ gauge field fluxes.

One way to construct microscopic models
with topological order is to use the so-called wire construction. The
idea, following Kane and his collaborators,
\cite{Mukhopadhyay01,Kane02,Teo14,Kane17} is to couple elementary
building blocks that realize a conformal-field theory (CFT) in
$(1+1)$-dimensional spacetime [by construction this building block
cannot be gapped into a phase supporting topological order in
$(1+1)$-dimensional spacetime] so as to realize an incompressible
phase of matter in $(d+1)$-dimensional spacetime that supports
topological order.  Although, the diagnostic for topological order is
a degeneracy of the ground-state manifold that depends on the genus of
compactified space, it is more convenient to use a stronger
(a sufficient but not necessary) condition for topological order, namely,
the existence of protected gapless boundary states that are localized
on the $(d-1)$-dimensional boundaries of $d>1$-dimensional space. 
It is then suggested to weakly couple these 
building blocks so as to gap the bulk while leaving 
the boundaries gapless. 
A generic coupling between these building
blocks will not do that, for such a coupling can yield
three possible outcomes.
First, the resulting phase of matter in $(d+1)$-dimensional spacetime may be
gapless and ordered. This is what
happens when antiferromagnetic spin-1/2 chains are coupled so as to
realize an antiferromagnetic square lattice.
\cite{Chakravarty96,Sierra96,Sierra97} Second, the resulting phase of
matter in $(d+1)$-dimensional spacetime may be gapful,
but without topological order.  This is what happens when
antiferromagnetic spin-1/2 chains are weakly coupled pairwise so as to
realize a stacking of two-leg ladders which, in turn, are even more
weakly coupled pairwise.  \cite{Syljuasen97}
We are interested in the third outcome, namely, when the resulting
phase of matter in $(d+1)$-dimensional spacetime is incompressible and
supports topological order.  Which outcome is realized is determined
by the choice of the couplings between the building blocks,
that is by the energetics.

In this paper we will be dealing with a model in  two-dimensional space.
In this case, 
a sufficient but not necessary condition for topological order
is that in the infrared limit 
(i) the first and last building blocks acquire 
a nonvanishing yet reduced central charge when their direct coupling
is forbidden  by locality
(open boundary conditions along the stacking direction),
whereas (ii) the ground state is fully gapped when their direct
coupling is compatible with locality
(closed boundary conditions along the stacking direction).
This situation is pictured in Fig.\
\ref{Fig: open versus closed BCS}.
Each block represents some given non-chiral CFT
in $(1+1)$-dimensional spacetime.
The stacking direction of the blocks is oriented by the arrow.
The coloring red and blue represents the left- and right-movers from the CFT,
respectively. It is possible to gap out a pair of movers of opposite chirality
belonging to two consecutive blocks by
coupling in a local way the right movers from a block
to the left movers of the nearest-neighbor block along the stacking direction.
This leaves the left movers from the first block and the right movers
from the last block gapless in panel (a) from Fig.\
\ref{Fig: open versus closed BCS}, whereas all states are gapped
when periodic boundary conditions
are imposed as in panel (b) from Fig.\
\ref{Fig: open versus closed BCS}.
The challenge is to realize 
Fig.\ \ref{Fig: open versus closed BCS}
by appealing only to local couplings between the microscopic degrees of
freedom 
such as  lattice electrons or magnetic moments.
This challenge was met for all symmetry classes from the ten-fold way
in Ref.\ \onlinecite{Neupert14},
where it was shown that five of them can support Abelian topological order
(ATO) upon choosing local (electronic) interactions between consecutive blocks.

\begin{figure*}[t]
\begin{center}
\includegraphics[width=0.75\textwidth]{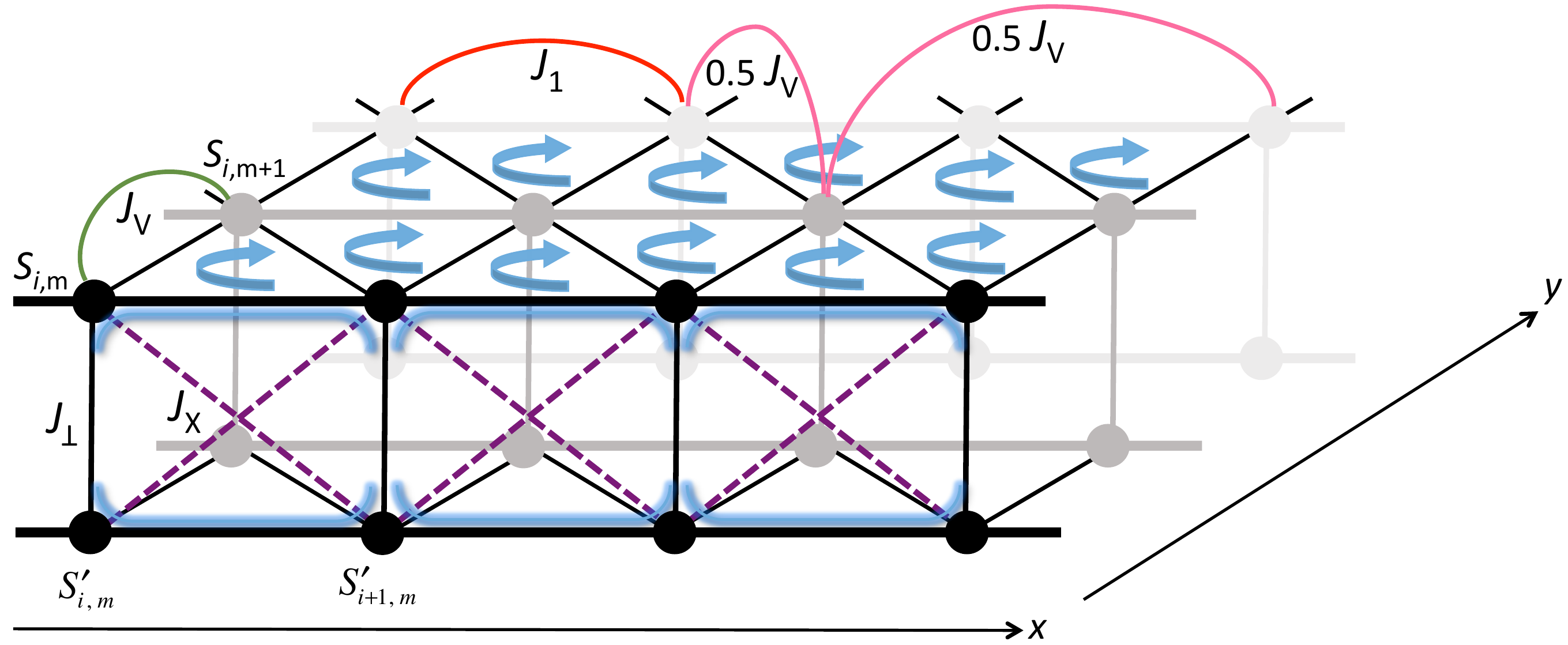}
\\
\caption{(Color online)
Coupled quantum spin-1/2 two-leg ladders that realize the Ising
topological order in two-dimensional space. The intra-ladder couplings
$J^{\,}_{1}$,
$J^{\,}_{\perp}$,
$J^{\,}_{\times}=-J^{\,}_{\perp}/2$,
and
$J^{\,}_{U}$
(represented by the blue curly bracket)
are defined in Eq.\ (\ref{eq: def H ladder}).
The inter-ladder couplings
$J^{\,}_{\vee}$
(represented by the green bond),
$J^{\,}_{\vee}/2$
(represented by the magenta bond),	
and
$J^{\,}_{\chi}$ (represented by the blue arrows)
are defined in Eq.\ (\ref{eq: def couplings between consecutive ladders}).
The lattice geometry can also be thought of as that of a bilayer of
two square lattices.
\label{Fig: 2Dlattice}
         }
\end{center}
\end{figure*}

A proposal to realize a spin liquid supporting chiral edge states with
non-integer valued central charges was given in Refs.\
\onlinecite{Huang16a}
and
\onlinecite{Huang17}.
It is the fractional part to the chiral central charge of the edge states
that signals the non-Abelian topological order (NATO).
This proposal relies on local interactions within and between consecutive
blocks from Fig.\ \ref{Fig: open versus closed BCS}
that are both marginally relevant and compete with each other.
Consequently, it could not be proven that all states are gapped when
periodic boundary conditions are imposed as in panel (b)
from Fig.\ \ref{Fig: open versus closed BCS}.
The purpose of this paper is to modify the field theory studied
in Ref.\ \onlinecite{Huang17}
to rule out the possibility that the flow to strong coupling
in Ref.\ \onlinecite{Huang17},
when periodic boundary conditions are imposed as in panel (b)
from Fig.\ \ref{Fig: open versus closed BCS},
delivers a gapless phase of matter.
This modification makes the theory amenable to a mean-field approximation
that predicts two gapped phases separated by a gap-closing transition
when the couplings between consecutive blocks are
chosen to be marginally relevant.
One gapped phase supports ATO.
The other gapped phase supports NATO.
We also find a third gapless phase, a two-dimensional sliding
Luttinger phase when the couplings between consecutive blocks are
chosen to be marginally irrelevant.

\subsection{Results and outline}

As was the case with Ref.\ \onlinecite{Huang17},
we shall take the blocks from
Fig.\ \ref{Fig: open versus closed BCS}
to realize an Ising CFT on the boundary, i.e., 
a CFT with central charge $c=1/2$.
However, unlike in Ref.\ \onlinecite{Huang17} where 
this Ising CFT was driven by marginal perturbations
to a CFT with central charge $c=2$, in this paper the Ising CFT is driven by
a strongly relevant perturbation.
This distinction gives a much better control on the
strong coupling fixed point that realizes the Ising NATO
when two consecutive blocks are coupled through interactions.

Throughout this paper, we are mostly preoccupied with the analysis of the 
field theory corresponding to Fig.\ \ref{Fig: open versus closed BCS}.
The choice of a microscopic theory delivering the Ising criticality is
dictated by simplicity at the level of CFT rather than 
by simplicity on the microscopic level.
This microscopic theory is a quantum spin-1/2 ladder,
whose low-energy effective field theory is depicted
by any one of the single square box colored in red and blue
in Fig.\ \ref{Fig: open versus closed BCS}.
This was also the case in Ref.\ \onlinecite{Huang17}.
However, instead of relying on two-body spin-1/2 interactions
which are reduced to marginal current-current interactions at low energies, 
as was the case in Ref.\ \onlinecite{Huang17},
we shall rely in this paper on four-body spin-1/2 interactions
which are  reduced to a mass term for three out of the four
gapless Majorana fields that encode
the critical theory of two decoupled antiferromagnetic
quantum spin-1/2 chains. Once a single spin-1/2 ladder is tuned to the Ising
critical point, we couple the ladders
as was done in Ref.\ \onlinecite{Huang17}.
The resulting lattice model is depicted
in Fig.\ \ref{Fig: 2Dlattice}.
Each ladder viewed from the side in Fig.\ \ref{Fig: 2Dlattice}.
is represented by a square box colored in red and blue
in Fig.\ \ref{Fig: open versus closed BCS}
at low energies.

The lattice model for a single spin-1/2 ladder is defined
in Sec.\ \ref{sec: A single two-leg ladder}.
The lattice model for a one-dimensional array
of coupled spin-1/2 ladders is defined
in Sec.\ \ref{sec: Coupled two-leg ladders}.
Its continuum limit is derived and shown to agree with
the Majorana Hamiltonian (\ref{eq: desired Majorana theory}).

The continuum limit is derived under the assumption
that one can eliminate couplings between the most relevant fields
on consecutive ladders and neglect those between more distant
ladders. Then, at low energies, the coupled quantum spin-1/2 ladders in Fig.\
\ref{Fig: 2Dlattice}
admit an effective description in terms of an
interacting quantum field-theory with four Majorana fields per ladder.
Dealing with the fermionic field theory,
one has to remember that its Hilbert space
is greater than the one of the original spin model.
In particular, it allows states created by odd numbers 
of Majorana fermions per ladder.
There are no such states in the
spin model. This fermionic field theory
is the starting point captured by
Eq.\ (\ref{eq: desired Majorana theory})
from Sec.\ \ref{sec: The Majorana field theory}.
In this mapping the Majorana fields carry a flavor index
$\mathtt{m}=1,\cdots,n$ that labels the quantum spin-1/2 ladders.
The kinetic energy of the Majorana fields is encoded by a
Wess-Zumino-Novikov-Witten (WZNW) action
$\widehat{\mathcal{H}}^{\,}_{\mathrm{WZNW}}$.
This kinetic energy ignores all inter-ladder
interactions and treats any one of the ladders as two
decoupled antiferromagnetic quantum spin-1/2 chains,
each of which is at an $SU(2)^{\,}_{1}$ quantum critical point.
The intra-ladder interactions between the quantum spin-1/2
turn at low energies into bare masses
$m^{\,}_{\mu}\in\mathbb{R}$ ($\mu=0,1,2,3$) for
each Majorana field.
The inter-ladder interactions between the quantum spin-1/2
turn at low energies into an $O(4)$-symmetric interaction
that couples Majorana fields belonging to two
consecutive ladders. This interaction resembles the
Gross-Neveu interaction, and we shall call it 
a Gross-Neveu-like interaction.

\begin{figure}[t]
\begin{center}
\includegraphics[width=0.45\textwidth]{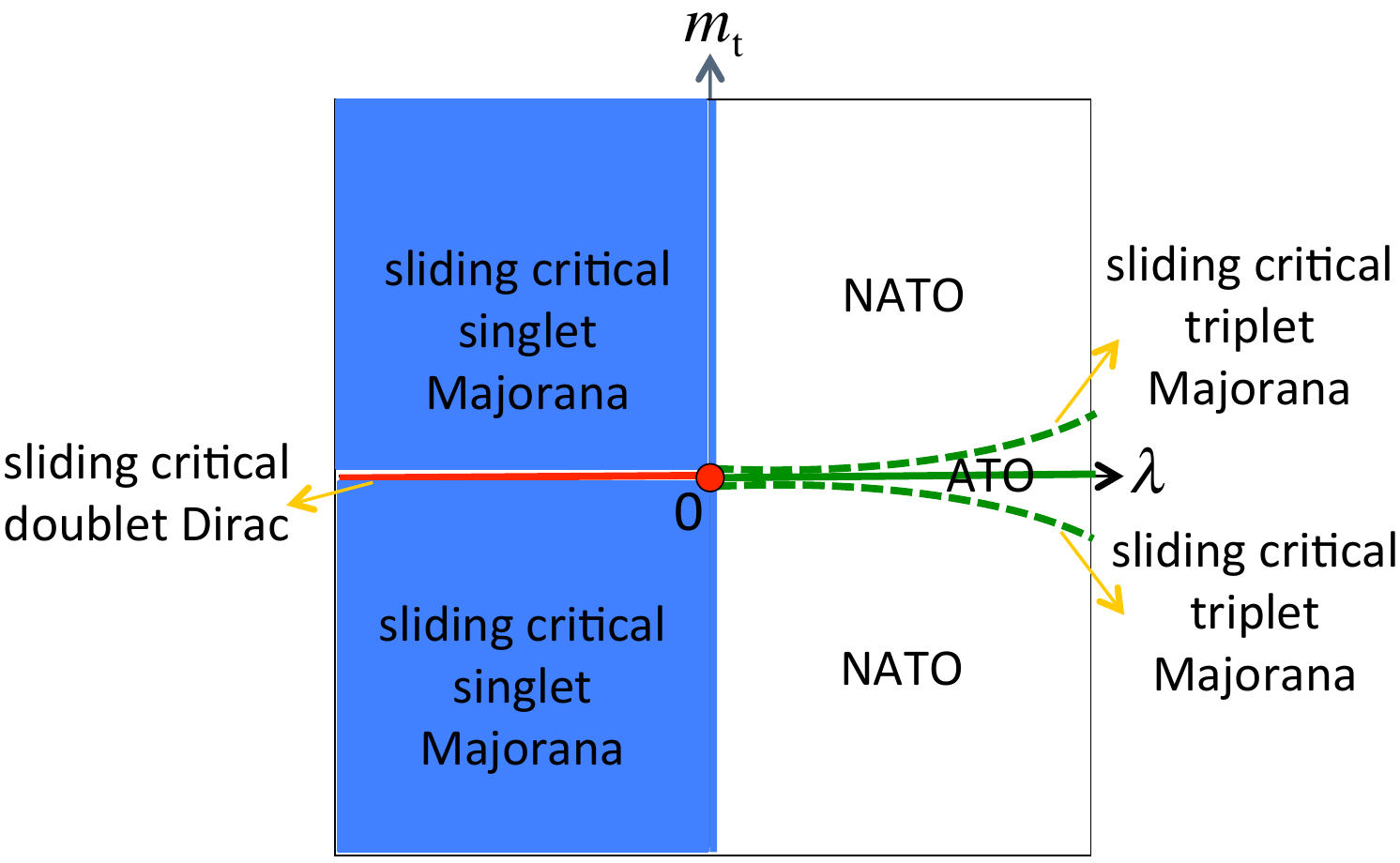}
\caption{(Color online)
Conjectured phase diagram of the theory
(\ref{eq: desired Majorana theory}) with $m^{\,}_{0}=0$ and
$m^{\,}_{a}=m^{\,}_{\mathrm{t}}$ for $a=1,2,3$.
\label{Fig: phase diagram topological order a}
         }
\end{center}
\end{figure}

We treat the $O(4)$-symmetric Gross-Neveu-like interaction
by two alternative methods. 
In Sec.\ \ref{subsec: Hubbard-Stratonovich transformation}, we use the
mean field procedure based on decoupling of the four-fermion interaction
by means of the Hubbard-Stratonovich transformation. 
In Secs.\ \ref{sec: RPA} and
\ref{sec: Two-dimensional Majorana fermions, one-dimensional solitons},
we use the approach which is based on combination of 
non-perturbative results extracted from the exact solution
of the $O(4)$-symmetric Gross-Neveu model and Random Phase approximation. 
The phase diagram
from Fig.\ \ref{Fig: phase diagram topological order a}
is conjectured from a mean-field approximation
that we derive in the reminder of
Sec.\ \ref{sec: The mean-field approach}.
In Fig.\ \ref{Fig: phase diagram topological order a},
$\lambda$ denotes the coupling of the non-Abelian
current-current interactions between consecutive blocks.
This interaction is either marginally irrelevant for negative $\lambda$
or marginally relevant for positive $\lambda$.
The mean-field phase diagram
in Fig.\ \ref{Fig: phase diagram topological order b}
is parametrized by $m^{\,}_{\mathrm{t}}$ and the
mean-field value of the spectral gap  $|\phi^{\,}(\lambda)|/2$
under the assumption that the so-called singlet Majorana is gapless,
$m^{\,}_{0}\equiv m^{\,}_{\mathrm{s}}=0$,
while a triplet of Majoranas have the
isotropic mass $m^{\,}_{a}\equiv m^{\,}_{\mathrm{t}}$ for $a=1,2,3$.
There exist mean-field critical lines that correspond to the condition
$|\phi^{\,}(\lambda)|/2=|m^{\,}_{\mathrm{t}}|$
along which the mean-field Majorana gap vanishes.
The regions $|\phi^{\,}(\lambda)|/2>|m^{\,}_{\mathrm{t}}|$
and $|\phi^{\,}(\lambda)|/2<|m^{\,}_{\mathrm{t}}|$
correspond to phases of matter supporting NATO and ATO, respectively.

From the mean-field phase diagram
in Fig.\ \ref{Fig: phase diagram topological order b},
we conjecture the phase diagram in Fig.\ 
\ref{Fig: phase diagram topological order a} that is parametrized by the
inter-ladder interaction with the uniform coupling $\lambda$ and by
the triplet mass $m^{\,}_{\mathrm{t}}$. The bare value of the
triplet mass is a function of the microscopic magnetic couplings
of any one of the ladders.
For $\lambda>0$, the $O(4)$-symmetric Gross-Neveu-like interaction
guarantees a non-vanishing value for the mean-field $\phi(\lambda)$.
On the other hand, for $\lambda<0$, the $O(4)$ Gross-Neveu-like interaction
also guarantees that the mean-field $\phi(\lambda)$ vanishes.  The line 
$m^{\,}_{\mathrm{t}}=0$ is exactly solvable and we use this solution in
\ref{sec: RPA}. The dashed green line in Fig.\
\ref{Fig: phase diagram topological order a}
corresponds to the mean-field transition line.
The phases NATO and ATO in
Fig.\ \ref{Fig: phase diagram topological order a}
correspond to the mean-field regions
$|\phi(\lambda)|/2>|m^{\,}_{\mathrm{t}}|$ and
$|\phi(\lambda)|/2<|m^{\,}_{\mathrm{t}}|$, respectively.

The nonvanishing mean-field values for $\phi(\lambda)$ follow from 
integrating over the Majorana fields in
Sec.\ \ref{subsec: Integrating out the Majorana fields}
and deriving a mean field  equation obeyed by $\phi(\lambda)$
in Sec.\ \ref{subsec: Mean-field gap equations}.

In Secs.\ \ref{sec: RPA} and
\ref{sec: Two-dimensional Majorana fermions, one-dimensional solitons},
we establish the form of the bulk excitation
spectrum.  It consists of Majorana fermions propagating in two spatial
dimensions and visons excitations which can propagate only along
chains. As far as we are aware this is the only microscopic model
(besides the Kitaev one) where such particles have been rigorously
obtained.

As we have mentioned above,
the lattice model for a single spin-1/2 ladder is defined
in Sec.\ \ref{sec: A single two-leg ladder}, where we also derive its 
continuum limit.
The lattice model for a one-dimensional array
of coupled spin-1/2 ladders is defined
in Sec.\ \ref{sec: Coupled two-leg ladders}.
We then discuss its continuum limit, which is 
the Majorana Hamiltonian (\ref{eq: desired Majorana theory}).
We conclude with a summary in Sec.\ \ref{sec: Summary}.

\section{Majorana field theory}
\label{sec: The Majorana field theory}

\subsection{Definition}

We begin with
\begin{subequations}
\label{eq: desired Majorana theory}
\begin{align}
&
\widehat{\mathcal{H}}\:=
\widehat{\mathcal{H}}^{\,}_{0}
+
\widehat{\mathcal{H}}^{\,}_{\mathrm{intra}-\mathrm{ladder}}
+
\widehat{\mathcal{H}}^{\,}_{\mathrm{inter}-\mathrm{ladder}},
\label{eq: desired Majorana theory a}
\\
&
\widehat{\mathcal{H}}^{\,}_{0}=
\sum^{n}_{\mathtt{m}=1}
\sum^{3}_{\mu=0}
\frac{\mathrm{i}}{2}
v^{\,}_{\mu}
\left(
\widehat{\chi}^{\mu}_{\mathrm{L},\mathtt{m}}
\partial^{\,}_{x}
\widehat{\chi}^{\mu}_{\mathrm{L},\mathtt{m}}
-
\widehat{\chi}^{\mu}_{\mathrm{R},\mathtt{m}}
\partial^{\,}_{x}
\widehat{\chi}^{\mu}_{\mathrm{R},\mathtt{m}}
\right),
\label{eq: desired Majorana theory b}
\\
&
\widehat{\mathcal{H}}^{\,}_{\mathrm{intra}-\mathrm{ladder}}=
\sum^{n}_{\mathtt{m}=1}
\sum^{3}_{\mu=0}
\mathrm{i}\,m^{\,}_{\mu}\,
\widehat{\chi}^{\mu}_{\mathrm{L},\mathtt{m}}
\widehat{\chi}^{\mu}_{\mathrm{R},\mathtt{m}},
\label{eq: desired Majorana theory c}
\\
&
\widehat{\mathcal{H}}^{\,}_{\mathrm{inter}-\mathrm{ladder}}=
\sum^{n-1}_{\mathtt{m}=1}
\frac{\lambda}{4}
\left(
\sum^{3}_{\mu=0}
\widehat{\chi}^{\mu}_{\mathrm{L},\mathtt{m}}\,
\widehat{\chi}^{\mu}_{\mathrm{R},\mathtt{m}+1}\,
\right)^{2},
\label{eq: desired Majorana theory d}
\end{align}
where the velocities $v^{\,}_{\mu}$, the masses $m^{\,}_{\mu}$, and the
coupling $\lambda$ are all real valued. The quantum fields obey
the Majorana equal-time anti-commutators
\begin{equation}
\left\{
\widehat{\chi}^{\mu}_{\mathrm{M},\mathtt{m}}(x),
\widehat{\chi}^{\mu'}_{\mathrm{M}',\mathtt{m}'}(x')
\right\}=
\delta^{\,}_{\mathrm{M}\mathrm{M}'}\,
\delta^{\,}_{\mathtt{m}\mathtt{m}'}\,	
\delta^{\,}_{\mu\mu'}\,
\delta(x-x'),
\end{equation}
\end{subequations}
where $\mu=0,1,2,3$ 
labels a quartet of Majorana fields,
$\mathrm{M}=\mathrm{L},\mathrm{R}$
denotes the left- and right-movers,
and  
$\mathtt{m},\mathtt{m}'=1,\cdots,n$
is the ladder index.
Hamiltonian (\ref{eq: desired Majorana theory}) has the following
symmetries.

First, the $\mu$-resolved fermion parity is conserved owing to the
symmetry of Hamiltonian (\ref{eq: desired Majorana theory})
under the Ising-like transformation
\begin{equation}
\widehat{\chi}^{\mu}_{\mathrm{M},\mathtt{m}}(x)\mapsto
\sigma^{\mu}\,
\widehat{\chi}^{\mu}_{\mathrm{M},\mathtt{m}}(x),
\qquad
\sigma^{\mu}=\pm1,
\label{eq: symmetries a}
\end{equation}
for any $\mu=0,\cdots,3$,
$\mathrm{M}=\mathrm{L},\mathrm{R}$,
$\mathtt{m}=1,\cdots,n$,
and
$0\leq x\leq L^{\,}_{x}$.

Second, Hamiltonian (\ref{eq: desired Majorana theory})
is invariant under the $\mathtt{m}$-resolved $\mathbb{Z}^{\,}_{2}$
transformation by which
\begin{equation}
\widehat{\chi}^{\mu}_{\mathrm{M},\mathtt{m}}(x)\mapsto
\sigma^{\,}_{\mathtt{m}}\,
\widehat{\chi}^{\mu}_{\mathrm{M},\mathtt{m}}(x),
\qquad
\sigma^{\,}_{\mathtt{m}}=\pm1,
\label{eq: symmetries c}
\end{equation}
for any $\mu=0,\cdots,3$,
$\mathrm{M}=\mathrm{L},\mathrm{R}$,
$\mathtt{m}=1,\cdots,n$,
and
$0\leq x\leq L^{\,}_{x}$.

We observe that
$\widehat{\mathcal{H}}^{\,}_{0}$
defined by Eq.\ (\ref{eq: desired Majorana theory b})
is $O(4)$ symmetric if $v^{\,}_{\mu}\equiv v$ is independent of $\mu$,
$\widehat{\mathcal{H}}^{\,}_{\mathrm{intra}-\mathrm{ladder}}$ 
defined by Eq.\ (\ref{eq: desired Majorana theory c})
is $O(4)$ symmetric if $m^{\,}_{\mu}\equiv m$ is independent of $\mu$,
and $\widehat{\mathcal{H}}^{\,}_{\mathrm{inter}-\mathrm{ladder}}$
defined by Eq.\ (\ref{eq: desired Majorana theory d})
is $O(4)$ symmetric.
This global $O(4)=\mathbb{Z}^{\,}_{2}\times SO(4)$
symmetry encodes the global
$\mathbb{Z}^{\,}_{2}\times SU(2)\times SU(2)$
symmetry of the microscopic inter-ladder interactions
depicted in Fig.\ \ref{Fig: 2Dlattice}, as will be explained
in more details in Secs.\ \ref{sec: A single two-leg ladder} and
\ref{sec: Coupled two-leg ladders}.

\begin{figure}[t]
\begin{center}
\includegraphics[width=0.5\textwidth]{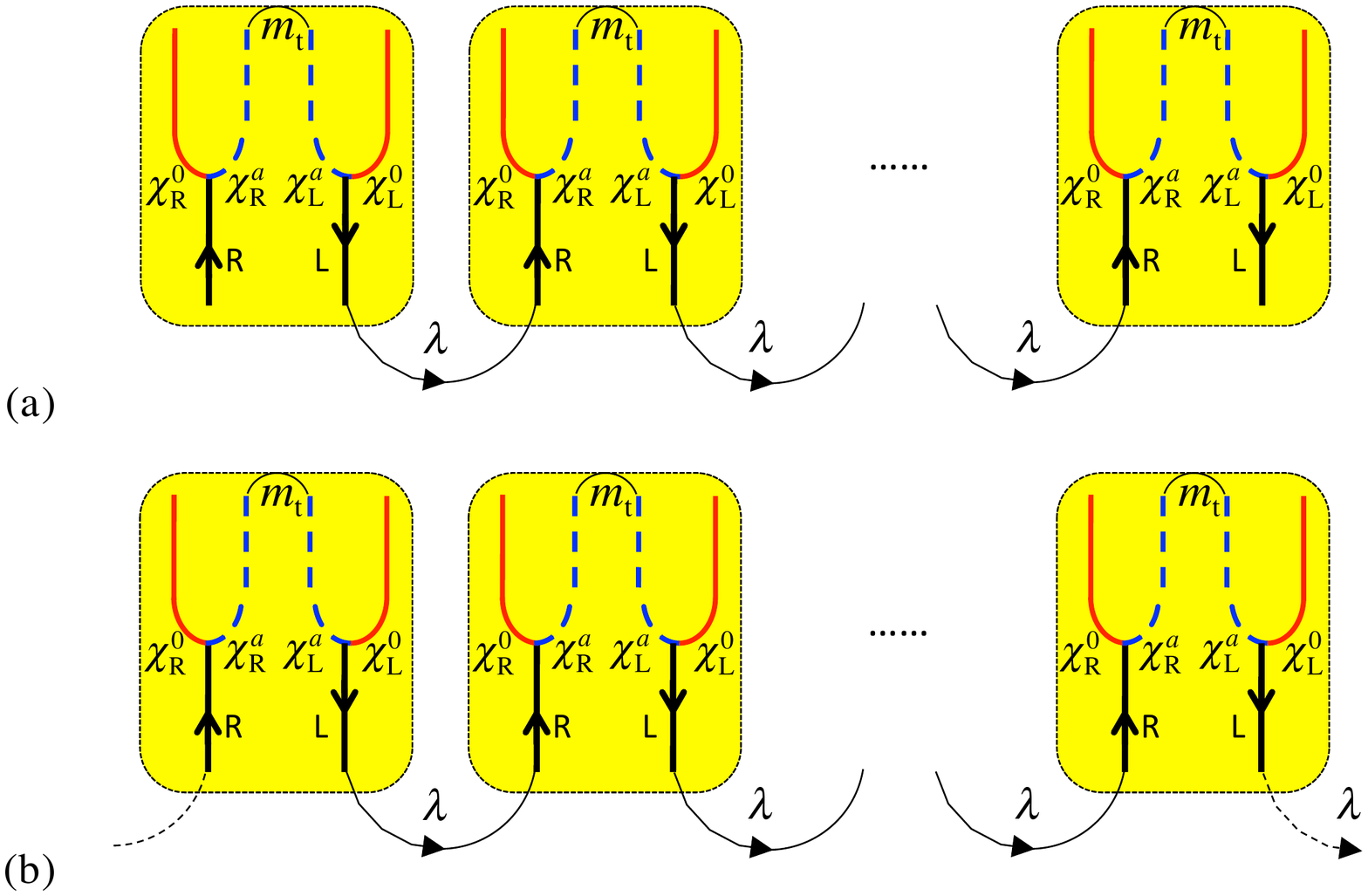}
\caption{(Color online)
(a)
A pictorial representation of the theory (\ref{eq: desired Majorana theory}) 
with $m^{\,}_{0}=0$ and
$m^{\,}_{a}=m^{\,}_{\mathrm{t}}$ for $a=1,2,3$
when open boundary conditions (OBC) are imposed along the $y$ direction.
(b)
A pictorial representation of the theory (\ref{eq: desired Majorana theory}) 
with $m^{\,}_{0}=0$ and
$m^{\,}_{a}=m^{\,}_{\mathrm{t}}$ for $a=1,2,3$
when periodic boundary conditions (PBC) are imposed along the $y$ direction.
\label{Fig: bundle}
         }
\end{center}
\end{figure}

\subsection{Limiting cases}
\label{subsec: Limiting cases}

In this subsection,
we consider the following limiting cases for the theory defined by
Eq.\ (\ref{eq: desired Majorana theory}) under the assumptions that
\begin{equation}
\begin{split}
&
v^{\,}_{0}\equiv v^{\,}_{\mathrm{s}},\qquad
m^{\,}_{0}\equiv m^{\,}_{\mathrm{s}}=0,
\\
&
v^{\,}_{a}\equiv v^{\,}_{\mathrm{t}},\qquad
m^{\,}_{a}\equiv m^{\,}_{\mathrm{t}},\qquad
a=1,2,3.
\label{eq: singlet and triplet anisotropy assumption}
\end{split}
\end{equation}
A cartoon picture of the theory (\ref{eq: desired Majorana theory}) 
with these assumptions
is depicted in Fig.~\ref{Fig: bundle}(a).	

\textbf{Case $\lambda=0$ and $m^{\,}_{\mathrm{t}}\neq0$.} 
There are $n$ gapless helical Majorana fields 
$\widehat{\chi}^{0}_{\mathrm{L},\mathtt{m}}$
and
$\widehat{\chi}^{0}_{\mathrm{R},\mathtt{m}}$ for $\mathtt{m}=1,\cdots,n$
that propagate in opposite directions in each bundle $\mathtt{m}$.
The two-dimensional system is critical in the singlet Majorana sector
where it realizes a sliding Luttinger phase.
This case corresponds to the vertical axis of the conjectured phase diagram
in Fig.\
\ref{Fig: phase diagram topological order a}.	

\textbf{Case $\lambda\neq0$ and $m^{\,}_{\mathrm{t}}=0$.} 
The Hamiltonian (\ref{eq: desired Majorana theory}) simplifies to
\begin{subequations}
\label{eq: desired Majorana theory mt=0}
\begin{align}
&
\widehat{\mathcal{H}}
\:=
\widehat{\mathcal{H}}^{\,}_{\mathrm{edge-states},\mathtt{m}=1}
+
\widehat{\mathcal{H}}^{\,}_{\mathrm{edge-states},\mathtt{m}=n}
+
\sum^{n-1}_{m=1}
\widehat{\mathcal{H}}^{\,}_{\mathrm{GN},\mathtt{m}},
\label{eq: desired Majorana theory mt=0 a}
\\
&
\widehat{\mathcal{H}}^{\,}_{\mathrm{edge-states},\mathtt{m}=1}\:=
\sum^{3}_{\mu=0}
\frac{\mathrm{i}}{2}
v^{\,}_{\mu}
\left(
-
\widehat{\chi}^{\mu}_{\mathrm{R},\mathtt{m}=1}
\partial^{\,}_{x}
\widehat{\chi}^{\mu}_{\mathrm{R},\mathtt{m}=1}
\right),
\label{eq: desired Majorana theory mt=0 b}
\\
&
\widehat{\mathcal{H}}^{\,}_{\mathrm{edge-states},\mathtt{m}=n}\:=
\sum^{3}_{\mu=0}
\frac{\mathrm{i}}{2}
v^{\,}_{\mu}
\left(
+
\widehat{\chi}^{\mu}_{\mathrm{L},\mathtt{m}=n}
\partial^{\,}_{x}
\widehat{\chi}^{\mu}_{\mathrm{L},\mathtt{m}=n}
\right),
\label{eq: desired Majorana theory mt=0 c}
\\
&
\widehat{\mathcal{H}}^{\,}_{\mathrm{GN},\mathtt{m}}\:=
\sum^{3}_{\mu=0}
\frac{\mathrm{i}}{2}
v^{\,}_{\mu}
\left(
\widehat{\chi}^{\mu}_{\mathrm{L},\mathtt{m}}
\partial^{\,}_{x}
\widehat{\chi}^{\mu}_{\mathrm{L},\mathtt{m}}
-
\widehat{\chi}^{\mu}_{\mathrm{R},\mathtt{m}+1}
\partial^{\,}_{x}
\widehat{\chi}^{\mu}_{\mathrm{R},\mathtt{m}+1}
\right)
\nonumber\\
&\hspace{4em}
+
\frac{\lambda}{4}
\left(
\sum^{3}_{\mu=0}
\widehat{\chi}^{\mu}_{\mathrm{L},\mathtt{m}}\,
\widehat{\chi}^{\mu}_{\mathrm{R},\mathtt{m}+1}\,
\right)^{2}.
\label{eq: desired Majorana theory mt=0 d}
\end{align}
\end{subequations}
Here,
the four-Majorana interaction
in Eq.\ (\ref{eq: desired Majorana theory mt=0 d})
is an $O(4)$-symmetric interaction of the Gross-Neveu type.

The $\mathtt{m}$-resolved symmetry (\ref{eq: symmetries c}) of
Hamiltonian (\ref{eq: desired Majorana theory mt=0})
is enhanced to the invariance under
the $\mathrm{M}$- and $\mathtt{m}$-resolved
$\mathbb{Z}^{\,}_{2}$ transformation
\begin{equation}
\begin{split}
&
\widehat{\chi}^{\mu}_{\mathrm{M},\mathtt{m}}(x)\mapsto
\sigma^{\,}_{\mathrm{M},\mathtt{m}}\,
\widehat{\chi}^{\mu}_{\mathrm{M},\mathtt{m}}(x),
\qquad
\sigma^{\,}_{\mathrm{M},\mathtt{m}}=\pm1,
\end{split}
\label{eq: symmetries c chiral}
\end{equation}
for any $\mu=0,\cdots,3$,
$\mathrm{M}=\mathrm{L},\mathrm{R}$,
$\mathtt{m}=1,\cdots,n$,
and
$0\leq x\leq L^{\,}_{x}$.
Indeed, whereas any transformation
(\ref{eq: symmetries c chiral})
changes
\begin{equation}
\widehat{\phi}^{\,}_{\mathrm{m},\mathrm{m}+1}\:=
\lambda\,  
\sum^{3}_{\mu=0}
\mathrm{i}
\widehat{\chi}^{\mu}_{\mathrm{L},\mathtt{m}}\,
\widehat{\chi}^{\mu}_{\mathrm{R},\mathtt{m}+1}
\label{eq: definition of bond order parameter}
\end{equation}
according to the rule
\begin{equation}
\widehat{\phi}^{\,}_{\mathrm{m},\mathrm{m}+1}\mapsto
\sigma^{\,}_{\mathrm{L},\mathtt{m}}\,
\sigma^{\,}_{\mathrm{R},\mathtt{m}+1}\,
\widehat{\phi}^{\,}_{\mathrm{m},\mathrm{m}+1},
\label{eq: symmetries c chiral on widehat phi}
\end{equation}
it leaves 
$\widehat{\phi}^{2}_{\mathrm{m},\mathrm{m}+1}$ unchanged.
Any one of these $\mathrm{M}$- and $\mathtt{m}$-resolved symmetries
obeying the conditions
$\sigma^{\,}_{\mathrm{L},\mathtt{m}}\,
\sigma^{\,}_{\mathrm{R},\mathtt{m}+1}=-1$
and either $\sigma^{\,}_{\mathrm{L},\mathtt{m}}=-\sigma^{\,}_{\mathrm{R},\mathtt{m}}$
or $\sigma^{\,}_{\mathrm{L},\mathtt{m}+1}=-\sigma^{\,}_{\mathrm{R},\mathtt{m}+1}$
for some $\mathtt{m}$
is broken if any one of the masses $m^{\,}_{\mu}$ is non-vanishing.
This enhanced symmetry relative to the symmetry
(\ref{eq: symmetries c})
reflects the fact that the limit with all masses
$m^{\,}_{\mu}$ vanishing is nothing but
$n$ decoupled Hamiltonians, each of which represents
a pair of interacting non-chiral Majorana fields evolving in
$(1+1)$-dimensional spacetime.

In this limit, Hamiltonian (\ref{eq: desired Majorana theory mt=0})
is known~\cite{Witten78} to be integrable and gapped (gapless)
when $v^{\,}_{\mu}=v$ with $\mu=0,1,2,3$ and
$\lambda>0$ $(\lambda<0)$. 
Thus, we should further distinguish between the following two cases.

\textit{Case $\lambda>0$.} 
The bulk is gapped with the four gapless chiral Majorana edge modes
$\widehat{\chi}^{\mu}_{\mathrm{R},\mathtt{m}=1}$
and
$\widehat{\chi}^{\mu}_{\mathrm{L},\mathtt{m}=n}$.
Since the chiral central charge of each edge is two,
the corresponding bulk hosts an ATO.
This case corresponds to the positive horizontal-axis 
(represented by the solid green line) of the conjectured phase diagram
in Fig.\
\ref{Fig: phase diagram topological order a}.	

\textit{Case $\lambda<0$.} 
The four pairs of gapless helical Majorana fields
$\widehat{\chi}^{\mu}_{\mathrm{L},\mathtt{m}}$
and
$\widehat{\chi}^{\mu}_{\mathrm{R},\mathtt{m}}$
with $\mu=0,1,2,3$
are freely propagating in each ladder $\mathtt{m}=1,\cdots,n$.
The two-dimensional bulk is critical 
and shares the same universality class as a sliding Luttinger phase.
This case corresponds to the negative horizontal-axis 
(represented by the solid red line) of the conjectured phase diagram
in Fig.\
\ref{Fig: phase diagram topological order a}.	

\textbf{Case $\lambda<0$ and $m^{\,}_{\mathrm{t}}\neq0$.} 
We conjecture that,
since the Gross-Neveu interaction with $\lambda<0$ is marginally irrelevant,
the resulting theory is the same as the case of 
$\lambda=0$ and $m^{\,}_{\mathrm{t}}\neq0$.  
This case corresponds to the blue colored region of
the conjectured phase diagram
in Fig.\
\ref{Fig: phase diagram topological order a}.	

\textbf{Case $\lambda>0$ and $m^{\,}_{\mathrm{t}}\neq0$.} 
We conjecture the competition between two phases,
an ATO phase and a NATO phase separated by a bulk gap closing transition
(represented by the dashed green lines in 
Fig.\
\ref{Fig: phase diagram topological order a}).	
This conjecture will be verified within a mean-field approximation.

\begin{figure}[t]
\begin{center}
\includegraphics[width=0.45\textwidth]{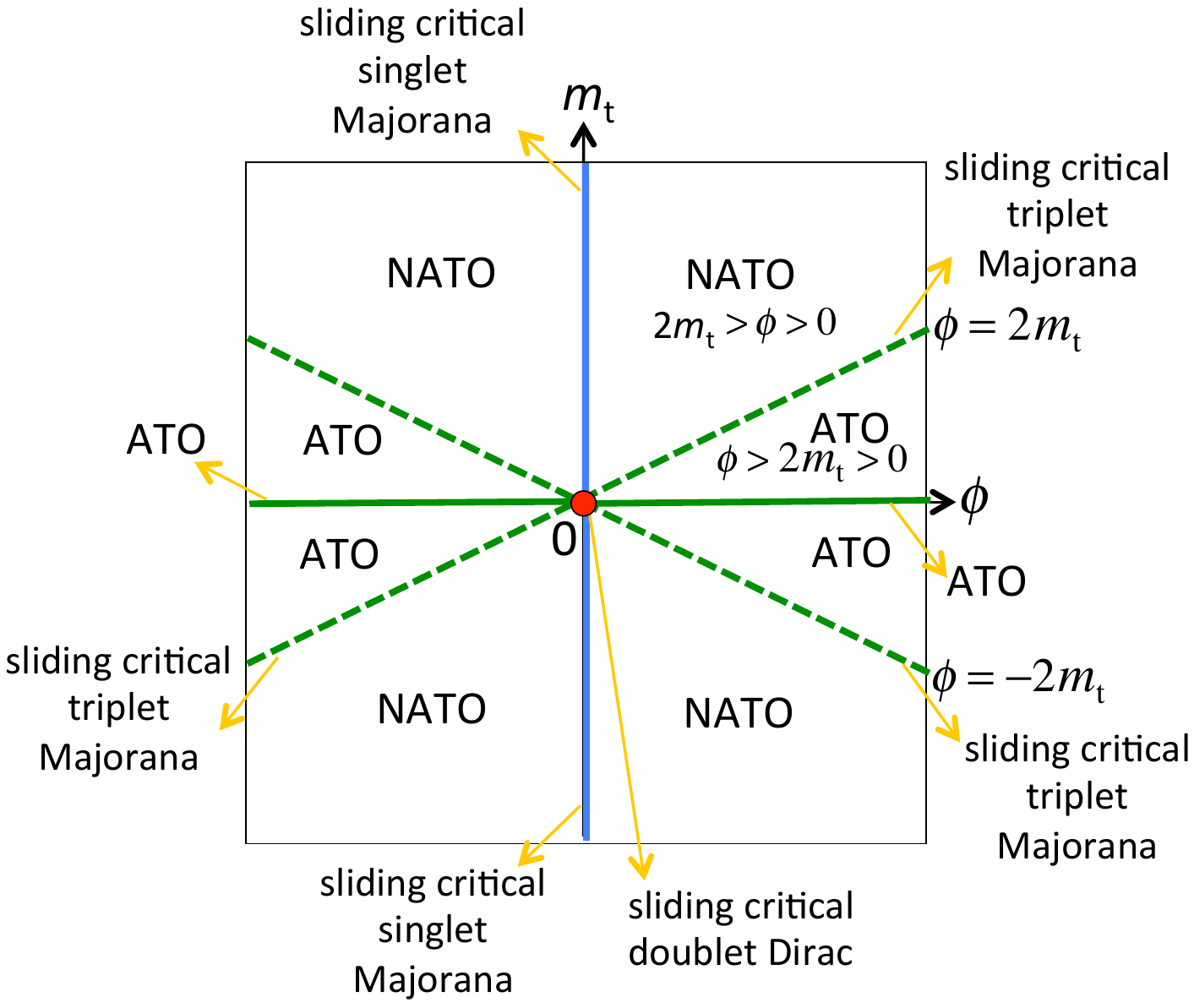}
\caption{(Color online)
The mean-field phase diagram
presented in terms of $m^{\,}_{\mathrm{t}}$ and the
mean-field value $\phi$ under the condition of $m^{\,}_{0}=0$
and $m^{\,}_{a}=m^{\,}_{\mathrm{t}}$ for $a=1,2,3.$
\label{Fig: phase diagram topological order b}
         }
\end{center}
\end{figure}

\section{Mean-field approximation}
\label{sec: The mean-field approach}

We are going to carry out a mean-field calculation from which 
we deduce the mean-field phase diagram
in Fig.\ 
\ref{Fig: phase diagram topological order b}.
Afterwards, we establish the conjectured phase diagram in
Fig.\
\ref{Fig: phase diagram topological order a}.

Our strategy does not rely on the $O(4)$ symmetry of the
Gross-Neveu-like interaction, it can generically be broken by
anisotropic singlet ($v^{\,}_{\mathrm{s}}$) and triplet
($v^{\,}_{\mathrm{t}}$) velocities, i.e.,
\begin{equation}
v^{\,}_{\mathrm{s}}
\neq
v^{\,}_{\mathrm{t}}.
\end{equation}

If we decouple the Gross-Neveu-like interaction in an $O(4)$-symmetric
way through a scalar field $\phi$, then a uniform and non-vanishing
expectation value for $\phi$ provides the singlet and triplet
Majoranas with an $O(4)$-symmetric mean-field mass.

In the process of solving the mean-field gap equation
(\ref{eq: the saddle-point equation final log ms neq 0}),	
we shall be primarily interested with the case $m^{\,}_{\mathrm{s}}=0$
for which the decoupled ladders are fine-tuned to an Ising critical
point.  In Fig.\ \ref{Fig: phase diagram topological order a}, we
identify the regions from the $\lambda-m^{\,}_{\mathrm{t}}$ plane for
which the mean-field single-particle singlet gap is non-vanishing when
periodic boundary conditions (PBC) are imposed.

\subsection{Hubbard-Stratonovich transformation}
\label{subsec: Hubbard-Stratonovich transformation}

We proceed with some manipulations on the partition function
\begin{equation}
Z\:=
\mathrm{Tr}\,
\exp
\left(
-\beta\,\int\limits_{0}^{L^{\,}_{x}}\mathrm{d}x\,\widehat{\mathcal{H}}
\right),
\end{equation}
where $\beta$ is the inverse temperature, the trace
is over the Fock space spanned by the Majorana fields,
and $\widehat{\mathcal{H}}$ was defined in Eq.\
(\ref{eq: desired Majorana theory}).
We can manipulate the inter-ladder current-current interaction
(\ref{eq: desired Majorana theory d})
by introducing an auxiliary scalar field.
This we do using the path-integral representation
of the partition function.

We work in two-dimensional
Euclidean spacetime and use the path-integral representation
of our model. Periodic boundary conditions are imposed in space across the
rectangle of area $L^{\,}_{x}\times L^{\,}_{y}$.
We shall denote by $\mathfrak{a}^{\,}_{y}\equiv1/\Lambda^{\,}_{y}$ the separation
between two consecutive ladders. We shall denote by
$\mathfrak{a}^{\,}_{x}\equiv1/\Lambda^{\,}_{x}$
the ultraviolet cutoff along the ladders.
The boundary conditions along the imaginary-time segment $[0,\beta[$
are periodic for bosonic fields and antiperiodic for Grassmann-valued fields.
The model is defined by
\begin{subequations}
\label{eq: def partition fct with Hubbard-Stratonovich decoupling}
\begin{align}
&
Z\:=
\int\mathcal{D}[\phi]
\int\mathcal{D}[\chi^{0},\chi^{1},\chi^{2},\chi^{3}]\,
e^{-S},
\label{eq: def partition fct with Hubbard-Stratonovich decoupling a}
\\
&
S\:=
\int\limits_{0}^{\beta}\mathrm{d}\tau\,
\int\limits_{0}^{L^{\,}_{x}}\mathrm{d}x\,
\sum_{\mathtt{m}=1}^{L^{\,}_{y}/\mathfrak{a}^{\,}_{y}}
\left(
\mathcal{L}^{\,}_{\chi,\mathtt{m}}
+
\mathcal{L}^{\,}_{\phi,\mathtt{m}}
+
\mathcal{L}^{\,}_{\chi,\phi,\mathtt{m}}
\right),
\label{eq: def partition fct with Hubbard-Stratonovich decoupling b}
\\
&
\mathcal{L}^{\,}_{\chi,\mathtt{m}}\:=
\frac{1}{2}
\sum^{3}_{\mu=0}
\Big[
\chi^{\mu}_{\mathrm{L},\mathtt{m}}
\left(\partial^{\,}_{\tau}+\mathrm{i}v^{\,}_{\mu}\partial^{\,}_{x}\right)
\chi^{\mu}_{\mathrm{L},\mathtt{m}}
\nonumber\\
&\hspace{6em}
+
\chi^{\mu}_{\mathrm{R},\mathtt{m}}
\left(\partial^{\,}_{\tau}-\mathrm{i}v^{\,}_{\mu}\partial^{\,}_{x}\right)
\chi^{\mu}_{\mathrm{R},\mathtt{m}}
\Big],
\label{eq: def partition fct with Hubbard-Stratonovich decoupling c}
\\
&
\mathcal{L}^{\,}_{\phi,\mathtt{m}}\:=
\frac{1}{4\,\lambda}(\phi^{\,}_{\mathtt{m},\mathtt{m}+1})^{2},
\label{eq: def partition fct with Hubbard-Stratonovich decoupling d}
\\
&
\mathcal{L}^{\,}_{\chi,\phi,\mathtt{m}}\:=
\sum_{\mu=0}^{3}
\mathrm{i}\,
m^{\,}_{\mu}
\chi^{\mu}_{\mathrm{L},\mathtt{m}}\,
\chi^{\mu}_{\mathrm{R},\mathtt{m}}
\nonumber\\
&
\hphantom{
\mathcal{L}^{\,}_{\chi,\phi,\mathtt{m}}\:=
         }
+
\sum_{\mu=0}^{3}
\frac{1}{2}
\left(
-\mathrm{i}\chi^{\mu}_{\mathrm{L},\mathtt{m}}\,\chi^{\mu}_{\mathrm{R},\mathtt{m}+1}
\right)
\phi^{\,}_{\mathtt{m},\mathtt{m}+1}.
\label{eq: def partition fct with Hubbard-Stratonovich decoupling e}
\end{align}
\end{subequations}
Here, the engineering dimensions of the Majoranas are
$\mathrm{length}^{-1/2}$,
the engineering dimensions of the auxiliary bosonic fields are
$\mathrm{length}^{-1}$,
and the engineering dimensions of the couplings $\lambda$
are
$\mathrm{length}^{0}$. 
The action
(\ref{eq: def partition fct with Hubbard-Stratonovich decoupling b})
has the following
symmetries.

First, the $\mu$-resolved Majorana
parity is conserved owing to the
symmetry of $S$
defined in Eq.\
(\ref{eq: def partition fct with Hubbard-Stratonovich decoupling b})
under the Ising-like transformation
\begin{equation}
\chi^{\mu}_{\mathrm{M},\mathtt{m}}(\tau,x)\mapsto
\sigma^{\mu}\,
\chi^{\mu}_{\mathrm{M},\mathtt{m}}(\tau,x),
\qquad
\sigma^{\mu}=\pm1,
\label{eq: symmetries Majorana Grassmann a}
\end{equation}
for any $\mu=0,\cdots,3$,
$\mathrm{M}=\mathrm{L},\mathrm{R}$,
$\mathtt{m}=1,\cdots,n$,
$0\leq\tau\leq\beta$,
and
$0\leq x\leq L^{\,}_{x}$.

Second, action
(\ref{eq: def partition fct with Hubbard-Stratonovich decoupling b})
is invariant under the $\mathtt{m}$-resolved Ising-like transformation
\begin{equation}
\begin{split}
&
\chi^{\mu}_{\mathrm{M},\mathtt{m}}(\tau,x)\mapsto
\sigma^{\,}_{\mathtt{m}}\,
\chi^{\mu}_{\mathrm{M},\mathtt{m}}(\tau,x),
\qquad
\sigma^{\,}_{\mathtt{m}}=\pm1,
\\
&
\phi^{\,}_{\mathtt{m},\mathtt{m}+1}(\tau,x)\mapsto
\sigma^{\,}_{\mathtt{m}}\,
\sigma^{\,}_{\mathtt{m}+1}\,
\phi^{\,}_{\mathtt{m},\mathtt{m}+1}(\tau,x),
\end{split}
\label{eq: symmetries Majorana Grassmann c}
\end{equation}
for any $\mu=0,\cdots,3$,
$\mathrm{M}=\mathrm{L},\mathrm{R}$,
$\mathtt{m}=1,\cdots,n$,
$0\leq\tau\leq\beta$,
and
$0\leq x\leq L^{\,}_{x}$.

The $\mathtt{m}$-resolved symmetry (\ref{eq: symmetries Majorana Grassmann c})
of the action
(\ref{eq: def partition fct with Hubbard-Stratonovich decoupling b})
is enhanced in the massless limit $m^{\,}_{\mu}=0$ for $\mu=0,1,2,3$
to the $\mathrm{M}$- and $\mathtt{m}$-resolved symmetry under the transformation
\begin{equation}
\begin{split}
&
\chi^{\mu}_{\mathrm{M},\mathtt{m}}(\tau,x)\mapsto
\sigma^{\,}_{\mathrm{M},\mathtt{m}}\,
\chi^{\mu}_{\mathrm{M},\mathtt{m}}(\tau,x),
\qquad
\sigma^{\,}_{\mathrm{M},\mathtt{m}}=\pm1,
\\
&
\phi^{\,}_{\mathtt{m},\mathtt{m}+1}(\tau,x)\mapsto
\sigma^{\,}_{\mathrm{L},\mathtt{m}}\,
\sigma^{\,}_{\mathrm{R},\mathtt{m}+1}\,
\phi^{\,}_{\mathtt{m},\mathtt{m}+1}(\tau,x),
\end{split}
\label{eq: symmetries Majorana Grassmann c chiral}
\end{equation}
for any $\mu=0,\cdots,3$,
$\mathrm{M}=\mathrm{L},\mathrm{R}$,
$\mathtt{m}=1,\cdots,n$,
$0\leq\tau\leq\beta$,
and
$0\leq x\leq L^{\,}_{x}$.
Any non-vanishing mass $m^{\,}_{\mu}$
reduces the $\mathrm{M}$- and $\mathtt{m}$-resolved symmetry of the action
(\ref{eq: def partition fct with Hubbard-Stratonovich decoupling b})
to the $\mathtt{m}$-resolved symmetry
(\ref{eq: symmetries Majorana Grassmann c}).

\subsection{Mean-field Majorana single-particle Hamiltonian}
\label{subsec: The mean-field Majorana single particle Hamiltonian}

To proceed, we assume that the scalar fields
are independent of spacetime and of the index $\mathtt{m}$, i.e.,
\begin{equation}
\phi^{\,}_{\mathtt{m},\mathtt{m}+1}(\tau,x)\equiv \phi,
\qquad
S^{\,}_{\phi}=
\beta\,L^{\,}_{x}\,\frac{L^{\,}_{y}}{\mathfrak{a}^{\,}_{y}}
\frac{1}{4\lambda}\phi^{2}.
\end{equation}
This assumption implies translation symmetry in spacetime.
Hence, we introduce the Fourier transformations 
\begin{subequations}
\begin{equation}
\begin{split}
\chi^{\mu}_{\mathrm{M},\mathtt{m}}(\tau,x)=&\, 
\sqrt{\frac{\mathfrak{a}^{\,}_{y}}{\beta\,L^{\,}_{x}\,L^{\,}_{y}}}\!
\sum_{\omega,k^{\,}_{x},k^{\,}_{y}}\!
e^{-\mathrm{i}(k^{\,}_{x}\,x+k^{\,}_{y}\,\mathtt{m}\,\mathfrak{a}_{y}-\omega\,\tau)}
\chi^{\mu}_{\mathrm{M},\omega,\bm{k}}
\end{split}
\label{eq: Fourier transformation m direction a}
\end{equation}
with the reality condition
\begin{equation}
\chi^{\mu*}_{\mathrm{M},\omega,\bm{k}}=\chi^{\mu}_{\mathrm{M},-\omega,-\bm{k}}
\label{eq: Fourier transformation m direction b}
\end{equation}
\end{subequations}
for
$\mu=0,1,2,3$,
$\mathrm{M}=\mathrm{L},\mathrm{R}$,
and
$\mathtt{m}=1,\cdots,L^{\,}_{y}/\mathfrak{a}^{\,}_{y}$.
We shall make use of the identity
\begin{equation}
\begin{split}
&
\int\limits_{0}^{\beta}\mathrm{d}\tau
\int\limits_{0}^{L^{\,}_{x}}\mathrm{d}x\,
\sum_{\mathtt{m}=1}^{L^{\,}_{y}/\mathfrak{a}^{\,}_{y}}\,
\chi^{\mu}_{\mathrm{L},\mathtt{m}}\,
\chi^{\mu}_{\mathrm{R},\mathtt{m}+1} 
=
\\
&\qquad\qquad\qquad\qquad
\sum^{\,}_{\omega,\bm{k}}\,
e^{-\mathrm{i}k^{\,}_{y}\,\mathfrak{a}^{\,}_{y}}\,
\chi^{\mu}_{\mathrm{L},-\omega,-\bm{k}}\,
\chi^{\mu}_{\mathrm{R},\omega,\bm{k}}
\end{split}
\end{equation}
for any $\mu=0,1,2,3$.
We should emphasize that we have imposed periodic boundary condition along the
$y$-direction
\begin{equation}
\chi^{\mu}_{\mathrm{M},n+1}\equiv\chi^{\mu}_{\mathrm{M},1},
\qquad \mathrm{M}=\mathrm{L},\mathrm{R},
\end{equation}
when we perform the Fourier transformation.
This amounts to extending the upper limit for the summation
from $n-1$ to $n$ in the original inter-ladder Hamiltonian
(\ref{eq: desired Majorana theory d}).
This choice of boundary conditions is depicted in Fig.\ \ref{Fig: bundle}(b).
If so,
\begin{subequations}
\begin{widetext}
\begin{align}
S^{\,}_{\chi}
+
S^{\,}_{\chi,\phi}\equiv&\,
\int\limits_{0}^{\beta}\mathrm{d}\tau
\int\limits_{0}^{L^{\,}_{x}}\mathrm{d}x\,
\sum_{\mathtt{m}=1}^{L^{\,}_{y}/\mathfrak{a}^{\,}_{y}}\,
\left(
\mathcal{L}^{\,}_{\chi,\mathtt{m}}
+
\mathcal{L}^{\,}_{\chi,\phi,\mathtt{m}}
\right)
\nonumber\\
=&\,
\sum^{\,}_{\omega,\bm{k}}\, 
\sum^{3}_{\mu=0}
\frac{1}{2}
\begin{pmatrix}
\chi^{\mu}_{\mathrm{R},-\omega,-\bm{k}}
\chi^{\mu}_{\mathrm{L},-\omega,-\bm{k}}
\end{pmatrix}
\begin{pmatrix}
\mathrm{i}\omega-v^{\,}_{\mu}k^{\,}_{x} 
&
-
\mathrm{i}
\left(
m^{\,}_{\mu}
-
e^{+\mathrm{i}k^{\,}_{y}\,\mathfrak{a}^{\,}_{y}}\,
\frac{1}{2}\phi
\right)
\\
\mathrm{i}
\left(
m^{\,}_{\mu}
-
e^{-\mathrm{i}k^{\,}_{y}\,\mathfrak{a}^{\,}_{y}}\,\,
\frac{1}{2}\phi
\right)
&
\mathrm{i}\omega+v^{\,}_{\mu}k^{\,}_{x}
\end{pmatrix}\,
\begin{pmatrix}
\chi^{\mu}_{\mathrm{R},\omega,\bm{k}}
\\
\chi^{\mu}_{\mathrm{L},\omega,\bm{k}}
\end{pmatrix}.
\end{align}
\end{widetext}
The mean-field Majorana single-particle Hamiltonian is defined by
\begin{equation}
\widehat{H}^{\mathrm{MF}}_{\bm{k}}\:=
\sum^{3}_{\mu=0}
\widehat{H}^{\mathrm{MF}}_{\mu,\bm{k}},
\end{equation}
where
\begin{align}
\widehat{H}^{\mathrm{MF}}_{\mu,\bm{k}}\:=&\,
\frac{1}{2}
\begin{pmatrix}
-v^{\,}_{\mu}k^{\,}_{x}
&
-\mathrm{i}
\left(
m^{\,}_{\mu}
-
e^{+\mathrm{i}k^{\,}_{y}\mathfrak{a}^{\,}_{y}}\frac{\phi}{2}
\right)
\\
\mathrm{i}
\left(
m^{\,}_{\mu}
-
e^{-\mathrm{i}k^{\,}_{y}\mathfrak{a}^{\,}_{y}}\frac{\phi}{2}
\right)
&
v^{\,}_{\mu}k^{\,}_{x}
\end{pmatrix}.
\label{eq: mean-field Majorana single-particle Hamiltonian}
\end{align}
\end{subequations}

Thus, there are eight branches of mean-field excitations with the dispersions
(under the assumption that $\phi$ is real valued)
\begin{equation}
\varepsilon^{\,}_{\mu,\pm}(k^{\,}_{x},k^{\,}_{y})=
\pm
\frac{1}{2}
\sqrt{
v^{2}_{\mu}k^{2}_{x}
+
m^{2}_{\mu}
+
\frac{\phi^{2}}{4}
-
m^{\,}_{\mu}\phi
\cos\left(k^{\,}_{y}\mathfrak{a}^{\,}_{y}\right)
	},
\end{equation}
for $\mu=0,\cdots,3.$
The mean-field gaps are non-vanishing if and only if
\begin{equation}
m^{\,}_{\mu}
-
e^{\pm\mathrm{i}k^{\,}_{y}\,\mathfrak{a}^{\,}_{y}}\,\frac{\phi}{2}
\neq0.
\end{equation}
More specifically,
the mean-field Majorana gap around $(k^{\,}_{x}=0,k^{\,}_{y}=0)$ and $(k^{\,}_{x}=0,k^{\,}_{y}=\pi)$ 
are, for $\mu=0,\cdots,3,$
\begin{subequations}
\begin{align}
\varepsilon^{\,}_{\mu,+}(0,0)
-
\varepsilon^{\,}_{\mu,-}(0,0)=
\left|m^{\,}_{\mu}-\frac{\phi}{2}\right|,
\label{eq: gap function a}
\end{align}
and
\begin{align}
\varepsilon^{\,}_{\mu,+}(0,\pi)
-
\varepsilon^{\,}_{\mu,-}(0,\pi)=
\left|m^{\,}_{\mu}+\frac{\phi}{2}\right|,
\label{eq: gap function b}
\end{align}
\end{subequations}
respectively. 
The mean-field gap (\ref{eq: gap function a}) and (\ref{eq: gap function b}) 
is the smallest gap when $\mathrm{sgn}(m^{\,}_{\mu}\phi)=+$
and $\mathrm{sgn}(m^{\,}_{\mu}\phi)=-$, respectively.

\begin{figure}[t]
\begin{center}
(a)
\includegraphics[width=0.45\textwidth]{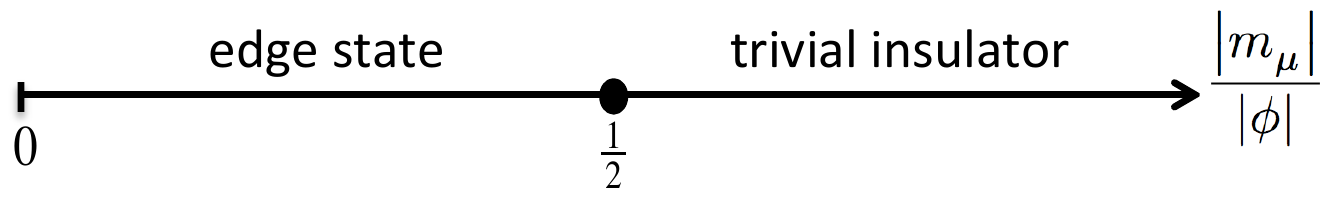}
\\
(b)
\includegraphics[width=0.45\textwidth]{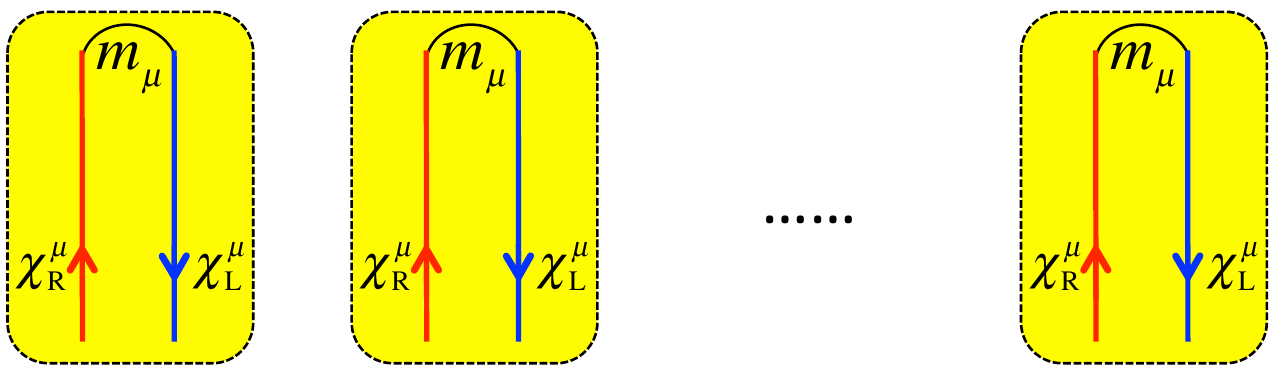}
\\
(c)
\includegraphics[width=0.45\textwidth]{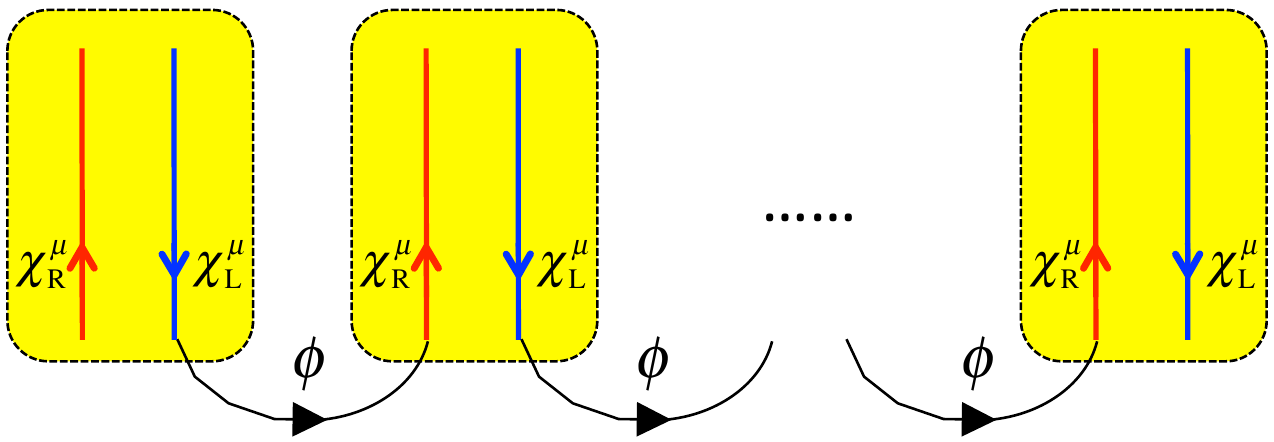}
\caption{(Color online)
  (a)
Flavor($\mu$)-resolved
phase diagram for the single-particle mean-field Hamiltonian
as a function of $|m^{\,}_{\mu}|/|\phi|$.
The topologically trivial insulating phase in the limit
$|m^{\,}_{\mu}|/|\phi|=\infty$ 
is depicted in panel (b).
The topologically non-trivial insulating phase in the limit
$|m^{\,}_{\mu}|/|\phi|=0$ 
is depicted in panel (c).
A phase transition should occur when $|m^{\,}_{\mu}|/|\phi|$ is of order 1/2.
(b)
When $\phi=0$ and $m^{\,}_{\mu}\neq0$,
the single-particle mean-field Hamiltonian is gapped
by pairing left- and right-moving Majorana modes in one ladder at a time.
By construction there is no edge state.
This is the topologically trivial insulator.
(c)
When $m^{\,}_{\mu}=0$ and $\phi\neq0$, not all Majorana
modes are paired.  A pair of Majorana modes with opposite chiralities
remains free to propagate in the first and last ladder.
This is the topologically non-trivial insulator.
\label{Fig: topological criteria single flavor}
         }
\end{center}
\end{figure}

For any non-vanishing mean-field Majorana gap $\Delta^{\,}_{\mu}$
\begin{equation}
\Delta^{\,}_{\mu}\:=
\left||m^{\,}_{\mu}|-\frac{|\phi|}{2}\right|,
\label{eq: gap function}
\end{equation}
the flavor $\mu$ realizes an insulating phase.
Whether this insulating phase is trivial
(no protected edge state when OBC are imposed
along the $y$-direction)
or non-trivial
(existence of protected edge states when OBC are imposed
along the $y$-direction)
depends on the relative magnitude of
$|m^{\,}_{\mu}|$ with respect to the mean-field value $|\phi|/2$.
The flavor $\mu$ realizes a topologically trivial insulating phase if 
\begin{subequations}
\label{eq: topologial criteria}
\begin{equation}
|m^{\,}_{\mu}|>\frac{|\phi|}{2},
\label{eq: topologial criteria a}
\end{equation}
while it realizes a topologically non-trivial insulating phase if
\begin{equation}
|m^{\,}_{\mu}|<\frac{|\phi|}{2}.
\label{eq: topologial criteria b}
\end{equation}
\end{subequations}
The criteria (\ref{eq: topologial criteria})
for the topological non-trivial and trivial phases
can be understood as follows.
In the limit $|m^{\,}_{\mu}|/|\phi|=\infty$, 
the single-particle mean-field Hamiltonian is gapped
by pairing left- and right-moving Majorana modes in one ladder at a time.
By construction there is no edge state.
This is the topologically trivial insulator.
In the opposite limit of $|m^{\,}_{\mu}|/|\phi|=0$, not all Majorana
modes are paired. 
A pair of Majorana modes with opposite chiralities
remains free to propagate in the first and last ladder.
A phase transition should occur when $|m^{\,}_{\mu}|/|\phi|$ is of order 1/2.
Figure \ref{Fig: topological criteria single flavor}
captures the essence of this criterion.

Once it is established that $\phi$ is non-vanishing,
the resulting central charge of the edge states depends on
how many $m^{\,}_{\mu}$ for $\mu=0,1,2,3$
satisfy the topologically non-trivial condition
(\ref{eq: topologial criteria b}).
For instance, if one (three) out of the four $m^{\,}_{\mu}$
satisfies Eq.\ (\ref{eq: topologial criteria b}),
then the central charge of the edge state is $1/2\,(3/2)$.
We conclude that the gapped bulk hosts NATO.
Similarly, if two (four) out of the four $m^{\,}_{\mu}$
satisfy Eq.\ (\ref{eq: topologial criteria b}),
then the central charge of the edge states is $2\,(4)$.
We conclude that the gapped bulk hosts ATO.

We close this discussion by observing
that the mean-field single-particle Hamiltonian
(\ref{eq: mean-field Majorana single-particle Hamiltonian}) 
was studied recently by Kane \textit{et.~al.}
in Ref.\ \onlinecite{Kane17} [see their Eq.\ (58)]
from a different perspective, namely that of
a wire construction for paired states in the FQHE
at an even-denominator filling fraction $\nu$,
say $\nu=1/2$.

\subsection{Integrating out the Majorana fields}
\label{subsec: Integrating out the Majorana fields}

In what follows, we only consider the case 
$v^{\,}_{0}\equiv v^{\,}_{\mathrm{s}}$,
$m^{\,}_{0}\equiv m^{\,}_{\mathrm{s}}$,
$v^{\,}_{a}\equiv v^{\,}_{\mathrm{t}}$,
and 
$m^{\,}_{a}\equiv m^{\,}_{\mathrm{t}}$ for any $a=1,2,3.$	
The extension to the case of arbitrary values for 
$v^{\,}_{\mu}$ and $m^{\,}_{\mu}$
does not present major difficulties.

Integration over the Majorana fields delivers the product of two Pfaffians.
There is a Pfaffian
that arises from integrating over the singlet $\chi^{0}$'s,
and another Pfaffian
that arises from integrating over the triplet $\chi^{a}$'s.
It follows that
\begin{subequations}
\label{eq: define effective action}
\begin{align}
&
Z\propto
\int\mathcal{D}[\phi]\,
e^{-S'},
\\
&
S'\:=
S^{\,}_{\phi}
+
S^{\,}_{\mathrm{F}},
\\
&
S^{\,}_{\phi}\:=
\frac{\beta\,L^{\,}_{x}\,L^{\,}_{y}}{\mathfrak{a}^{\,}_{y}}
\frac{\phi^{2}}{4\lambda},
\\
&
S^{\,}_{\mathrm{F}}\:= 
\nonumber\\
&\,
-
\frac{1}{2}
\sum^{\,}_{\omega,\bm{k}}
\left[
\log
\left(
-
\omega^{2}
-
v^{2}_{\mathrm{s}}k^{2}_{x}
-
m^{2}_{\mathrm{s}}
-
\frac{\phi^{2}}{4}
+
m^{\,}_{\mathrm{s}}\phi
\cos\frac{k^{\,}_{y}}{\Lambda^{\,}_{y}}
\right)
\right.
\nonumber\\
&\,
\left.
+3
\log
\left(
-
\omega^{2}
-
v^{2}_{\mathrm{t}}k^{2}_{x}
-
m^{2}_{\mathrm{t}}
-
\frac{\phi^{2}}{4}
+
m^{\,}_{\mathrm{t}}\phi
\cos\frac{k^{\,}_{y}}{\Lambda^{\,}_{y}}
\right)
\right].
\label{eq: define effective action d}
\end{align}
Here, we have introduce the momentum cutoff
\begin{equation}
\Lambda^{\,}_{y}\:=\frac{1}{\mathfrak{a}^{\,}_{y}}.
\end{equation}
\end{subequations}
The action $S^{\prime}$ controls the global symmetries
of the theory. It is invariant under the global Ising-like
($\mathbb{Z}^{\,}_{2}$)
transformation defined by
\begin{subequations}
\begin{equation}
\phi\mapsto-\phi
\end{equation}
if we compensate this change of sign with the change of variable
\begin{equation}
k^{\,}_{y}
\,\mapsto k^{\,}_{y}+\pi\Lambda^{\,}_{y}
\end{equation}
\end{subequations}
in the summation over $k^{\,}_{y}$.

\subsection{Mean-field gap equations}
\label{subsec: Mean-field gap equations}

The saddle-point equation
\begin{equation}
0\equiv
\frac{\mathfrak{a}^{\,}_{y}}{\beta\,L^{\,}_{x}\,L^{\,}_{y}}\,
\frac{\partial S'}{\partial\phi}
\end{equation}
are then explicitly given by
\begin{widetext}
\begin{align}
&
0=
\frac{1}{2\,\lambda}
\phi
-
\frac{\mathfrak{a}^{\,}_{y}}{\beta\,L^{\,}_{x}\,L^{\,}_{y}}\,
\sum^{\,}_{\omega,\bm{k}}
\left[
\frac{1}{2}
\frac{
\frac{1}{2}\phi
-m^{\,}_{\mathrm{s}}\,\cos\frac{k^{\,}_{y}}{\Lambda^{\,}_{y}}
     }
     {
\omega^{2}
+v^{2}_{\mathrm{s}}k^{2}_{x}
+m^{2}_{\mathrm{s}}
+\frac{1}{4}\phi^{2}
-m^{\,}_{\mathrm{s}}\,\phi\,
\cos\frac{k^{\,}_{y}}{\Lambda^{\,}_{y}}
     }
+
\frac{3}{2}
\frac{
\frac{1}{2}\phi
-m^{\,}_{\mathrm{t}}\,\cos\frac{k^{\,}_{y}}{\Lambda^{\,}_{y}}
     }
     {
\omega^{2}
+v^{2}_{\mathrm{t}}k^{2}_{x}
+m^{2}_{\mathrm{t}}
+\frac{1}{4}\phi^{2}
-m^{\,}_{\mathrm{t}}\,\phi\,
\cos\frac{k^{\,}_{y}}{\Lambda^{\,}_{y}}
     }
\right].
\label{eq: action first order differential}
\end{align}
\end{widetext}
We observe that Eq.\
(\ref{eq: action first order differential})
is invariant under 
\begin{equation}
\begin{split}
&
\phi
\,\mapsto -\phi,
\\
&
m^{\,}_{\mathrm{s}}
\,\mapsto -m^{\,}_{\mathrm{s}},
\\
&
m^{\,}_{\mathrm{t}}
\,\mapsto -m^{\,}_{\mathrm{t}}.
\end{split}
\end{equation}
It is also invariant under
\begin{subequations}
\begin{equation}
m^{\,}_{\mathrm{s}}\mapsto
-m^{\,}_{\mathrm{s}},
\qquad
m^{\,}_{\mathrm{t}}
\mapsto
-m^{\,}_{\mathrm{t}},
\end{equation}
if we compensate this change of sign with the change of variable
\begin{equation}
k^{\,}_{y}
\,\mapsto k^{\,}_{y}+\pi\Lambda^{\,}_{y}
\end{equation}
\end{subequations}
in the summation over $k^{\,}_{y}$.
The same is true of the partition function defined in Eq.\
(\ref{eq: define effective action}).
  
In the limit $\beta\to\infty$, $L^{\,}_{x}\to\infty$, and $L^{\,}_{y}\to\infty$
(zero temperature and thermodynamic limit), the sums become integrals
in three-dimensional spacetime. 
Power counting predicts that those momentum integrals
are logarithmically divergent in the ultraviolet. A momentum cutoff is thus
needed to evaluate those integrals.
It is chosen to be
$|k^{\,}_{x}|\leq\pi\Lambda^{\,}_{x}$
and
$|k^{\,}_{y}|\leq\pi\Lambda^{\,}_{y}$.
All integrals over the Matsubara frequencies are performed before the
momentum integrals by application of the Residue theorem.
To this end, the identity
\begin{equation}
\int\limits_{-\infty}^{+\infty}\frac{\mathrm{d}\omega}{2\pi}\,
\frac{
a^{2}
     }
     {
\omega^{2}+b^{2}
     }
=
\frac{1}{2}
\frac{
a^{2}
     }
     {
\sqrt{b^{2}}
     },
\qquad a,b\in\mathbb{R},
\label{eq: integral over omega}
\end{equation}
is used.
The remaining integral over $k^{\,}_{x}$ is of the form
\begin{equation}
\int\limits_{0}^{b}\mathrm{d}x\,\frac{1}{\sqrt{x^{2}+a^{2}}}=
\mathrm{arcsinh}\left(\frac{b}{a}\right),
\qquad
0<a,b.
\label{eq: integral over kx}
\end{equation}
Finally, the remaining integral over $k^{\,}_{y}$ can be simplified by
changing variable
\begin{equation}
q\:=\frac{k^{\,}_{y}}{\Lambda^{\,}_{y}}.
\end{equation}

In summary, the saddle-point equation has become the single integral
\begin{widetext}
\begin{align}
0=&\, 
\frac{\phi}{2\pi\Lambda^{\,}_{x}}
-
\frac{\lambda}{4\pi}
\left[
\frac{1}{v^{\,}_{\mathrm{s}}}
\int\limits_{-\pi}^{+\pi}\frac{\mathrm{d}q}{2\pi}
\left(
\frac{\phi}{2\pi\Lambda^{\,}_{x}}
-
\frac{2m^{\,}_{\mathrm{s}}}{2\pi\Lambda^{\,}_{x}}
\,\cos q
\right)
\times
\mathrm{arcsinh}
\left(
\frac{
	2\pi\Lambda^{\,}_{x}v^{\,}_{\mathrm{s}}
	}
	{
\sqrt{
4m^{2}_{\mathrm{s}}
+\phi^{2}
-4m^{\,}_{\mathrm{s}}\,\phi\,
\cos q
     }
    }
    \right)
\right]
-
\frac{3\lambda}{4\pi}
\left[
\mathrm{s}\to\,\mathrm{t}
\right].
\label{eq: the saddle-point equation final ms neq 0}
\end{align}
\end{widetext}
Equation\ (\ref{eq: the saddle-point equation final ms neq 0})
is a non-linear equation for one unknown $\frac{\phi}{2\pi\Lambda^{\,}_{x}}$. 
It can only be solved numerically for arbitrary value of $\lambda$,
$\frac{m^{\,}_{\mathrm{t}}}{2\pi\Lambda^{\,}_{x}}$,
and $\frac{m^{\,}_{\mathrm{s}}}{2\pi\Lambda^{\,}_{x}}$
(set $v^{\,}_{\mathrm{s}}=v^{\,}_{\mathrm{t}}\equiv1$)
if no further approximation is imposed.
Nevertheless, it is still useful to look at two limiting cases
of the saddle-point equation
(\ref{eq: the saddle-point equation final ms neq 0}).

\textbf{Case  $\lambda=0$.}
The solution for $\phi$ is simply
\begin{equation}
\phi=0.
\label{eq: the saddle-point equation under lambda=0 condition}
\end{equation}

\textbf{Case $m^{\,}_{\mathrm{s}}=m^{\,}_{\mathrm{t}}=0$}.
Assuming $\phi\neq0,$
Eq.~(\ref{eq: the saddle-point equation final ms neq 0}) simplifies to 
(set $v^{\,}_{\mathrm{s}}=v^{\,}_{\mathrm{t}}\equiv1$)
\begin{equation}
\begin{split}
1
=&\,
\frac{\lambda}{\pi}\,
\mathrm{arcsinh}
\left(
\frac{2\pi\Lambda^{\,}_{x}}{|\phi|}
\right).
\end{split}
\label{eq: the saddle-point equation under mt=0 final}
\end{equation}
Since 
$\mathrm{arcsinh}
\left(
\frac{2\pi\Lambda^{\,}_{x}}{|\phi|}
\right)$
is positive, we must require $\lambda>0$ to find the solution of $\phi$ 
from (\ref{eq: the saddle-point equation under mt=0 final}).
Non-vanishing solutions for $\phi$ are
\begin{equation}
\begin{split}
|\phi|
=&\,
2\pi\Lambda^{\,}_{x}\,
\frac{1}
	 {
\mathrm{sinh}
\left(
\frac{\pi}{\lambda}
\right)
	 },\quad \lambda>0.
\end{split}
\label{eq: solution of phi under mt=0}
\end{equation}

\subsection{Approximate mean-field gap equations}
\label{subsec: approximated mean-field gap equations}

Insertion of the asymptotic expansion
\begin{equation}
\mathrm{arcsinh}(x)
\approx
\ln\left(2x\right)
+\mathcal{O}(x^{-2})
\label{eq: identity between arcsinh and ln leading order}
\end{equation}
into the saddle-point equation
(\ref{eq: the saddle-point equation final ms neq 0})
gives
\begin{subequations}
\begin{widetext}
\begin{equation}
\begin{split}
0
=&\, 
\frac{\phi}{2\pi\Lambda^{\,}_{x}}
-
\frac{\lambda}{4\pi}
\left[
\frac{1}{v^{\,}_{\mathrm{s}}}
\int\limits_{-\pi}^{+\pi}\frac{\mathrm{d}q}{2\pi}
\left(
\frac{\phi}{2\pi\Lambda^{\,}_{x}}
-
\frac{2m^{\,}_{\mathrm{s}}}{2\pi\Lambda^{\,}_{x}}
\,\cos q
\right)
\times
\ln
\left(
\frac{
	4\pi\Lambda^{\,}_{x}v^{\,}_{\mathrm{s}}
	}
	{
\sqrt{
4m^{2}_{\mathrm{s}}
+\phi^{2}
-4m^{\,}_{\mathrm{s}}\,\phi\,
\cos q
     }
    }
    \right)
\right]
-
\frac{3\lambda}{4\pi}
\left[
\mathrm{s}\to\,\mathrm{t}
\right]
\end{split}
\label{eq: the saddle-point equation final log ms neq 0 befor integral}
\end{equation}
\end{widetext}
with the conditions
\begin{equation}
0\leq
\sqrt{
4m^{2}_{\mathrm{s}}
+\phi^{2}
-4m^{\,}_{\mathrm{s}}\,\phi\,
\cos q
     }
\ll2\pi\Lambda^{\,}_{x}v^{\,}_{\mathrm{s}}
\end{equation}
and
\begin{equation}
0\leq
\sqrt{
4m^{2}_{\mathrm{t}}
+\phi^{2}
-4m^{\,}_{\mathrm{t}}\,\phi\,
\cos q
     }
\ll2\pi\Lambda^{\,}_{x}v^{\,}_{\mathrm{t}}.
\end{equation}
\end{subequations}

The integrals in Eq.\
(\ref{eq: the saddle-point equation final log ms neq 0 befor integral})
can be carried out. There follows
\begin{subequations}
\begin{widetext}
\begin{equation}
\begin{split}
0
=&\, 
\frac{\phi}{2\pi\Lambda^{\,}_{x}}
-
\frac{\lambda}{4\pi}
\frac{1}{v^{\,}_{\mathrm{s}}}
\Biggl\{
\frac{\phi}{2\pi\Lambda^{\,}_{x}}
\times
\ln
\left(
\frac{
	4\sqrt{2}\pi\Lambda^{\,}_{x}v^{\,}_{\mathrm{s}}
	}
	{
\sqrt{
4m^{2}_{\mathrm{s}}
+\phi^{2}
+|4m^{2}_{\mathrm{s}}-\phi^{2}|
     }
    }
    \right)
\\
&
\hspace{7em}
-\frac{1}{4}
\left(
\frac{\phi}{2\pi\Lambda^{\,}_{x}}
\right)^{-1}
\left[
\left(
\frac{\phi}{2\pi\Lambda^{\,}_{x}}
\right)^{2}
+
\left(
\frac{2m^{\,}_{\mathrm{s}}}{2\pi\Lambda^{\,}_{x}}
\right)^{2}
-
\left|
\left(
\frac{\phi}{2\pi\Lambda^{\,}_{x}}
\right)^{2}
-
\left(
\frac{2m^{\,}_{\mathrm{s}}}{2\pi\Lambda^{\,}_{x}}
\right)^{2}
\right|
\right]
\Biggr\}
\\
&\,
-
\frac{3\lambda}{4\pi}
\frac{1}{v^{\,}_{\mathrm{t}}}
\left\{
\mathrm{s}\to\,\mathrm{t}
\right\}
\end{split}
\label{eq: the saddle-point equation final log ms neq 0}
\end{equation}
\end{widetext}
with the conditions
\begin{equation}
0\leq
2
\left|
m^{\,}_{\mathrm{s}}
\right|
+
\left|
\phi
\right|
\ll2\pi\Lambda^{\,}_{x}v^{\,}_{\mathrm{s}}
\end{equation}
and
\begin{equation}
0\leq
2
\left|
m^{\,}_{\mathrm{t}}
\right|
+
\left|
\phi
\right|
\ll2\pi\Lambda^{\,}_{x}v^{\,}_{\mathrm{t}}.
\end{equation}
\end{subequations}

From now on, we treat the case $m^{\,}_{\mathrm{s}}=0$ for which
\begin{subequations}
\label{eq: the saddle-point equation final log}
\begin{widetext}
\begin{equation}
\begin{split}
0
=&\,
\frac{\phi}{2\pi\Lambda^{\,}_{x}}
- 
\frac{\lambda}{4\pi}
\frac{1}{v^{\,}_{\mathrm{s}}}
\frac{\phi}{2\pi\Lambda^{\,}_{x}}
\times
\ln
\left(
\frac{
	4\pi\Lambda^{\,}_{x}v^{\,}_{\mathrm{s}}
	}
	{
	|\phi|
    }
\right)
\\
&\,
-
\frac{3\lambda}{4\pi}
\frac{1}{v^{\,}_{\mathrm{t}}}
\Biggl\{
\frac{\phi}{2\pi\Lambda^{\,}_{x}}
\times
\ln
\left(
\frac{
	4\sqrt{2}\pi\Lambda^{\,}_{x}v^{\,}_{\mathrm{t}}
	}
	{
\sqrt{
4m^{2}_{\mathrm{t}}
+\phi^{2}
+|4m^{2}_{\mathrm{t}}-\phi^{2}|
     }
    }
    \right)
\\
&
\hspace{5em}
-\frac{1}{4}
\left(
\frac{\phi}{2\pi\Lambda^{\,}_{x}}
\right)^{-1}
\left[
\left(
\frac{\phi}{2\pi\Lambda^{\,}_{x}}
\right)^{2}
+
\left(
\frac{2m^{\,}_{\mathrm{t}}}{2\pi\Lambda^{\,}_{x}}
\right)^{2}
-
\left|
\left(
\frac{\phi}{2\pi\Lambda^{\,}_{x}}
\right)^{2}
-
\left(
\frac{2m^{\,}_{\mathrm{t}}}{2\pi\Lambda^{\,}_{x}}
\right)^{2}
\right|
\right]
\Biggr\}
\end{split}
\label{eq: the saddle-point equation final log a}
\end{equation}
\end{widetext}
with the conditions
\begin{equation}
0\leq
|\phi|
\ll2\pi\Lambda^{\,}_{x}v^{\,}_{\mathrm{s}}
\label{eq: the saddle-point equation final log b}
\end{equation}
and
\begin{equation}
0\leq
2
\left|
m^{\,}_{\mathrm{t}}
\right|
+
\left|
\phi
\right|
\ll2\pi\Lambda^{\,}_{x}v^{\,}_{\mathrm{t}}.
\label{eq: the saddle-point equation final log c}
\end{equation}
\end{subequations}
Equation\ (\ref{eq: the saddle-point equation final log})
is solved for the following four cases.

\textbf{Case $m^{\,}_{\mathtt{t}}=0$.} 
Assuming $\phi\neq0$,
Eq.\ (\ref{eq: the saddle-point equation final log}) simplifies to 
(set $v^{\,}_{\mathrm{s}}=v^{\,}_{\mathrm{t}}\equiv1$)
\begin{subequations}
\begin{equation}
\begin{split}
1
=&\,
\frac{\lambda}{\pi}\,
\ln
\left(
\frac{4\pi\Lambda^{\,}_{x}}{|\phi|}
\right)
\end{split}
\label{eq: the saddle-point equation under mt=0 final log}
\end{equation}
with 
\begin{equation}
0\leq
|\phi|
\ll2\pi\Lambda^{\,}_{x}.
\end{equation}
\end{subequations}
Since 
$\ln
\left(
x
\right)$
is positive for $x>1$, we must require $\lambda>0$
to find the solution of $\phi$ 
from (\ref{eq: the saddle-point equation under mt=0 final log}).
Hence, a solution with a non-vanishing
$|\phi|/(2\pi\Lambda^{\,}_{x})$ is
\begin{equation}
\begin{split}
\frac{|\phi|}{2\pi\Lambda^{\,}_{x}}
=&\,
2\times
e^{-\frac{\pi}{\lambda}},\quad \lambda>0.
\end{split}
\label{eq: solution of phi under mt=0 log}
\end{equation}
This is the usual weak-coupling BCS gap.

\textbf{Case $|\phi|<2|m^{\,}_{\mathrm{t}}|$.}
It follows that $4m^{2}_{\mathtt{t}}-\phi^{2}>0$. 
Assuming $\phi\neq0$,
Eq.\ (\ref{eq: the saddle-point equation final log})
simplifies to 
(set $v^{\,}_{\mathrm{s}}=v^{\,}_{\mathrm{t}}\equiv1$)
\begin{subequations}
\begin{equation}
\begin{split}
1
=&\, 
\frac{\lambda}{4\pi}
\ln
\left(
\frac{
	4\pi\Lambda^{\,}_{x}
	}
	{
	|\phi|
    }
\right)
+
\frac{3\lambda}{4\pi}
\ln
\left(
\frac{
	2\pi\Lambda^{\,}_{x}
	}
	{
|m^{\,}_{\mathrm{t}}|
    }
    \right)
-\frac{3\lambda}{8\pi}
\end{split}
\label{eq: the saddle-point equation under phi smaller than 2mt final log}
\end{equation}
with 
\begin{equation}
0\leq
|\phi|
\ll2\pi\Lambda^{\,}_{x}
\end{equation}
and
\begin{equation}
0\leq
2
\left|
m^{\,}_{\mathrm{t}}
\right|
+
\left|
\phi
\right|
\ll2\pi\Lambda^{\,}_{x}.
\end{equation}
\end{subequations}
Hence, a solution with a non-vanishing
$|\phi|/(2\pi\Lambda^{\,}_{x})$ is
\begin{equation}
\begin{split}
\frac{|\phi|}{2\pi\Lambda^{\,}_{x}}
=
2
\times\,e^{-3/2}
\times
\left(
\frac{|m^{\,}_{\mathrm{t}}|}{2\pi\Lambda^{\,}_{x}}
\right)^{-3}
\times\,e^{-4\pi/\lambda}.
\end{split}
\label{eq: solution of phi under phi smaller than 2mt log}
\end{equation}
Increasing $|m^{\,}_{\mathrm{t}}|$ decreases $|\phi|$.
Increasing $\lambda>0$ decreases $|\phi|$.
There is a competition between
$\lambda>0$ and $|m^{\,}_{\mathrm{t}}|$.

\textbf{Case $|\phi|>2|m^{\,}_{\mathrm{t}}|$.} 
It follows that $4m^{2}_{\mathtt{t}}-\phi^{2}<0$. 
Assuming $\phi\neq0$, 
Eq.\ (\ref{eq: the saddle-point equation final log})
simplifies to (set $v^{\,}_{\mathrm{s}}=v^{\,}_{\mathrm{t}}\equiv1$)
\begin{subequations}
\begin{equation}
\begin{split}
1
=&\, 
\frac{\lambda}{\pi}
\ln
\left(
\frac{
	4\pi\Lambda^{\,}_{x}
	}
	{
	|\phi|
    }
\right)
-
\frac{3\lambda}{2\pi}
\left(
\frac{m^{\,}_{\mathrm{t}}}{\phi}
\right)^{2}
\end{split}
\label{eq: the saddle-point equation under phi larger than 2mt final log}
\end{equation}
with 
\begin{equation}
0\leq
|\phi|
\ll2\pi\Lambda^{\,}_{x}
\end{equation}
and
\begin{equation}
0\leq
2
\left|
m^{\,}_{\mathrm{t}}
\right|
+
\left|
\phi
\right|
\ll2\pi\Lambda^{\,}_{x}.
\end{equation}
\end{subequations}
Hence, a solution with a non-vanishing
$|\phi|/(2\pi\Lambda^{\,}_{x})$ is
\begin{equation}
\frac{|\phi|}{2\pi\Lambda^{\,}_{x}}\,
e^{+\frac{3}{2}\left(\frac{m^{\,}_{\mathrm{t}}}{\phi}\right)^{2}}=
2\,e^{-\pi/\lambda}.
\label{eq: solution of phi under phi larger than 2mt log}
\end{equation}

\textbf{Case $\phi=\pm2m^{\,}_{\mathtt{t}}$.} 
Assuming $\phi\neq0$,
Eq.\ (\ref{eq: the saddle-point equation final log})
simplifies to  (set $v^{\,}_{\mathrm{s}}=v^{\,}_{\mathrm{t}}\equiv1$)
\begin{subequations}
\begin{equation}
\begin{split}
1
=&\,
\frac{\lambda}{\pi}\,
\ln
\left(
\frac{4\pi\Lambda^{\,}_{x}}{|\phi|}
\right)
-
\frac{3\lambda}{8\pi}
\end{split}
\label{eq: the saddle-point equation under phi=2mt log final}
\end{equation}
with 
\begin{equation}
0\leq
2|\phi|
\ll2\pi\Lambda^{\,}_{x}.
\end{equation}
\end{subequations}
Hence, a solution with a non-vanishing
$|\phi|/(2\pi\Lambda^{\,}_{x})$ is
\begin{equation}
\frac{|\phi|}{2\pi\Lambda^{\,}_{x}}=
\frac{2|m^{\,}_{\mathrm{t}}|}{2\pi\Lambda^{\,}_{x}}
=
2\times
e^{-3/8}\times
e^{-\frac{\pi}{\lambda}}.
\label{eq: solution of phi under phi=2mt log}
\end{equation}

\subsection{Hessian at the saddle points}
\label{subsec: Hessian at the saddle points}

We are going to compute the Hessian of the effective potential defined by $S'$
in Eq.\ (\ref{eq: define effective action}).
To this end, define
\begin{equation}
V^{\,}_{\mathrm{eff}}\:=
\frac{\mathfrak{a}^{\,}_{y}}{\beta\,L^{\,}_{x}\,L^{\,}_{y}}S'. 
\end{equation}
We begin with the saddle points of $V^{\,}_{\mathrm{eff}}$
for $m^{\,}_{\mathrm{s}}=0$ within logarithmic accuracy.
They are simply given by the right-hand side of 
Eq.\ (\ref{eq: the saddle-point equation final log a}).
Next, we turn our attention to the second-order derivative of
$V^{\,}_{\mathrm{eff}}$,
\begin{subequations}
\label{eq: Veff second-order derivatives log approx}
\begin{widetext}
\begin{align}
\frac{\partial^{2}\,V^{\,}_{\mathrm{eff}}}{\partial\phi^{2}}=&
\frac{1}{2\lambda}
-
\frac{1}{8\pi}
\frac{1}{v^{\,}_{\mathrm{s}}}
\Biggl[
\ln
\left(
\frac{
	4\pi\Lambda^{\,}_{x}v^{\,}_{\mathrm{s}}
	}
	{
	|\phi|
    }
    \right)
-1
\Biggr]
\nonumber\\
&\,
-
\frac{3}{8\pi}
\frac{1}{v^{\,}_{\mathrm{t}}}
\Biggl\{\,
\ln
\left(
\frac{
	4\sqrt{2}\pi\Lambda^{\,}_{x}v^{\,}_{\mathrm{t}}
	}
	{
\sqrt{
4m^{2}_{\mathrm{t}}
+\phi^{2}
+|4m^{2}_{\mathrm{t}}-\phi^{2}|
     }
    }
    \right)
-1
\nonumber\\
&\hspace{4em}
+
\frac{1}{4}
\left(
\frac{\phi}{2\pi\Lambda^{\,}_{x}}
\right)^{-2}
\left[
\left(
\frac{\phi}{2\pi\Lambda^{\,}_{x}}
\right)^{2}
+
\left(
\frac{2m^{\,}_{\mathrm{t}}}{2\pi\Lambda^{\,}_{x}}
\right)^{2}
-
\left|
\left(
\frac{\phi}{2\pi\Lambda^{\,}_{x}}
\right)^{2}
-
\left(
\frac{2m^{\,}_{\mathrm{t}}}{2\pi\Lambda^{\,}_{x}}
\right)^{2}
\right|
\right]
\Biggr\}
\end{align}
\end{widetext}
with
\begin{equation}
0\leq
|\phi|
\ll2\pi\Lambda^{\,}_{x}v^{\,}_{\mathrm{s}}
\end{equation}
and
\begin{equation}
0\leq
2
\left|
m^{\,}_{\mathrm{t}}
\right|
+
\left|
\phi
\right|
\ll2\pi\Lambda^{\,}_{x}v^{\,}_{\mathrm{t}}.
\end{equation}
\end{subequations}
There are four cases to consider.

\textbf{Case $m^{\,}_{\mathtt{t}}=0$.}
Insertion of Eq.\ (\ref{eq: solution of phi under mt=0 log})
into Eq.\ (\ref{eq: Veff second-order derivatives log approx})
gives (set $v^{\,}_{\mathrm{s}}=v^{\,}_{\mathrm{t}}\equiv1$)
\begin{equation}
\begin{split}
\left.
\frac{\partial^{2}\,V^{\,}_{\mathrm{eff}}}{\partial\phi^{2}}
\right|^{\,}_{\mathrm{saddle}}
=
\frac{1}{2\pi}
>0.
\end{split}
\label{eq: Veff second-order derivatives under mt=0 final log approx}
\end{equation}
Solution (\ref{eq: solution of phi under mt=0 log})
is a local minima of the effective potential.

\textbf{Case $|\phi|<2|m^{\,}_{\mathrm{t}}|$.}
Insertion of Eq.\
(\ref{eq: solution of phi under phi smaller than 2mt log})
into
Eq.\ (\ref{eq: Veff second-order derivatives log approx})
gives
(set $v^{\,}_{\mathrm{s}}=v^{\,}_{\mathrm{t}}\equiv1$)
\begin{align}
\frac{\partial^{2}\,V^{\,}_{\mathrm{eff}}}{\partial\phi^{2}}
\Big|_{\mathrm{saddle}}
=&\,
\frac{1}{8\pi}
>0.
\end{align}
Solution
(\ref{eq: solution of phi under phi smaller than 2mt log})
is a local minima of the effective potential.

\textbf{Case $|\phi|>2|m^{\,}_{\mathrm{t}}|$.} 
Insertion of Eq.\
(\ref{eq: solution of phi under phi larger than 2mt log})
into Eq.\
(\ref{eq: Veff second-order derivatives log approx})
gives
(set $v^{\,}_{\mathrm{s}}=v^{\,}_{\mathrm{t}}\equiv1$)
\begin{align}
\left.
\frac{\partial^{2}\,V^{\,}_{\mathrm{eff}}}{\partial\phi^{2}}
\right|^{\,}_{\mathrm{saddle}}
=&\,
\frac{1}{2\pi}
-
\frac{3}{2\pi}
\left(
\frac{m^{\,}_{\mathrm{t}}}{\phi}
\right)^{2}>0.
\end{align}
Solution (\ref{eq: solution of phi under phi larger than 2mt log})
is a local minima of the effective potential.

\textbf{Case $\phi=\pm2m^{\,}_{\mathtt{t}}$.}
Insertion of Eq.\
(\ref{eq: solution of phi under phi=2mt log})
into Eq.\
(\ref{eq: Veff second-order derivatives log approx})
gives
(set $v^{\,}_{\mathrm{s}}=v^{\,}_{\mathrm{t}}\equiv1$)
\begin{align}
\left.
\frac{\partial^{2}\,V^{\,}_{\mathrm{eff}}}{\partial\phi^{2}}
\right|^{\,}_{\mathrm{saddle}}
=
\frac{1}{8\pi}
>0.
\end{align}
Solution (\ref{eq: solution of phi under phi=2mt log})
with $\lambda>0$ is a local minima of the effective potential.

\subsection{Interpretation}
\label{subsec: Behavior of the approximate mean-field solution}

To proceed, we recall the definition of the mean-field Majorana gap
(\ref{eq: gap function}) 
\begin{subequations}
\label{eq: mean-field Majorana gap}
\begin{equation}
\Delta^{\,}_{0}\equiv\Delta^{\,}_{\mathrm{s}}\:=
\frac{|\phi|}{2}
\label{eq: mean-field Majorana gap a}
\end{equation}
for the singlet Majorana field with $m^{\,}_{\mathrm{s}}=0$, and
\begin{equation}
\Delta^{\,}_{a}\equiv\Delta^{\,}_{\mathrm{t}}\:=
\left||m^{\,}_{\mathrm{t}}|-\frac{|\phi|}{2}\right|,
\quad a=1,2,3
\label{eq: mean-field Majorana gap b}
\end{equation}
\end{subequations}
for the triplet of Majorana fields, and
the corresponding topological criteria
(\ref{eq: topologial criteria}).
One observes that the singlet Majorana gap $\Delta^{\,}_{\mathrm{s}}$ 
(\ref{eq: mean-field Majorana gap a})
is non-vanishing as long as $\phi\neq0$.

The approximate mean-field solution
given by Eqs.\
(\ref{eq: solution of phi under mt=0 log}),
(\ref{eq: solution of phi under phi smaller than 2mt log}),
(\ref{eq: solution of phi under phi larger than 2mt log}),
and
(\ref{eq: solution of phi under phi=2mt log})
when
$|m^{\,}_{\mathrm{t}}|=0$,
$|\phi|<2|m^{\,}_{\mathrm{t}}|$, 
$|\phi|>2|m^{\,}_{\mathrm{t}}|$, 
and
$|\phi|=2|m^{\,}_{\mathrm{t}}|$, respectively,
imply the mean-field phase diagram shown in
Fig.~\ref{Fig: phase diagram topological order b}.
More specifically, we first look at the line $m^{\,}_{\mathrm{t}}=0$,
along which we have a non-vanishing value of $\phi$.
This corresponds to a phase with (mean-field) ATO, 
for which the boundary realizes a CFT with central charge $2$
as both the singlet and triplet of chiral Majorana edge states are gapless.
We also find that $|\phi|$ reaches its maximum value when
$m^{\,}_{\mathrm{t}}=0$ for a given $\lambda >0$.
The generic trend is that $|\phi|$ decreases as $|m^{\,}_{\mathrm{t}}|$
increases. If we increase $|m^{\,}_{\mathrm{t}}|$ a little away from 0,
$|\phi|$ decreases a little. However,
the ATO phase is robust, for the mean-field bulk gap 
$\Delta^{\,}_{\mathrm{s}}$ and $\Delta^{\,}_{\mathrm{t}}$
(\ref{eq: mean-field Majorana gap}) 
remain non-vanishing.
We have to increase $|m^{\,}_{\mathrm{t}}|$
until it satisfies $2|m^{\,}_{\mathrm{t}}|=|\phi|$
for the mean-field triplet bulk gap
$\Delta^{\,}_{\mathrm{t}}$
(\ref{eq: mean-field Majorana gap b}) 
to close. Only then can the
(mean-field) ATO phase be destroyed.
The triplet bulk gap 
$\Delta^{\,}_{\mathrm{t}}$
reopens when $2|m^{\,}_{\mathrm{t}}|>|\phi|$,
however the triplet of chiral Majorana edge states are now gapped,
leaving only a singlet of massless chiral Majorana edge states.
This mean-field phase supports (mean-field) NATO,
for which the boundary realizes a CFT with central charge $1/2$.
In the large $|m^{\,}_{\mathrm{t}}|$ limit,
the value of $|\phi|$ is further suppressed 
[see Eq.\ (\ref{eq: solution of phi under phi smaller than 2mt log})]. 
However, the mean-field bulk gap 
$\Delta^{\,}_{\mathrm{s}}$ and $\Delta^{\,}_{\mathrm{t}}$
remain gapped,
whatever the small but non-vanishing value of $|\phi|$ is.

The assumption that the singlet mass $m^{\,}_{\mathrm{s}}$
vanishes in order to derive the non-vanishing solutions
(\ref{eq: solution of phi under mt=0 log}),
(\ref{eq: solution of phi under phi smaller than 2mt log}),
(\ref{eq: solution of phi under phi larger than 2mt log}),
and
(\ref{eq: solution of phi under phi=2mt log})
to the gap equation (\ref{eq: the saddle-point equation final log})
is not essential as long as a non-vanishing $m^{\,}_{\mathrm{s}}$
is smaller in magnitude than the saddle-point $|\phi/2|$.
This is to say that the ATO and NATO phases for $\lambda>0$
and $m^{\,}_{\mathrm{s}}$=0 extend to non-vanishing yet 
not too strong $|\phi/2|>|m^{\,}_{\mathrm{s}}|>0$. The ATO and NATO
phases do not require a precise tuning of the two-leg ladders
to their Ising critical point provided the detuning is smaller
in magnitude than $|\phi/2|$.

\section{Beyond mean-field theory:
Dimensional crossover from a random phase approximation}
\label{sec: RPA} 

The mean-field approximation of Sec.\ \ref{sec: The mean-field approach}
is done in two-dimensional space. It posits that
all excitations belong to a quartet of point-like particles obeying
the Majorana equal-time algebra. However, the line
$m^{\,}_{\mathrm{t}}=m^{\,}_{\mathrm{s}}=0$ when $v^{\,}_{\mu}=v$ for $\mu=0,1,2,3$
in Fig.\ \ref{Fig: phase diagram topological order a}
corresponds to an integrable model for which this is not the case.
As was alluded to below Eq.\
(\ref{eq: desired Majorana theory mt=0}),
Hamiltonian (\ref{eq: desired Majorana theory mt=0})
with $\lambda>0$ is a massive theory in which the
quartet of Majoranas do not exist as sharp (coherent) excitations,
i.e., none of the components
\begin{align}
&
G^{\mathrm{1d}}_{\mu,\mathrm{M},\mathtt{m};\mu'\mathrm{M}',\mathtt{m}'}(\omega,k^{\,}_{x})\:=
\nonumber\\
&\qquad\qquad
-
\langle0|
\widehat{\chi}^{\mu}_{\mathrm{M},\mathtt{m}}(\omega,k^{\,}_{x})\,
\widehat{\chi}^{\mu'}_{\mathrm{M}',\mathtt{m}'}(-\omega,-k^{\,}_x)\,
|0\rangle
\end{align}
(the ket $|0\rangle$ denotes the ground state)
support poles. Here,
$\omega$ is a fermionic Matsubara frequency,
$k^{\,}_{x}$ is a one-dimensional momentum,
$\mu,\mu'=0,1,2,3$ refer to the index for the quartet of
Majorana fields,
$\mathrm{M},\mathrm{M}'=\mathrm{L},\mathrm{R}$ refer to the left- and
right-moving components of the Majorana fields, and $\mathtt{m},\mathtt{m}'=1,\cdots,n$
refer to the index of the ladders. 

The line
$m^{\,}_{\mathrm{t}}=m^{\,}_{\mathrm{s}}=0$ when $v^{\,}_{\mu}=v$ for $\mu=0,1,2,3$
in Fig.\ \ref{Fig: phase diagram topological order a}
consists of decoupled one-dimensional Gross-Neveu
Hamiltonians with $O(4)$ symmetry, recall Eq.\
(\ref{eq: desired Majorana theory mt=0 d}),
each one of which has the Lagrangian density
\begin{subequations}
\label{eq: desired Majorana theory RPA}
\begin{align}
&
\widehat{\mathcal{L}}^{\,}_{\mathrm{GN}}\:=
\widehat{\mathcal{L}}^{\,}_{0}
+
\widehat{\mathcal{H}}^{\,}_{\mathrm{int}},
\label{eq: desired Majorana theory RPA a}
\\
&
\widehat{\mathcal{L}}^{\,}_{0}=
\frac{1}{2}
\sum^{3}_{\mu=0}
\left[
\widehat{\chi}^{\mu}_{\mathrm{L}}
\left(
\partial^{\,}_{\tau}
+
\mathrm{i}v\,
\partial^{\,}_{x}
\right)
\widehat{\chi}^{\mu}_{\mathrm{L}}
+
\widehat{\chi}^{\mu}_{\mathrm{R}}
\left(
\partial^{\,}_{\tau}
-
\mathrm{i}v\,
\partial^{\,}_{x}
\right)
\widehat{\chi}^{\mu}_{\mathrm{R}}
\right],
\label{eq: desired Majorana theory RPA b}
\\
&
\widehat{\mathcal{H}}^{\,}_{\mathrm{int}}=
\frac{\lambda}{4}
\left(
\sum^{3}_{\mu=0}
\widehat{\chi}^{\mu}_{\mathrm{L}}\,
\widehat{\chi}^{\mu}_{\mathrm{R}}\,
\right)^{2}.
\label{eq: desired Majorana theory RPA c}
\end{align}
\end{subequations}
where the velocity $v$ and the
coupling $\lambda$ are all real valued.

We are going to extract the single-particle Green function
for the Majorana fermions with the Hamiltonian
(\ref{eq: desired Majorana theory RPA})
using non-perturbative results valid for integrable systems.
We will then treat a non-vanishing mass $m^{\,}_{\mathrm{t}}\neq0$
non-perturbatively within a Random Phase Approximation (RPA).

It is known \cite{Witten78} that the $O(4)$ GN defined by the
Lagrangian density (\ref{eq: desired Majorana theory RPA})
is equivalent to two independent copies of the sine-Gordon model.
We identify the first copy as the spin-sector and the second copy
as the charge sector for interacting spin-1/2 electrons.
In turn, the creation
$\widehat{\psi}^{\dag}_{\mathrm{M},\sigma}$
and annihilation
$\widehat{\psi}^{\,}_{\mathrm{M},\sigma}$
operators for the electrons
are related to the Majorana fermions by
\begin{subequations}
\label{appeq: define Dirac fermions}
\begin{align}
&
\widehat{\psi}^{\dag}_{\mathrm{M},\uparrow}
\equiv 
\frac{1}{\sqrt{2}}
\left(
\widehat{\chi}^{1}_{\mathrm{M}}
-
\mathrm{i}
\widehat{\chi}^{2}_{\mathrm{M}}
\right),
\qquad
\widehat{\psi}^{\,}_{\mathrm{M},\uparrow}
\equiv 
\frac{1}{\sqrt{2}}
\left(
\widehat{\chi}^{1}_{\mathrm{M}}
+
\mathrm{i}
\widehat{\chi}^{2}_{\mathrm{M}}
\right),
\\
&
\widehat{\psi}^{\dag}_{\mathrm{M},\downarrow}
\equiv 
\frac{1}{\sqrt{2}}
\left(
\widehat{\chi}^{3}_{\mathrm{M}}
-
\mathrm{i}
\widehat{\chi}^{0}_{\mathrm{M}}
\right),
\qquad
\widehat{\psi}^{\,}_{\mathrm{M},\downarrow}
\equiv 
\frac{1}{\sqrt{2}}
\left(
\widehat{\chi}^{3}_{\mathrm{M}}
+
\mathrm{i}
\widehat{\chi}^{0}_{\mathrm{M}}
\right),
\end{align}
\end{subequations}
where $\mathrm{M}=\mathrm{L},\mathrm{R}$ and $\sigma=\uparrow,\downarrow$.
By relying on Abelian bosonization rules, the
$O(4)$ GN Lagrangian density (\ref{eq: desired Majorana theory RPA})
becomes
\begin{subequations}
\label{eq: def double SG theory}
\begin{align}
&
\widehat{\mathcal{L}}^{\,}_{\mathrm{GN}}= 
\widehat{\mathcal{L}}^{\,}_{\mathrm{GN},\mathrm{s}}
+
\widehat{\mathcal{L}}^{\,}_{\mathrm{GN},\mathrm{c}},
\\
&
\widehat{\mathcal{L}}^{\,}_{\mathrm{GN},\mathrm{s}}= 
\frac{1}{2}
\left[
v^{-1}_{\mathrm{s}}
\left(
\partial^{\,}_{\tau}\widehat{\varphi}^{\,}_{\mathrm{s}}
\right)^{2}
+
v^{\,}_{\mathrm{s}}
\left(
\partial^{\,}_{x}\widehat{\varphi}^{\,}_{\mathrm{s}}
\right)^{2}
\right]
-
\frac{\lambda}{4}\,	
\cos(\beta\,\widehat{\varphi}^{\,}_{\mathrm{s}}),
\\
&
\widehat{\mathcal{L}}^{\,}_{\mathrm{GN},\mathrm{c}}= 
\frac{1}{2}
\left[
v^{-1}_{\mathrm{c}}
\left(
\partial^{\,}_{\tau}\widehat{\varphi}^{\,}_{\mathrm{c}}
\right)^{2}
+
v^{\,}_{\mathrm{c}}
\left(
\partial^{\,}_{x}\widehat{\varphi}^{\,}_{\mathrm{c}}
\right)^{2}
\right]
-
\frac{\lambda}{4}\,	
\cos(\beta\,\widehat{\varphi}^{\,}_{\mathrm{c}}),
\end{align}
with
\begin{equation}
v^{\,}_{\mathrm{s}}=v^{\,}_{\mathrm{c}}\equiv v
\end{equation}
and
\begin{equation}
\beta=
\sqrt{
\frac{8\pi}{1+\frac{\lambda}{2\pi}}
	 }.
\end{equation}
\end{subequations}
Equation (\ref{eq: def double SG theory})
is also derived in Sec.\
\ref{sec: Coupled two-leg ladders}
starting from the spin-1/2 lattice model depicted in
Fig.\ \ref{Fig: 2Dlattice}.

The quantum critical point $\lambda=0$
($\beta^{2}=8\pi$) supports an
$\widehat{su}(2)^{\,}_{1}\oplus \widehat{su}(2)^{\,}_{1}$
current algebra. When $\lambda>0$ ($\beta^{2}<8\pi$)
each cosine interaction becomes marginally relevant,
a spectral gap opens up, and soliton-like excitations (kinks) by which
the asymptotic expectation values of
$\widehat{\varphi}^{\,}_{\mathrm{a}}(x,\tau)$
with $\mathrm{a}=\mathrm{s},\mathrm{c}$
at $x=-\infty$ and $x=+\infty$ changes by $\pm2\pi/\beta$
over a region of size $1/M$
can be thought of as massive particles with the mass $M$
a function of the deviation $8\pi-\beta^{2}>0$.
At $\beta^{2}=4\pi$, 
$\widehat{\mathcal{L}}^{\,}_{\mathrm{GN},\mathrm{a}}$
is a non-interacting massive Dirac theory for both
$\mathrm{a}=\mathrm{s}$ and $\mathrm{a}=\mathrm{c}$.
When $\beta^{2}<4\pi$, breather modes supplement the kinks
as massive point-like excitations.

If the real-valued scalar field
$\widehat{\varphi}^{\,}_{\mathrm{a}}(\tau,x)$
is decomposed into left- and right-moving parts according to the rule
\begin{subequations}
\label{eq: Mandelstam rep for electron operator}
\begin{equation}
\widehat{\varphi}^{\,}_{\mathrm{a}}(\tau,x)=
\widehat{\varphi}^{\,}_{\mathrm{a},\mathrm{R}}(\tau+\mathrm{i}x)
+
\widehat{\varphi}^{\,}_{\mathrm{a},\mathrm{L}}(\tau-\mathrm{i}x),
\end{equation}
it is then possible to use the Mandelstam representation
\begin{equation}
\widehat{\psi}^{\,}_{\mathrm{M},\sigma}\:=
\frac{\eta^{\,}_{\sigma}}{\sqrt{2\pi}}\,
e^{\mathrm{i}\sqrt{2\pi}\,\widehat{\varphi}^{\,}_{\mathrm{c},\mathrm{M}}}\,
e^{\mathrm{i}f^{\,}_{\sigma}\sqrt{2\pi}\,\widehat{\varphi}^{\,}_{\mathrm{s},\mathrm{M}}}
\end{equation}
with
$\mathrm{M}=\mathrm{L},\mathrm{R}$,
$\sigma=\uparrow,\downarrow$,
$f^{\,}_{\uparrow}=-f^{\,}_{\downarrow}=1$,
and
$\eta^{\,}_{\sigma}$ the Klein factors fulfilling
\begin{equation}
\left\{
\eta^{\,}_{\sigma},\eta^{\,}_{\sigma'}
\right\}=
2\delta^{\,}_{\sigma,\sigma'}.
\end{equation}
\end{subequations}
The chiral vertex operator
$\exp(\pm\mathrm{i}\sqrt{2\pi}\,\widehat{\varphi}^{\,}_{\mathrm{a},\mathrm{M}})$
carries the Lorentz spin
\begin{equation}
s\:=\pm1/4,
\end{equation}
i.e., under the rotation
\begin{equation}
\tau+\mathrm{i}x\mapsto
e^{\mathrm{i}\alpha}(\tau+\mathrm{i}x)
\end{equation}
of two-dimensional Euclidean space,
it is multiplied by the phase $\exp(\pm\mathrm{i}\alpha/4)$.
The chiral electron annihilation operator,
which must carry the Lorentz spin $s=1/2$,
is glued by taking the product of two chiral vertex operators,
each of which carries the Lorentz spin $s=1/4$,
according to Eq.\ (\ref{eq: Mandelstam rep for electron operator}).

To calculate the two-point correlation functions for the
chiral Majorana fields,
they are first expressed in terms of two-point functions
for the chiral electron fields using Eq.\
(\ref{appeq: define Dirac fermions}).
The Mandelstam representation
(\ref{eq: Mandelstam rep for electron operator}) is then used
to represent the two-point Green's functions for the Majorana fields
in terms of two-point functions for the chiral vertex operators.
Finally, the two-point functions for the chiral vertex operators are
calculated using the form factors of the massive integrable theory
defined by the Lagrangian density
(\ref{eq: def double SG theory}).

In a relativistically invariant massive integrable theory
in two-dimensional Euclidean space, all multiparticle states
are the kets
\begin{subequations}
\label{eq: summary massive integrable theories}
\begin{equation}
|\theta^{\,}_{n},\cdots,\theta^{\,}_{1}\rangle
^{\,}_{\epsilon^{\,}_{n},\cdots,\epsilon^{\,}_{1}}
\end{equation}
with the many-body energy
\begin{equation}
\sum^{n}_{j=1}
E^{\,}_{\epsilon^{\,}_{j}}(\theta^{\,}_{j}),
\qquad
E^{\,}_{\epsilon^{\,}_{j}}(\theta^{\,}_{j})\:=
M\,\cosh\theta^{\,}_{j},
\end{equation}
the many-body momentum
\begin{equation}
\sum^{n}_{j=1}
P^{\,}_{\epsilon^{\,}_{j}}(\theta^{\,}_{j}),
\qquad
P^{\,}_{\epsilon^{\,}_{j}}(\theta^{\,}_{j})\:=
\frac{M}{v}\,\sinh\theta^{\,}_{j},
\end{equation}
where $\theta^{\,}_{i}$ denotes the rapidity
of a single-particle state with the quantum number $\epsilon^{\,}_{i}$.
They are pairwise orthogonal and orthogonal to the ground state
$|0\rangle$ with the resolution of the identity
\begin{equation}
\begin{split}
1=&\,
|0\rangle\langle0|
+
\sum^{+\infty}_{n=1}
\sum_{\epsilon^{\,}_{i}}
\int\limits^{+\infty}_{-\infty}
\frac{\mathrm{d}\theta^{\,}_{1}\cdots \mathrm{d}\theta^{\,}_{n}}{(2\pi)^{n}n!}
\\
&\,
\times
|\theta^{\,}_{n},\cdots,\theta^{\,}_{1}\rangle
^{\,}_{\epsilon^{\,}_{n},\cdots,\epsilon^{\,}_{1}}
{}^{\epsilon^{\,}_{1},\cdots,\epsilon^{\,}_{n}}
\langle\theta^{\,}_{1},\cdots,\theta^{\,}_{n}|.
\end{split}
\label{eq: resolution of identity}
\end{equation}
\end{subequations}
The two-point functions for the chiral vertex operators are
calculated using an integral representation for the form factor
\begin{equation}
\left\langle0\left|
e^{\pm\mathrm{i}\sqrt{2\pi}\,\widehat{\varphi}^{\,}_{\mathrm{a},\mathrm{M}}}
\right|\theta^{\,}_{n},\cdots,\theta^{\,}_{1}\right\rangle
^{\,}_{\epsilon^{\,}_{n},\cdots,\epsilon^{\,}_{1}}
\end{equation}
due to Ref.\ \onlinecite{Lukyanov01}.
Following Refs.\
\onlinecite{Essler01} and \onlinecite{Essler05a}
we will truncate the resolution of the identity
(\ref{eq: resolution of identity})
to the order $n=1$ when evaluating the form factors for
the electron operators.

The one-particle form factors for the pair of chiral vertex operators
$\exp(\mathrm{i}\sqrt{2\pi}\widehat{\varphi}^{\,}_{\mathrm{a},\mathrm{M}})$ 
between the vacuum and a state supporting a single soliton are
\begin{subequations}
\label{eq: fixing matrix elements}
\begin{align}
&
\left\langle0\left|
e^{\mathrm{i}\sqrt{2\pi}\,\widehat{\varphi}^{\,}_{\mathrm{a},\mathrm{R}}}
\right|\theta\right\rangle\approx
\sqrt{Z^{\,}_{0}}\,
\left(2\pi M v^{-1}\right)^{1/4}\,
e^{+\theta/4},
\\
&
\left\langle0\left|
e^{\mathrm{i}\sqrt{2\pi}\,\widehat{\varphi}^{\,}_{\mathrm{a},\mathrm{L}}}
\right|\theta\right\rangle\approx
\sqrt{Z^{\,}_{0}}\,
\left(2\pi M v^{-1}\right)^{1/4}\,
e^{-\theta/4}.
\end{align}
The dependence on the rapidity $\theta$ is fixed by Lorentz invariance,
whereas the positive constant $Z^{\,}_{0}$ is not fixed by symmetry,
but was calculated in Ref.\ \onlinecite{Lukyanov01}
to be
\begin{equation}
Z^{\,}_{0}\approx 0.92.
\end{equation}
\end{subequations}
In the same work it was demonstrated that most of the spectral
weight is contained in the emission of a single kink. For example,
about 80 percent of the spectral weight in the
spectral functions entering the Majorana two-point functions
(\ref{eq: 1D Green's function})
originate from the emission of a single kink.
After substituting these matrix elements
into the Lehmann expansion for the Majorana
two-point functions in $(1+1)$-dimensional Euclidean space, we obtain,
for any $\mu=0,1,2,3$ and after setting $v=1$,
\begin{widetext}
\begin{subequations}
\label{eq: G1d left right components}
\begin{align}
&
G^{\mathrm{1d}}_{\mathrm{LL}}(\tau,x)\:=
-
\langle0|
\widehat{\chi}^{\mu}_{\mathrm{L}}(\tau,x)\,
\widehat{\chi}^{\mu}_{\mathrm{L}}(0,0)
|0\rangle=
-
Z^{2}_{0}\,
\left(\frac{\tau+\mathrm{i}x}{\tau-\mathrm{i}x}\right)^{1/2}
\left[
\int\limits_{-\infty}^{+\infty}\frac{\mathrm{d}\theta}{2\pi}
e^{-\theta/2}\,
e^{-M\,\rho\,\cosh\theta}
\right]^{2}=
-
\frac{Z^{2}_{0}\,e^{-2M\,\rho}}{2\pi(\tau-\mathrm{i}x)},
\label{eq: G1d left right components a}
\\
&
G^{\mathrm{1d}}_{\mathrm{RR}}(\tau,x)\:=
-
\langle0|
\widehat{\chi}^{\mu}_{\mathrm{R}}(\tau,x)\,
\widehat{\chi}^{\mu}_{\mathrm{R}}(0,0)
|0\rangle
=
-
Z^{2}_{0}\,
\left(\frac{\tau-\mathrm{i}x}{\tau+\mathrm{i}x}\right)^{1/2}
\left[
\int\limits_{-\infty}^{+\infty}\frac{\mathrm{d}\theta}{2\pi}
e^{+\theta/2}\,
e^{-M\,\rho\,\cosh\theta}
\right]^{2}=
-
\frac{Z^{2}_{0}\,e^{-2M\,\rho}}{2\pi(\tau+\mathrm{i}x)},
\label{eq: G1d left right components b}
\\
&
G^{\mathrm{1d}}_{\mathrm{LR}}(\tau,x)=
G^{\mathrm{1d}}_{\mathrm{RL}}(\tau,x)\:=
-
\langle0|
\widehat{\chi}^{\mu}_{\mathrm{L}}(\tau,x)\,
\widehat{\chi}^{\mu}_{\mathrm{R}}(0,0)
|0\rangle=
-
Z^{2}_{0}\,M\,
\left[
\int\limits_{-\infty}^{+\infty}
\frac{\mathrm{d}\theta}{2\pi}\,
e^{-M\,\rho\,\cosh\theta}
\right]^{2}=
-
\frac{Z^{2}_{0}\,M\,K^{2}_{0}(M\,\rho)}{\pi^{2}},
\label{eq: G1d left right components c}
\end{align}
where
\begin{equation}
K^{\,}_{0}(z)\:=
\frac{1}{2}\int\limits^{+\infty}_{-\infty}\mathrm{d}t\,e^{-z\,\mathrm{cosh}z},
\qquad
\rho\:=\sqrt{\tau^{2}+x^{2}}.
\end{equation}
\end{subequations}
Fourier transformation to imaginary frequency ($\bar{\omega}$)
and momentum ($q$) space followed by the analytic continuation
$\bar{\omega}\to-\mathrm{i}\omega+0^{+}$ delivers 
the retarded two-point Green functions for the
chiral Majorana fields given by 
\begin{subequations}
\label{eq: 1D Green's function}
\begin{align}
&
G^{\mathrm{1d}}_{\mathrm{LL}}(\omega,q)\:=
\lim_{\bar{\omega}\to-\mathrm{i}\omega+0^{+}}
\int\mathrm{d}\tau\mathrm{d}x\,
e^{\mathrm{i}\bar{\omega}\,\tau-\mathrm{i}q\,x}\,
G^{\mathrm{1d}}_{\mathrm{LL}}(\tau,x)
\approx
\frac{
Z^{2}_{0}
     }
     {
\omega-q
     }
\left(
1
-
\frac{
1
     }
     {
\sqrt{1-s^{2}/(2M)^{2}}
     }
\right),
\label{eq: 1D Green's function a}
\\
&
G^{\mathrm{1d}}_{\mathrm{RR}}(\omega,q)\:=
\lim_{\bar{\omega}\to-\mathrm{i}\omega+0^{+}}
\int\mathrm{d}\tau\mathrm{d}x\,
e^{\mathrm{i}\bar{\omega}\,\tau-\mathrm{i}q\,x}\,
G^{\mathrm{1d}}_{\mathrm{RR}}(\tau,x)
\approx
\frac{
Z^{2}_{0}
     }
     {
\omega+q
     }
\left(
1-
\frac{
1
     }
     {
\sqrt{1-s^{2}/(2M)^{2}}
     }
\right),
\label{eq: 1D Green's function b}
\\
&
G^{\mathrm{1d}}_{\mathrm{LR}}(\omega,q)=
G^{\mathrm{1d}}_{\mathrm{RL}}(\omega,q)\:=
\lim_{\bar{\omega}\to-\mathrm{i}\omega+0^{+}}
\int\mathrm{d}\tau\mathrm{d}x\,
e^{\mathrm{i}\bar{\omega}\,\tau-\mathrm{i}q\,x}\,
G^{\mathrm{1d}}_{\mathrm{LR}}(\tau,x)
\approx
-
\frac{Z^{2}_{0}}{2M}\,
\frac{2}{\pi}\,
\frac{
\mathrm{arcsin}
\left(
s/2M
\right)
     }
     { 
[s/(2M)]\,\sqrt{1-[s/(2M)]^{2}}
     },
\label{eq: 1D Green's function c}
\end{align}
where 
\begin{equation}
s^{2}\:=\omega^{2}-q^{2}.
\end{equation}
\end{subequations}

\end{widetext}

Observe that the Green functions
(\ref{eq: 1D Green's function a})
and
(\ref{eq: 1D Green's function b})
are even functions of $M$, while the Green function
(\ref{eq: 1D Green's function c})
is an odd function of $M$. This latter fact follows from
the bond operator 
(\ref{eq: definition of bond order parameter})
being odd under any transformation
(\ref{eq: symmetries c chiral})
with
$\sigma^{\,}_{\mathrm{L},\mathtt{m}}\,\sigma^{\,}_{\mathrm{R},\mathtt{m}+1}=-1$.

Once we turn on a non-vanishing $m^{\,}_{\mathrm{t}}$, we restore true
two-dimensionality of space. In the spirit of the RPA for dimensional
crossovers from lower to higher dimensions, we make the RPA Ansatz for
the retarded Green's function in momentum space
\begin{subequations}
\begin{equation}
\widehat{G}^{\mathrm{2d\,RPA}}(\omega,k^{\,}_{x},k^{\,}_{y})\:=
\frac{1
     }
     {
\Big[\widehat{G}^{\mathrm{1d}}(\omega,k^{\,}_{x})\Big]^{-1}
-
\widehat{M}(k^{\,}_{y})
     },
\label{RPA}
\end{equation}
where
\begin{equation}
\widehat{G}^{\mathrm{1d}}(\omega,k^{\,}_{x})\:=
\begin{pmatrix}
G^{\mathrm{1d}}_{\mathrm{LL}}(\omega,k^{\,}_{x})
&
G^{\mathrm{1d}}_{\mathrm{LR}}(\omega,k^{\,}_{x})
\\
\\
G^{\mathrm{1d}}_{\mathrm{RL}}(\omega,k^{\,}_{x})
&
G^{\mathrm{1d}}_{\mathrm{RR}}(\omega,k^{\,}_{x})
\end{pmatrix}
\end{equation}
and the Fourier transform of the perturbation $\widehat{M}(k^{\,}_{y})$ given by
\begin{equation}
\widehat{M}(k^{\,}_{y})\:=
\begin{pmatrix}
0
&
m^{\,}_{\mu}\,e^{+\mathrm{i}k^{\,}_{y}}
\\
m^{\,}_{\mu}\,e^{-\mathrm{i}k^{\,}_{y}}
&
0
\end{pmatrix}.
\end{equation}
\end{subequations}

If the operator-valued denominator on the right-hand side of Eq.\ (\ref{RPA})
acquires first-order zeros as eigenvalues,
then this RPA predicts that a non-vanishing $m^{\,}_{\mu}$
turns the Majorana fields into well-defined quasi-particles.
The condition for this to happen is that
the determinant of the denominator on the right-hand side of
Eq.\ (\ref{RPA})
vanishes, namely
\begin{align}
0=
1
-
2\,
m^{\,}_{\mu}\,
G^{\mathrm{1d}}_{\mathrm{LR}}(\omega,k^{\,}_{x})\,
\cos k^{\,}_{y}
-
m^{2}_{\mu}\,
\det\widehat{G}^{\mathrm{1d}}(\omega,k^{\,}_{x}).
\end{align}

\noindent
Substituting the retarded Green's functions from
Eq.\ (\ref{eq: 1D Green's function}), we obtain the
dispersion for the triplet of Majoranas
(the singlet ones at $m^{\,}_{\mathrm{s}}=0$ do
not propagate, at least in this RPA formalism)
from solving
\begin{subequations}
\label{eq: RPA poles}
\begin{align}
\cos k^{\,}_{y}\approx
\frac{
1
-
\left(
\frac{Z^{2}_{0}\,m^{\,}_{\mathrm{t}}}{2M}
\right)^{2}
\Big[
g^{2}\left(\frac{s}{2M}\right)
-
f^{2}\left(\frac{s}{2M}\right)
\Big]
     }
     {
(-2)
\left(
\frac{Z^{2}_{0}\,m^{\,}_{\mathrm{t}}}{2M}
\right)\,
f\left(\frac{s}{2M}\right)
     },
\label{eq: RPA poles a}
\end{align}
where we have introduced the auxiliary functions
\begin{equation}
g(x)\:=
\frac{1}{x}\,
\left(1-\frac{1}{\sqrt{1-x^{2}}}\right),
\qquad
f(x)\:=
\frac{2}{\pi}\frac{\arcsin\left(x\right)}{x\,\sqrt{1-x^{2}}},
\label{eq: RPA poles b}
\end{equation}
with the limiting values
\begin{equation}
\lim_{x\to0}g(x)=0,
\qquad
\lim_{x\to0}f(x)=2/\pi,
\label{eq: RPA poles c}
\end{equation}
and the asymptotic expansion for $|x|\ll1$
\begin{align}
g(x)=
-
\frac{1}{2}x
-
\frac{3}{8}x^{3}
\cdots,
\quad
f(x)=
\frac{2}{\pi}
\left(
1
+
\frac{2}{3}x^{2}
+
\cdots
\right).
\label{eq: RPA poles d} 
\end{align}
\end{subequations}
We note that the RPA spectrum (\ref{eq: RPA poles a})
is invariant under the simultaneous transformation
\begin{equation}
k^{\,}_{y}
\mapsto 
k^{\,}_{y}\pm\pi,
\quad
m^{\,}_{\mathrm{t}}
\mapsto
-m^{\,}_{\mathrm{t}}.
\end{equation} 

For small $|m^{\,}_{\mathrm{t}}/M|\ll1$, we deduce from
Eq.\ (\ref{eq: RPA poles}) the relation
\begin{equation}
(s/2M)^{2}\approx
1
-
(Z^{2}_{0}\,m^{\,}_{\mathrm{t}}/M)^{2}\,\cos^{2}k^{\,}_{y},
\quad
|k^{\,}_{y}|<\pi/2.
\end{equation}
As it should be,
no RPA excitations can be found below the threshold $2M$
for the two-soliton continuum when $m^{\,}_{\mathrm{t}}=0$.
However, for any infinitesimal
$m^{\,}_{\mathrm{t}}\neq0$, one finds RPA excitations that are
dispersing along the $\mathtt{m}$ direction with the momentum $k^{\,}_{y}$
below the threshold $2M$.

\begin{figure}[t]
\begin{center}
\includegraphics[width=0.5\textwidth]{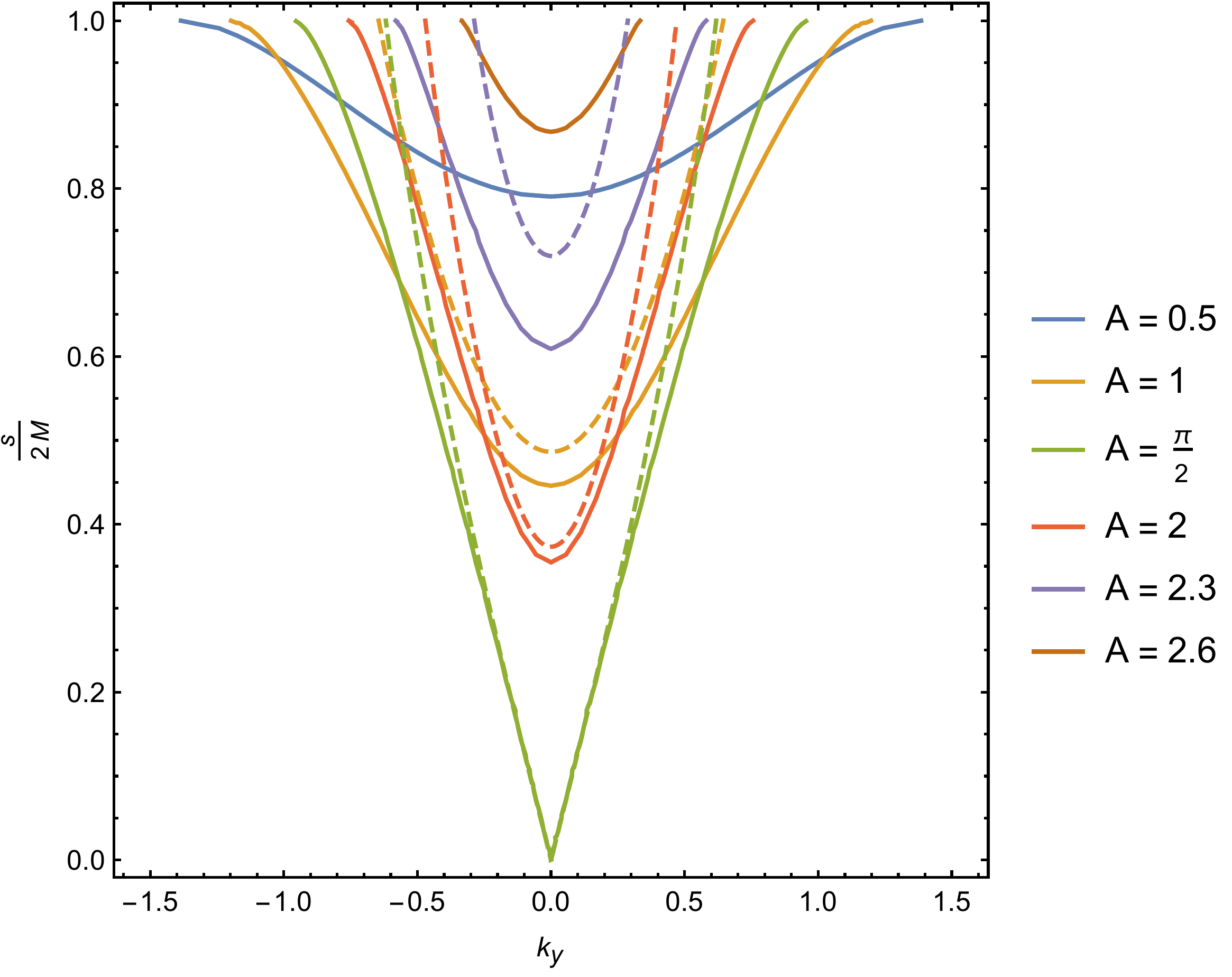}
\\
\caption{(Color online)
The solid lines are the dispersions $s/(2M)$ for the triplet of Majorana modes 
as a function of $k^{\,}_{y}$ 
that follows from solving Eq.\ (\ref{eq: RPA poles a})
for different values of
$A=-Z^{2}_{0}\,m^{\,}_{\mathrm{t}}/(2M)$.
The dashed lines are the corresponding dispersions
obtained from the approximate dispersion relation
(\ref{eq: RPA approximate dispersion relation}).
\label{Fig: Spectrum}
         }
\end{center}
\end{figure}

For arbitrary $|m^{\,}_{\mathrm{t}}/M|$,
one can solve Eq.\ (\ref{eq: RPA poles}) numerically,
thereby confirming the analytical results obtained for
$|m^{\,}_{\mathrm{t}}/M|\ll1$.
Figure \ref{Fig: Spectrum} displays the values of the pair
\begin{equation}
(k^{\,}_{y},s)\equiv(k^{\,}_{y},\sqrt{\omega^{2}-k^{2}_{x}})
\end{equation}
that solve
Eq.\ (\ref{eq: RPA poles a})
holding 
\begin{equation}
A\:=-\frac{Z^{2}_{0}\,m^{\,}_{\mathrm{t}}}{2M}>0
\end{equation}
fixed. By inspection of the dispersions
$(s,k^{\,}_{y})$
for different values of $A$,
we deduce the existence of a spectral gap
except for the special case when
\begin{equation}
A\to\frac{\pi}{2},
\end{equation}
which is nothing but the solution to
Eq.\ (\ref{eq: RPA poles a})
in the limit $k^{\,}_{y}\to0$ and $s\to0$, namely the solution to
\begin{equation}
0=\left(1-\frac{2}{\pi}\,A\right)^{2}.
\end{equation}
The condition
\begin{equation}
-\frac{Z^{2}_{0}\,m^{\,}_{\mathrm{t}}}{2{M}}=\frac{\pi}{2}
\label{eq: gap closing condition}
\end{equation}
is nothing but the RPA counterpart to the
mean-field transition from the ATO to the NATO phases
by which the number of Majorana edge states changes.
The numerical value of the condition (\ref{eq: gap closing condition})
with $Z^{2}_{0}\approx0.85$
is
\begin{equation}
\frac{|m^{\,}_{\mathrm{t}}|}{{M}}\approx3.7.
\end{equation}

When $|A-(\pi/2)|\ll1$, we can expand the right-hand side of Eq.\
(\ref{eq: RPA poles a}) in powers of $s/(2M)$
with the help of the asymptotic expansion (\ref{eq: RPA poles d}).
One finds the dispersion
\begin{subequations}
\begin{equation}
\begin{split}
\omega^{2}\approx&\,
k^{2}_{x}
+
\left[
\frac{32}{3\pi A}\cos k^{\,}_{y}
+
\left(
1
-
\frac{64}{3\pi^{2}}
\right)
\right]^{-1}  
\left(\frac{4M}{A}\right)^{2}
\\
&\,
\times
\left[
\sin^{2}k^{\,}_{y}
+
\left(\frac{2A}{\pi}-\cos k^{\,}_{y}\right)^{2}
\right].
\end{split}
\label{eq: RPA approximate dispersion relation}
\end{equation}
The squared mass
[take $A\approx\pi/2$ in the first square bracket and $k^{\,}_{x}=k^{\,}_{y}=0$
on the right-hand-side of Eq.\
(\ref{eq: RPA approximate dispersion relation})]
\begin{equation}
m^{2}_{\mathrm{RPA}}\approx
\left(\frac{8M}{\pi\,A}\right)^{2}
\left(A-\frac{\pi}{2}\right)^{2}\ll M^{2}
\end{equation}
\end{subequations}
for the triplet of Majorana fields follows.

At values of $A>\pi/2$ the gap increases fast. It should also be noted
that the dispersion does not include the entire Brillouin zone; there is a
critical value of $k^{\,}_{y}$ beyond which it crosses into the two-soliton
continuum above the energy threshold $2M$.

\section{Two-dimensional Majorana fermions, one-dimensional solitons}
\label{sec: Two-dimensional Majorana fermions, one-dimensional solitons}

Both the mean-field approach and the one based on combining
the exact solution for the Majorana two-point correlation functions
of the one-dimensional Gross-Neveu Hamiltonian
(\ref{eq: desired Majorana theory mt=0})
with the RPA tell us that the excitations of the
model of coupled wires obeying periodic boundary conditions
in all space directions can include Majorana modes.

In the limit $m^{\,}_{\mathrm{t}}=m^{\,}_{\mathrm{s}}=0$, the low
lying excitations of the Gross-Neveu Hamiltonian
(\ref{eq: desired Majorana theory mt=0}) are
exclusively made of solitons.
These solitons propagate along the $x$ direction only
(i.e., in one dimension only)
above the energy threshold $M$ introduced in
Sec.\ \ref{sec: RPA}. Remarkably, these
solitons are also present in the spectrum when a small in magnitude
$m^{\,}_{\mathrm{t}}\ne 0$ is added to the Gross-Neveu
Hamiltonian (\ref{eq: desired Majorana theory mt=0}), i.e., they are
not confined by the crossover to two-dimensional space  induced by
the coupling $m^{\,}_{\mathrm{t}}\ne 0$.
To arrive at this conclusion, we proceed as follows.

We are going to show that
the symmetry (\ref{eq: symmetries Majorana Grassmann c}),
which implies the conservation of the $\mathtt{m}$-resolved Majorana parity,
(i) cannot be spontaneously broken at any non-vanishing temperature $T>0$,
(ii) is spontaneously broken at zero temperature $T=0$.
The free-energy argument underlying claim (i) is that there are
gapped one-dimensional excitations 
of solitonic character in the many-body excitation spectrum
of Hamiltonian (\ref{eq: desired Majorana theory})
above the energy threshold $M$.
Their Boltzmann weight
at the temperature $T$ is of order $e^{-M/T}$ so that their
average separation is of order
\begin{equation}
\xi(T)\sim e^{M/T}.
\label{eq: average separation between solitons}
\end{equation}
This length scale thus diverges exponentially fast upon approaching
the zero-temperature limit at which long-range order associated to
the spontaneous symmetry breaking of
the symmetry (\ref{eq: symmetries Majorana Grassmann c})
occurs.

Absence of spontaneous symmetry breaking
of the symmetry (\ref{eq: symmetries Majorana Grassmann c})
at any non-vanishing temperature $T$
is a consequence of the local character of the
symmetry (\ref{eq: symmetries Majorana Grassmann c})
with respect to the label $\mathtt{m}$.
Spontaneous symmetry breaking
of the symmetry (\ref{eq: symmetries Majorana Grassmann c})
at $T=0$ results from the global nature
of the symmetry (\ref{eq: symmetries Majorana Grassmann c})
with respect to imaginary time $\tau$ and the coordinate $x$.

The proof of claim (i) goes as follows.
Integrating the Majorana fermions
in the partition function
(\ref{eq: def partition fct with Hubbard-Stratonovich decoupling})
endows the dynamical field $\phi^{\,}_{\mathtt{m},\mathtt{m}+1}(\tau,x)$
(that carries the engineering dimension of $\mathrm{length}^{-1}$)
with an effective action that must obey
the symmetry (\ref{eq: symmetries Majorana Grassmann c}).

The effective action for the dynamical field
$\phi^{\,}_{\mathtt{m},\mathtt{m}+1}(\tau,x)$
with a local Lagrangian density cannot contain a term such as
\begin{equation}
\mathcal{L}^{\,}_{\kappa}\:=
\kappa\,
\left[
\phi^{\,}_{\mathtt{m},\mathtt{m}+1}(\tau,x)
-
\phi^{\,}_{\mathtt{m}+1,\mathtt{m}+2}(\tau,x)
\right]^{2},
\end{equation}
whereas a term like
\begin{equation}
\mathcal{L}^{\,}_{\zeta}\:=
\frac{\zeta}{M^{2}}\,
\left[
\phi^{2}_{\mathtt{m},\mathtt{m}+1}(\tau,x)
-
\phi^{2}_{\mathtt{m}+1,\mathtt{m}+2}(\tau,x)
\right]^{2}
\end{equation}
is allowed by
the symmetry (\ref{eq: symmetries Majorana Grassmann c}).
Here, the couplings $\kappa$ and $\zeta$ 
are dimensionless.
If we define the length scale
\begin{equation}
\ell\:=\frac{1}{M}
\label{eq: size of domain wall}
\end{equation}
and assume that
$\phi^{\,}_{\mathtt{m},\mathtt{m}+1}$
has the two-soliton profile
\begin{equation}
\begin{split}
\phi^{\,}_{\mathtt{m},\mathtt{m}+1}(\tau,x)\propto&\,
\phi\,
\left[
\mathrm{arctan}\left(\frac{x+R}{\ell}\right)
\right.
\\
&
\left.
\qquad
-
\mathrm{arctan}\left(\frac{x-R}{\ell}\right)
\right],
\end{split}
\end{equation}
along the $x$ direction ($\phi>0$ is arbitrary),
we find that, if $R\gg\ell$, 
the action penalties are given by
\begin{equation}
S^{\,}_{\kappa}\sim
\frac{\kappa\,\phi^{2}}{T}\,
R
\label{eq: kappa panality}
\end{equation}
and
\begin{equation}
S^{\,}_{\zeta}\sim
\frac{\zeta\,\ell^{2}\,\phi^{4}}{T}\,\ell,
\label{eq: zeta panality}
\end{equation}
respectively. At any non-vanishing temperature $T>0$,
the action penalty (\ref{eq: kappa panality})
causes the linear confinement of the pair of solitons,
centered at $R$ and $-R$, respectively
At any non-vanishing temperature $T>0$,
the action penalty (\ref{eq: zeta panality})
is independent of the separation $R$ between the pair of solitons
centered at $R$ and $-R$, respectively, i.e., solitons are deconfined.
Thus, at any non-vanishing temperature $T>0$, the thermal fluctuations
that are encoded by the proliferation of solitons that interpolate
between all the symmetry sectors of
the symmetry (\ref{eq: symmetries Majorana Grassmann c})
about any mean-field that breaks
the symmetry (\ref{eq: symmetries Majorana Grassmann c})
restore this symmetry. 

The proof of claim (ii) goes as follows.
At zero temperature, there is no contribution from
the solitons owing to the finite energy of order of $M$ needed to create them.
The fact that the symmetry
(\ref{eq: symmetries Majorana Grassmann c})
is $\mathtt{m}$-resolved is inoperative when $T=0$.
On the other hand, the symmetry (\ref{eq: symmetries Majorana Grassmann c})
is global with respect to $\tau$ and $x$.
Because it is Ising like, the effective quantum action
at zero temperature can be thought of as a set of coupled
Landau-Ginzburg actions, each one of which describes
the classical Ising model in two-dimensional space
and is labeled by the directed bond $\langle\mathtt{m},\mathtt{m}+1\rangle$.
Their coupling is controlled by the Majorana mass $m^{\,}_{\mathrm{t}}$
for the triplet of Majorana fields (we are setting $m^{\,}_{\mathrm{s}}=0$).
Upon decoupling these classical Ising models in two-dimensional space 
by setting $m^{\,}_{\mathrm{t}}=0$, we know from Sec.\ \ref{sec: RPA}
that the Ising symmetry is spontaneously broken. Switching on
$m^{\,}_{\mathrm{t}}\neq0$
only reinforces this spontaneous breaking of the Ising symmetry
as the coupling induced by $m^{\,}_{\mathrm{t}}\neq0$ is not frustrating.
 
It is instructive to establish the degeneracy of the ground state
manifold that is spontaneously broken. In the limit
\begin{equation}
m^{\,}_{\mathrm{t}}=m^{\,}_{\mathrm{s}}=0,
\end{equation}
the Majorana modes decouple into the non-chiral pairs
$\chi^{\mu}_{\mathrm{L},\mathtt{m}}$
and
$\chi^{\mu}_{\mathrm{R},\mathtt{m}+1}$
with $\mu=0,1,2,3$, i.e., four flavors of Majorana fields
of opposite chiralities for each directed bond
$\langle\mathtt{m},\mathtt{m}+1\rangle$.
For each directed bond
$\langle\mathtt{m},\mathtt{m}+1\rangle$, the corresponding
Majorana fields are  strongly interacting through
a $O(4)$-symmetric Gross-Neveu interaction.
However, the Majorana fields belonging to distinct
directed bonds, say
$\langle\mathtt{m},\mathtt{m}+1\rangle$
and
$\langle\mathtt{m}+1,\mathtt{m}+2\rangle$,
are decoupled.
We may thus identify these pairs of interacting
non-chiral Majorana modes as one-dimensional bundles
labeled by the directed bond variable $\langle\mathtt{m},\mathtt{m}+1\rangle$.
Each bundle $\langle\mathtt{m},\mathtt{m}+1\rangle$ can be bosonized.
The interacting theory for the bundle
$\langle\mathtt{m},\mathtt{m}+1\rangle$
is characterized by the gap
\begin{equation}
\left\langle
e^{
\mathrm{i}
\sqrt{2\pi}
\left(
\widehat{\varphi}^{\,}_{\mathrm{L},\mathtt{m}}
+
\widehat{\varphi}^{\,}_{\mathrm{R},\mathtt{m}+1}
\right)
}
\right\rangle=
\pm
|M|^{1/2}.
\end{equation}
The sign ambiguity on the right-hand side signals the breaking of a global
$\mathbb{Z}^{\,}_{2}$ symmetry for \textit{each} bundle
$\langle\mathtt{m},\mathtt{m}+1\rangle$.
Correspondingly, the soliton-like excitations
are nothing but sine-Gordon solitons, i.e., domain walls
separating regions along the $x$ coordinate
with different signs of the order parameter.
From the point of view of the Majorana fermions
$\chi^{\mu}_{\mathrm{L},\mathtt{m}}$
and
$\chi^{\mu}_{\mathrm{R},\mathtt{m}+1}$
with $\mu=0,1,2,3$,
different vacua are connected by the gauge transformation
that changes a sign of either the left- or the right-moving Majorana fermion.
As follows from Eqs.\ (\ref{eq: fixing matrix elements}) and
(\ref{eq: G1d left right components c}), 
the Green's function $G^{\mathrm{1d}}_{\mathrm{LR}}(\tau,x)$ 
for a given bundle $\langle\mathtt{m},\mathtt{m}+1\rangle$
is proportional to
$\phi^{\,}_{\mathtt{m},\mathtt{m}+1}$
on this bundle. At the mean-field level,
$\phi^{\,}_{\mathtt{m},\mathtt{m}+1}$ is nothing but an order parameter that
breaks the symmetry of the Hamiltonian under
\begin{equation}
\chi^{\mu}_{\mathrm{L},\mathtt{m}}\,
\chi^{\mu}_{\mathrm{R},\mathtt{m}+1}\to
\sigma^{\,}_{\mathrm{L},\mathtt{m}}\,
\sigma^{\,}_{\mathrm{R},\mathtt{m}+1}\,
\chi^{\mu}_{\mathrm{L},\mathtt{m}}\,
\chi^{\mu}_{\mathrm{R},\mathtt{m}+1},
\
\sigma^{\,}_{\mathrm{L},\mathtt{m}},
\sigma^{\,}_{\mathrm{R},\mathtt{m}+1}=\pm1.
\end{equation}
In other words, the sign of the Green's function
(\ref{eq: 1D Green's function c})
is arbitrary. Choosing one sign breaks spontaneously
a two-fold degeneracy for the bundle
$\langle\mathtt{m},\mathtt{m}+1\rangle$.
Given that there are $n$ decoupled bundles of the form
$\langle\mathtt{m},\mathtt{m}+1\rangle$, given the periodic boundary
conditions identifying $\mathtt{m}$ with $\mathtt{m}+n$, one deduces
the degeneracy $2^{n}$ among all the possible symmetry breaking
ground states that can be spontaneously selected.

However, the true degeneracy to be broken spontaneously
in a system with periodic boundary conditions is
$2^{n-1}$ once we switch on
\begin{equation}
m^{\,}_{\mathrm{t}}\neq0
\end{equation}
while retaining $m^{\,}_{\mathrm{s}}=0$.
The symmetry (\ref{eq: symmetries Majorana Grassmann c})
allows us to freely change the sign of the mean-field order parameter
\begin{equation}
\phi^{\,}_{\mathtt{m}',\mathtt{m}'+1}(\tau,x)=
\sigma^{\,}_{\mathtt{m}'}\,
\sigma^{\,}_{\mathtt{m}'+1}\,\phi,
\quad
\mathtt{m}'=1,\cdots,n,
\quad
\phi>0,
\end{equation}
for each bundle $\langle\mathtt{m},\mathtt{m}+1\rangle$, as long as
the global condition
\begin{equation}
\prod_{\mathtt{m'}=1}^{n}
\phi^{\,}_{\mathtt{m}',\mathtt{m}'+1}(\tau,x)=
\pm
\phi^{n}
\label{eq: prod_of_sigmas}
\end{equation}
is satisfied. The sign on the right-hand side of
Eq.\ (\ref{eq: prod_of_sigmas}) is a gauge invariant quantity.
The global condition reduces the number of choices by
half, hence the $2^{n-1}$ ground state degeneracy when
$m^{\,}_{\mathrm{t}}\neq0$.  [Notice that, when
$m^{\,}_{\mathrm{s}}=m^{\,}_{\mathrm{t}}=0$, symmetry
(\ref{eq: symmetries Majorana Grassmann c chiral}) can be used
instead of symmetry (\ref{eq: symmetries Majorana Grassmann c}),
in which case condition (\ref{eq: prod_of_sigmas})
does not apply anymore.]

The symmetry (\ref{eq: symmetries Majorana Grassmann c})
thus implies that the sign of the Green's function
(\ref{eq: 1D Green's function c})
remains arbitrary even if the bundles
$\langle\mathtt{m},\mathtt{m}+1\rangle$
and
$\langle\mathtt{m}+1,\mathtt{m}+2\rangle$
are coupled by having $m^{\,}_{\mathrm{t}}\neq0$.
As we have explained above, it follows that the solitons,
whose existence is guaranteed from bosonization when $m^{\,}_{\mathrm{t}}=0$,
are not confined by the interactions induced by a
$m^{\,}_{\mathrm{t}}\neq0$.
On the other hand, the amplitude of the order parameter undergoes a change
in magnitude in a region of size (\ref{eq: size of domain wall})
around the soliton core.
The soliton energy is sensitive to any change of amplitude in the
order parameter. This is to say that solitons from different
bundles $\langle\mathtt{m},\mathtt{m}+1\rangle$
and $\langle\mathtt{m}+1,\mathtt{m}+2\rangle$
interact when $m^{\,}_{\mathrm{t}}\neq0$. Because solitons cost energy which magnitude is 
bounded from below by the energy scale of the order of $M$,
their average separation $\xi(T)$ is given by
Eq.\ (\ref{eq: average separation between solitons})
at any non-vanishing temperature $T>0$.
The divergence of $\xi(T)$ in
Eq.\ (\ref{eq: average separation between solitons}) is a signature of the onset of long-range order at $T=0$
that breaks spontaneously the symmetry
(\ref{eq: symmetries Majorana Grassmann c}).
Kitaev's honeycomb model
also has an exponentially large correlation length at
$T>0$ related to the thermal creation of local defects,
``visons'' that are localized on the plaquette of the honeycomb lattice.
In that model the defects are not mobile. In our case they are, although
their mobility is one dimensional.

This proof can be generalized to any perturbation local in the
spin operators, as those described in Secs.\
\ref{sec: A single two-leg ladder} and \ref{sec: Coupled two-leg ladders}.
The proof holds since such perturbations are invariant
with respect to a simultaneous change of sign of the left- and
right-moving Majoranas on a given two-leg ladder $\mathtt{m}$.
Our model has two sectors: the spin sector and the fermionic one.
In the latter sector, one is allowed to have operators
which include odd numbers of Majorana fermions on a given two-leg ladder
$\mathtt{m}$. In the spin sector this is not allowed. In the
microscopic derivation which starts with the lattice Hamiltonian of
spins as in Secs.~\ref{sec: A single two-leg ladder} and
\ref{sec: Coupled two-leg ladders}, we arrive to the spin sector only.
Hence, the $2^{n-1}$-degeneracy described above is not directly observable
in the spin sector of our model, that is in the subspace of the Hilbert space 
generated by the local spin operators.
  
\begin{figure*}[t]
\begin{center}
(a)
\includegraphics[width=0.45\textwidth]{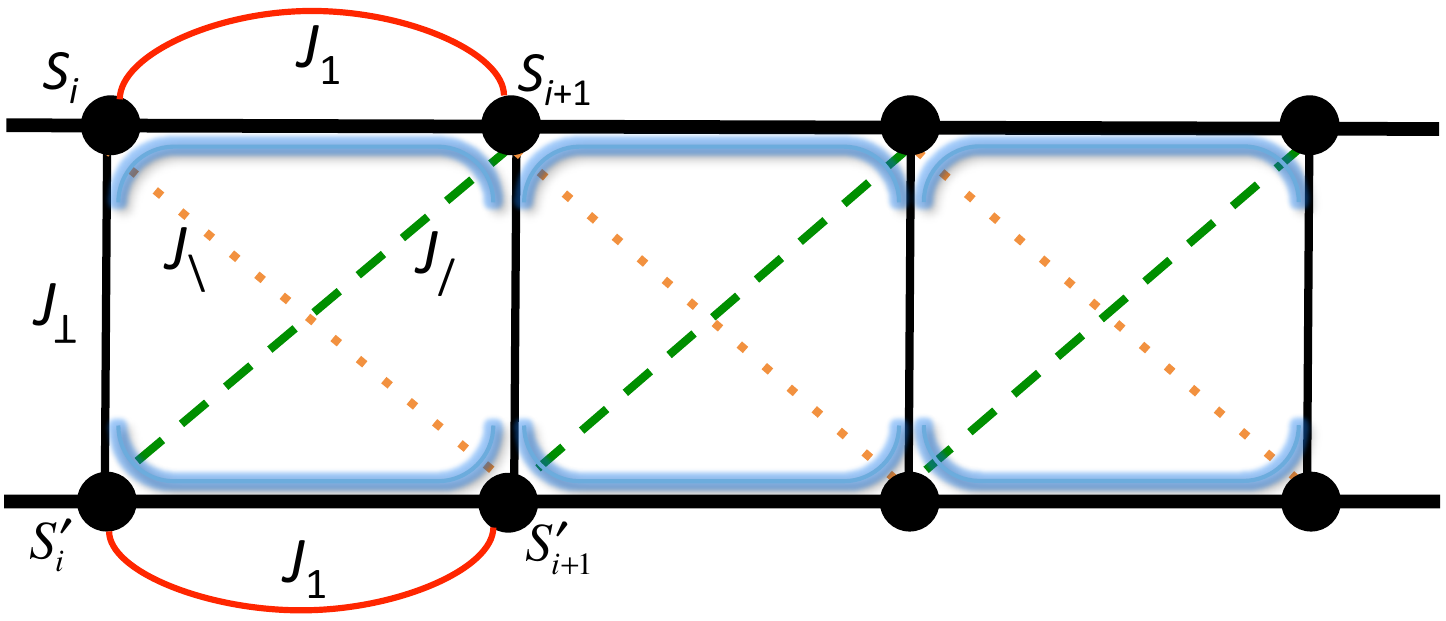}
\hspace{1em}
(b)
\includegraphics[width=0.45\textwidth]{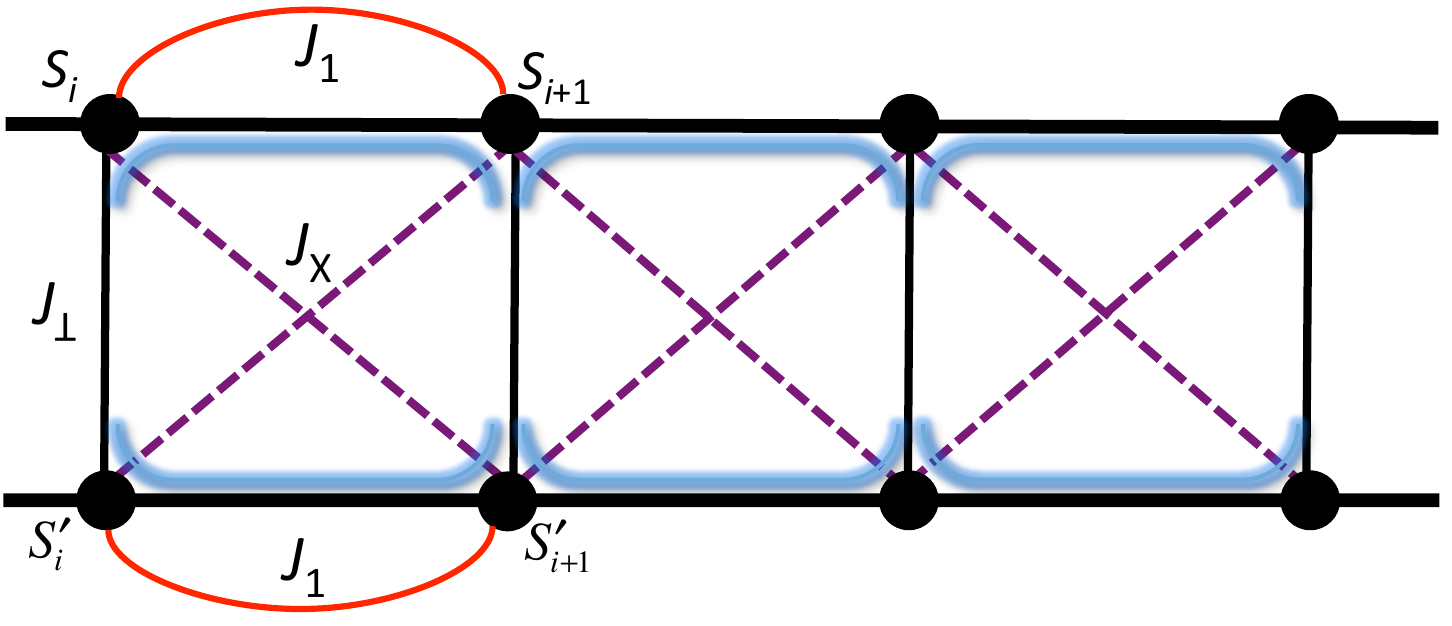}
\caption{(Color online)
(a)
Two quantum spin-1/2 chains can be arranged into a two-leg ladder.
The intra-chain couplings
$J^{\,}_{1}$
are defined in Eqs.\ (\ref{eq: def H ladder b})
and
(\ref{eq: def H ladder c}).
The inter-chain couplings
$J^{\,}_{\perp}$
(represented by the vertical black bond), 
$J^{\,}_{\backslash}$ (represented by the dashed orange bond), 
and
$J^{\,}_{/}$ (represented by the dashed green bond) 
are defined in Eqs.\ (\ref{eq: def H ladder d})
and
(\ref{eq: def H ladder e}).
The inter-chain four-spin coupling $J^{\,}_{U}$
(represented by the blue open-bracket) is defined in
Eq.\ (\ref{eq: def H ladder f}).
(b)
The special case of Fig.~\ref{Fig: ladder model}(a)
under the condition
(\ref{eq: condition for marginal term disappear}).
Here, $J^{\,}_{\times}=-J^{\,}_{\perp}/2$.
\label{Fig: ladder model}
         }
\end{center}
\end{figure*}

\section{A single two-leg ladder}
\label{sec: A single two-leg ladder}

\subsection{Microscopic lattice model and its continuum limit}
\label{subsec: The microscopic lattice model}

Consider the following quantum spin-1/2 Hamiltonian on a (two-leg) ladder
\begin{subequations}
\label{eq: def H ladder}
\begin{equation}
\widehat{H}^{\,}_{\mathrm{ladder}}\:=
\widehat{H}^{\,}_{\mathrm{leg}}
+
\widehat{H}^{\prime}_{\mathrm{leg}}
+
\widehat{H}^{\,}_{\mathrm{rung}}
+
\widehat{H}^{\,}_{\mathrm{cross}}
+
\widehat{H}^{\,}_{\mathrm{four-spin}}.
\label{eq: def H ladder a}
\end{equation}
The first leg of the ladder hosts the quantum spin-1/2 operators
$\widehat{\bm{S}}^{\,}_{i}$ 
on every site $i=1,\cdots,N$, where any two consecutive sites
is displaced by the lattice spacing $\mathfrak{a}$.
Similarly, the second leg of the ladder hosts the quantum spin-1/2 operators
$\widehat{\bm{S}}^{\prime}_{i'}$
on every site $i'=1,\cdots,N$.
Hamiltonians
$\widehat{H}^{\,}_{\mathrm{leg}}$
and
$\widehat{H}^{\prime}_{\mathrm{leg}}$
are a pair of decoupled quantum spin-1/2 antiferromagnetic
Heisenberg model at  criticality given by
\begin{equation}
\widehat{H}^{\,}_{\mathrm{leg}}\:=
\sum_{i=1}^{N}
J^{\,}_{1}\,
\widehat{\bm{S}}^{\,}_{i}
\cdot
\widehat{\bm{S}}^{\,}_{i+1}
\label{eq: def H ladder b}
\end{equation}
and 
\begin{equation}
\widehat{H}^{\prime}_{\mathrm{leg}}\:=
\sum_{i=1}^{N}
J^{\,}_{1}\,
\widehat{\bm{S}}^{\prime}_{i}
\cdot
\widehat{\bm{S}}^{\prime}_{i+1}
\label{eq: def H ladder c}
\end{equation}
with $J^{\,}_{1}\geq0$, respectively.
The quantum spin-1/2 operators on the two legs also interact 
through a $SU(2)$-symmetric Heisenberg exchange interaction for each rung
\begin{equation}
\widehat{H}^{\,}_{\mathrm{rung}}\:=
\sum_{i=1}^{N}
J^{\,}_{\perp}\,
\widehat{\bm{S}}^{\,}_{i}
\cdot
\widehat{\bm{S}}^{\prime}_{i}
\label{eq: def H ladder d}
\end{equation}
with $\mathrm{sgn}(J^{\,}_{\perp})$ arbitrary, 
a cross-type interaction for each plaquette
\begin{equation}
\widehat{H}^{\,}_{\mathrm{cross}}\:=
\sum_{i=1}^{N}
\left(
J^{\,}_{\backslash}\,
\widehat{\bm{S}}^{\,}_{i}
\cdot
\widehat{\bm{S}}^{\prime}_{i+1}
\,+\,
J^{\,}_{/}\,
\widehat{\bm{S}}^{\,}_{i+1}
\cdot
\widehat{\bm{S}}^{\prime}_{i}
\right)
\label{eq: def H ladder e}
\end{equation} 
with $\mathrm{sgn}(J^{\,}_{\backslash}\,J^{\,}_{/})$ arbitrary,
and
a four-spin interaction for each plaquette
\begin{equation}
\widehat{H}^{\,}_{\mathrm{four-spin}}\:=
\sum_{i}^{N}
J^{\,}_{U}\,
\left(
\widehat{\bm{S}}^{\,}_{i}
\cdot
\widehat{\bm{S}}^{\,}_{i+1}
\right)
\left(
\widehat{\bm{S}}^{\prime}_{i}
\cdot
\widehat{\bm{S}}^{\prime}_{i+1}
\right)
\label{eq: def H ladder f}
\end{equation}
\end{subequations}
with $\mathrm{sgn}(J^{\,}_{U})$ arbitrary.
Hamiltonian (\ref{eq: def H ladder}) has the following symmetries.

There is the global unitary $SU(2)$ symmetry generated
by the spin operator
\begin{subequations}
\label{eq: def global SU(2) symmetry generator}
\begin{equation}
\widehat{\bm{S}}^{\,}_{\mathrm{tot}}\:=
\widehat{\bm{S}}
+
\widehat{\bm{S}}^{\prime},
\end{equation}
where
\begin{equation}
\widehat{\bm{S}}\:=
\sum_{i=1}^{N}
\widehat{\bm{S}}^{\,}_{i},
\qquad
\widehat{\bm{S}}^{\prime}\:=
\sum_{i=1}^{N}
\widehat{\bm{S}}^{\prime}_{i}.
\end{equation}
\end{subequations}
We also note that the sum of Hamiltonians (\ref{eq: def H ladder b})
and (\ref{eq: def H ladder c})
has a global $SU(2)\times SU(2)$ symmetry generated independently by
$\widehat{\bm{S}}$ and $\widehat{\bm{S}}^{\prime}$, respectively.
This global symmetry is
broken down to the diagonal subgroup
with the generator (\ref{eq: def global SU(2) symmetry generator})
by the interactions
(\ref{eq: def H ladder d}),
(\ref{eq: def H ladder e}),
and
(\ref{eq: def H ladder f}).

There is the global anti-unitary symmetry under time reversal under which
\begin{equation}
\widehat{\bm{S}}^{\,}_{i}\mapsto
-\widehat{\bm{S}}^{\,}_{i},
\qquad  
\widehat{\bm{S}}^{\prime}_{i}\mapsto
-\widehat{\bm{S}}^{\prime}_{i},
\label{eq: def global Ising symmetry}
\end{equation}
for all $i=1,\cdots,N$.

When the condition
\begin{equation}
J^{\,}_{\backslash}=J^{\,}_{/}\equiv J^{\,}_{\times}
\label{eq: condition for symmetry under exchange unprimed and primed fields}
\end{equation}
holds, there are two additional involutive ($\mathbb{Z}^{\,}_{2}$) symmetries.

Under condition
(\ref{eq: condition for symmetry under exchange unprimed and primed fields}),
Hamiltonian (\ref{eq: def H ladder})
is invariant under the transformation
\begin{equation}
\widehat{\bm{S}}^{\,}_{i}
\mapsto
\widehat{\bm{S}}^{\prime}_{i},
\qquad
\widehat{\bm{S}}^{\prime}_{i}
\mapsto
\widehat{\bm{S}}^{\,}_{i},
\label{eq: def interchange upper and lower spins}
\end{equation}
for all $i=1,\cdots,N$.

Finally, if PBC are imposed together with the condition
(\ref{eq: condition for symmetry under exchange unprimed and primed fields}),
Hamiltonian $\widehat{H}^{\,}_{\mathrm{ladder}}$
is invariant under all lattice translations generated by
\begin{equation}
\widehat{\bm{S}}^{\,}_{i}
\mapsto
\widehat{\bm{S}}^{\,}_{i+1},
\qquad
\widehat{\bm{S}}^{\prime}_{i}
\mapsto
\widehat{\bm{S}}^{\prime}_{i+1},
\label{eq: trsf i to i+1 and i' to i'+1}
\end{equation}
for all $i=1,\cdots,N$.

Figure \ref{Fig: ladder model} depicts 
$\widehat{H}^{\,}_{\mathrm{ladder}}$ 
(\ref{eq: def H ladder}).
This Hamiltonian was studied in Refs.\ \onlinecite{Shelton96,Nersesyan_97}. 
We also refer the reader to Chapter 21 of Ref.~\onlinecite{Gogolin04}
and the Chapter 36 of Ref.\ \onlinecite{Tsvelik03}
for some aspects of $\widehat{H}^{\,}_{\mathrm{ladder}}$.

The naive continuum limit
$\widehat{\mathcal{H}}^{\,}_{\mathrm{ladder}}$
of $\widehat{H}^{\,}_{\mathrm{ladder}}$ 
defined by Eq.~(\ref{eq: def H ladder}) is a
$\widehat{su}(2)^{\,}_{1}\oplus\widehat{su}(2)^{\,}_{1}$ 
Wess-Zumino-Novikov-Witten (WZNW) model perturbed
by local interactions. For the upper leg,
it is obtained by making the replacements
\begin{subequations}
\begin{equation}
i\,\mathfrak{a}\to x,
\qquad
N\,\mathfrak{a}\to L,
\label{appeq: Ansatz for continuum limit aa}
\end{equation}
where the sites of the upper leg of the ladder
are $i=1,\cdots,N$ with $N$ even and
\begin{align}
&
\widehat{\bm{S}}^{\,}_{2i}\to
\mathfrak{a}\,
\left[
\widehat{\bm{J}}^{\,}_{\mathrm{L}}(x)
+
\widehat{\bm{J}}^{\,}_{\mathrm{R}}(x)
+
\widehat{\bm{n}}(x)
\right],
\label{appeq: Ansatz for continuum limit a}
\\
&
\widehat{\bm{S}}^{\,}_{2i+1}\to
\mathfrak{a}\,
\left[
\widehat{\bm{J}}^{\,}_{\mathrm{L}}(x)
+
\widehat{\bm{J}}^{\,}_{\mathrm{R}}(x)
-
\widehat{\bm{n}}(x)
\right],
\label{appeq: Ansatz for continuum limit b}
\\
&
\left(-1\right)^{i}\widehat{\bm{S}}^{\,}_{i}\cdot\widehat{\bm{S}}^{\,}_{i+1}\to
\mathfrak{a}\,\widehat{\varepsilon}(x),
\label{appeq: Ansatz for continuum limit c}
\end{align}
\end{subequations}
for all sites $i=1,\cdots,N/2$ of the upper leg, assuming that $N$ is even.
The left- and right- moving currents
$\widehat{\bm{J}}^{\,}_{\mathrm{L}}$
and
$\widehat{\bm{J}}^{\,}_{\mathrm{R}}$
generate the $\widehat{su}(2)^{\,}_{1}$ affine Lie algebra
of the $c=1$ quantum critical point of the
nearest-neighbor antiferromagnetic quantum spin-1/2 chain.
The fields $\widehat{\bm{n}}$ and $\widehat{\varepsilon}$
have anomalous scaling exponents $1/2$ at this quantum critical point.
The same replacements are done after adding a
prime to the sites and the quantum spin-1/2 hosted
by the lower leg of the ladder. Hereto,
the left- and right- moving currents
$\widehat{\bm{J}}^{\prime}_{\mathrm{L}}$
and
$\widehat{\bm{J}}^{\prime}_{\mathrm{R}}$
generate another $\widehat{su}(2)^{\,}_{1}$ affine Lie algebra,
while the fields $\widehat{\bm{n}}^{\prime}$
and $\widehat{\varepsilon}^{\prime}$
have the anomalous scaling dimensions 1/2 at this quantum critical point.
The perturbation to the 
$\widehat{su}(2)^{\,}_{1}\oplus\widehat{su}(2)^{\,}_{1}$ 
WZNW model with the conserved currents
$\widehat{\bm{J}}^{\,}_{\mathrm{L}}$,
$\widehat{\bm{J}}^{\,}_{\mathrm{R}}$,
$\widehat{\bm{J}}^{\prime}_{\mathrm{L}}$,
and
$\widehat{\bm{J}}^{\prime}_{\mathrm{R}}$
is 
\begin{subequations}
\label{eq: continuum limit square complete}
\begin{equation}
\begin{split}
\widehat{\mathcal{V}}(x)\:=&\,
g^{\,}_{nn}\,
\widehat{\bm{n}}(x)\cdot\widehat{\bm{n}}^{\prime}(x)
+
g^{\,}_{\varepsilon\varepsilon}\,
\widehat{\varepsilon}(x)\,\widehat{\varepsilon}^{\prime}(x)
\\
&
+
g^{\,}_{jj}
\left(
\widehat{\bm{J}}^{\,}_{\mathrm{L}}(x)\cdot\widehat{\bm{J}}^{\prime}_{\mathrm{L}}(x)
+
\widehat{\bm{J}}^{\,}_{\mathrm{R}}(x)\cdot\widehat{\bm{J}}^{\prime}_{\mathrm{R}}(x)
\right)
\\
&
+
g^{\,}_{jj}
\left(
\widehat{\bm{J}}^{\,}_{\mathrm{L}}(x)\cdot\widehat{\bm{J}}^{\prime}_{\mathrm{R}}(x)
+
\widehat{\bm{J}}^{\,}_{\mathrm{R}}(x)\cdot\widehat{\bm{J}}^{\prime}_{\mathrm{L}}(x)
\right)
\\
&
+
g^{\,}_{\text{tw},n}\,
\widehat{\bm{n}}(x)\cdot
\partial^{\,}_{x}\widehat{\bm{n}}^{\prime}(x)
+
g^{\,}_{\text{tw},\varepsilon}\,
\widehat{\varepsilon}(x)
\partial^{\,}_{x}\widehat{\varepsilon}^{\prime}(x)
\end{split}
\label{eq: continuum limit complete a}
\end{equation}
up to irrelevant local perturbations.
Here, the bare values of the couplings are
\begin{align}
&
g^{\,}_{nn}\equiv
\left(g^{\perp}_{nn} +g^{/}_{nn}+g^{\backslash}_{nn}\right)
\\
&
\hphantom{g^{\,}_{nn}}=
2\times\mathfrak{a}
\left(
J^{\,}_{\perp}-J^{\,}_{/}-J^{\,}_{\backslash}
\right)
\label{eq: continuum limit complete b}
\end{align}
for the inter-chain staggered magnetization coupling,
\begin{align}
&
g^{\,}_{\varepsilon\varepsilon}\equiv
\mathfrak{a}\,
J^{\,}_{U}
\label{eq: continuum limit complete c}
\end{align}
for the inter-chain dimerization coupling,
\begin{align}
&
g^{\,}_{jj}\equiv
\left(g^{\perp}_{jj}+g^{/}_{jj}+g^{\backslash}_{jj}\right)\,
\nonumber\\
&
\hphantom{g^{\,}_{jj}}
=
2\times\mathfrak{a}
\left(
J^{\,}_{\perp}+J^{\,}_{/}+J^{\,}_{\backslash}
\right)
\label{eq: continuum limit complete d}
\end{align}
for the inter-chain conformal current coupling,
\begin{align}
&
g^{\,}_{\text{tw},n}\equiv
\left(g^{/}_{\mathrm{tw},n}-g^{\backslash}_{\mathrm{tw},n}\right)\,
\nonumber\\
&
\hphantom{g^{\,}_{\text{tw},n}}
=
\mathfrak{a}^{2}
\left(
J^{\,}_{/}-J^{\,}_{\backslash}
\right)
\label{eq: continuum limit complete e}
\end{align}
for the inter-chain twisted magnetization coupling, and
\begin{align}
&
g^{\,}_{\text{tw},\varepsilon}
\equiv0
\label{eq: continuum limit complete f}
\end{align}
\end{subequations}
for the inter-chain twisted dimerization coupling.

The bare value of the inter-chain conformal current coupling
vanishes if %
~\cite{Shelton96}
\begin{subequations}
\label{eq: condition for marginal term disappear}
\begin{equation}
J^{\,}_{\perp}=
-\left(J^{\,}_{/}+J^{\,}_{\backslash}\right).
\label{eq: condition for marginal term disappear a}
\end{equation}
The bare value of the inter-chain twist magnetization coupling
vanishes if %
~\cite{Allen2000}
\begin{equation}
J^{\,}_{/}=
J^{\,}_{\backslash}
\equiv
J^{\,}_{\times}.
\label{eq: condition for marginal term disappear b}
\end{equation}
\end{subequations}
If we impose conditions
(\ref{eq: condition for marginal term disappear}),
then the effective local interaction
(\ref{eq: continuum limit square complete})
simplifies to
\begin{subequations}
\label{eq: fine tuned intra-ladder continuum limit}
\begin{equation}
\widehat{\mathcal{V}}^{\,}_{\mathrm{tuned}}(x)\:=
g^{\mathrm{tuned}}_{nn}\,
\widehat{\bm{n}}(x)\cdot\widehat{\bm{n}}^{\prime}(x)
+
g^{\,}_{\varepsilon\varepsilon}\,
\widehat{\varepsilon}(x)\,\widehat{\varepsilon}^{\prime}(x),
\label{eq: fine tuned intra-ladder continuum limit a}
\end{equation}
where
\begin{equation}
g^{\mathrm{tuned}}_{nn}\:=
4\times\mathfrak{a}\,J^{\,}_{\perp},
\qquad
g^{\,}_{\varepsilon\varepsilon}\:=
\mathfrak{a}\,J^{\,}_{U}.
\label{eq: fine tuned intra-ladder continuum limit b}
\end{equation}
\end{subequations}
We depict the model $\widehat{H}^{\,}_{\mathrm{ladder}}$ (\ref{eq: def H ladder}) 
under the condition (\ref{eq: condition for marginal term disappear})
in Fig.\ \ref{Fig: ladder model}(b). Its symmetry under
transformations
(\ref{eq: def interchange upper and lower spins})
or
(\ref{eq: trsf i to i+1 and i' to i'+1})
carries over in the continuum limit to the symmetry
by which unprimed and primed fields are exchanged or under sign reversal of
the staggered fields, respectively.

\subsection{Abelian bosonization}
\label{subsec: Abelian bosonization of the single ladder}

\begin{widetext}
To proceed, we follow Ref.\ \onlinecite{Shelton96}
and apply the Abelian bosonization rules on the tuned interaction density
(\ref{eq: fine tuned intra-ladder continuum limit}).
To this end, we consider the upper leg (lower leg) of the ladder
and introduce the pair of bosonic quantum fields
$\widehat{\phi}(t,x)$
and
$\widehat{\theta}(t,x)$
$\left(\widehat{\phi}^{\prime}(t,x)\right.$
and
$\left.\widehat{\theta}^{\prime}(t,x)\right)$
by demanding that they obey the equal-time algebra
\begin{subequations}
\label{eq: def Abelian bosonic fields for single ladder}
\begin{equation}
\left[ 
\widehat{\phi}(t,x), 
\widehat{\theta}(t,x')
\right]= 
-\frac{\mathrm{i}}{2}
\mathrm{sgn}(x-x'),
\qquad
\left[ 
\widehat{\phi}^{\prime}(t,x), 
\widehat{\theta}^{\prime}(t,x')
\right]= 
-\frac{\mathrm{i}}{2}
\mathrm{sgn}(x-x'),
\label{eq: def Abelian bosonic fields for single ladder a}
\end{equation}
for any $t\in\mathbb{R}$, and $0\leq x,x'\leq L^{\,}_{x}$.
The equal-time commutators between unprimed and primed fields are all vanishing.
The two pairs of bosonic fields are related to the staggered magnetization
and staggered dimerization by
\begin{equation}
\widehat{n}^{x}=
+
\frac{1}{\pi\mathfrak{a}}
\cos
\left(
\sqrt{2\pi}\,\widehat{\theta}   
\right),
\qquad
\widehat{n}^{y}=
+
\frac{1}{\pi\mathfrak{a}}
\sin
\left(
\sqrt{2\pi}\,\widehat{\theta}
\right),
\qquad
\widehat{n}^{z}=
-
\frac{1}{\pi\mathfrak{a}}
\sin
\left(
\sqrt{2\pi}\,\widehat{\phi}
\right),
\qquad
\widehat{\varepsilon}=
+
\frac{1}{\pi\mathfrak{a}}
\cos
\left(
\sqrt{2\pi}\,\widehat{\phi}
\right),
\label{eq: def Abelian bosonic fields for single ladder b}
\end{equation}
for the upper leg of the ladder and by
\begin{equation}
\widehat{n}^{\prime x}=
+
\frac{1}{\pi\mathfrak{a}}
\cos
\left(
\sqrt{2\pi}\,\widehat{\theta}^{\prime}   
\right),
\qquad
\widehat{n}^{\prime y}=
+
\frac{1}{\pi\mathfrak{a}}
\sin
\left(
\sqrt{2\pi}\,\widehat{\theta}^{\prime}
\right),
\qquad
\widehat{n}^{\prime z}=
-
\frac{1}{\pi\mathfrak{a}}
\sin
\left(
\sqrt{2\pi}\,\widehat{\phi}^{\prime}
\right),
\qquad
\widehat{\varepsilon}^{\prime}=
+
\frac{1}{\pi\mathfrak{a}}
\cos
\left(
\sqrt{2\pi}\,\widehat{\phi}^{\prime}
\right),
\label{eq: def Abelian bosonic fields for single ladder c}
\end{equation}
for the lower leg of the ladder.
\end{subequations}
After some algebra, we arrive at the Abelian bosonized representation
of the $\widehat{su}(2)^{\,}_{1}\oplus\widehat{su}(2)^{\,}_{1}$ 
WZNW model with the conserved currents
$\widehat{\bm{J}}^{\,}_{\mathrm{L}}$,
$\widehat{\bm{J}}^{\,}_{\mathrm{R}}$,
$\widehat{\bm{J}}^{\prime}_{\mathrm{L}}$,
and
$\widehat{\bm{J}}^{\prime}_{\mathrm{R}}$
perturbed by the intra-ladder tuned interaction density
(\ref{eq: fine tuned intra-ladder continuum limit})
that is given by the Hamiltonian density
\begin{subequations}
\label{eq: Abelian bosonized Hamiltonian for single ladder}
\begin{align}
&
\widehat{\mathcal{H}}\:=
\widehat{\mathcal{H}}^{\mathrm{upper}}_{\mathrm{leg}}
+
\widehat{\mathcal{H}}^{\mathrm{lower}}_{\mathrm{leg}}
+
\widehat{\mathcal{H}}^{\,}_{\mathrm{intra-ladder}},
\\
&
\widehat{\mathcal{H}}^{\mathrm{upper}}_{\mathrm{leg}}\:=
\frac{v}{2}
\left[
\widehat{\Pi}^{2}
+
\left(\partial^{\,}_{x}\widehat{\phi}\right)^{2}
\right],
\qquad
\widehat{\mathcal{H}}^{\mathrm{lower}}_{\mathrm{leg}}\:=
\frac{v}{2}
\left[
\widehat{\Pi}^{\prime 2}
+
\left(\partial^{\,}_{x}\widehat{\phi}^{\prime}\right)^{2}
\right],
\\
&
\widehat{\mathcal{H}}^{\,}_{\mathrm{intra-ladder}}\:=
-
\frac{g^{\mathrm{tuned}}_{nn}-g^{\,}_{\varepsilon\varepsilon}}{2(\pi\mathfrak{a})^{2}}\,
\cos
\left(\sqrt{2\pi}\left(\widehat{\phi}+\widehat{\phi}^{\prime}\right)\right)
+
\frac{g^{\mathrm{tuned}}_{nn}+g^{\,}_{\varepsilon\varepsilon}}{2(\pi\mathfrak{a})^{2}}
\cos
\left(\sqrt{2\pi}\left(\widehat{\phi}-\widehat{\phi}^{\prime}\right)\right)
+
\frac{g^{\mathrm{tuned}}_{nn}}{(\pi\mathfrak{a})^{2}}\,
\cos
\left(\sqrt{2\pi}\left(\widehat{\theta}-\widehat{\theta}^{\prime}\right)\right).
\end{align}
Here, we must supplement the equal-time algebra
(\ref{eq: def Abelian bosonic fields for single ladder a})
by the canonical bosonic equal-time commutators
\begin{equation}
\left[ 
\widehat{\phi}(t,x), 
\widehat{\Pi}(t,x')
\right]=
\mathrm{i}\delta(x-x'),
\qquad
\left[ 
\widehat{\phi}^{\prime}(t,x), 
\widehat{\Pi}^{\prime}(t,x')
\right]=
\mathrm{i}\delta(x-x'),
\end{equation}
with
\begin{equation}
\widehat{\Pi}(t,x')\:=
\left(v^{-1}\,\partial^{\,}_{t}\widehat{\phi}\right)(t,x'),
\qquad
\widehat{\Pi}^{\prime}(t,x;)\:=
\left(v^{-1}\,\partial^{\,}_{t}\widehat{\phi}^{\prime}\right)(t,x').
\end{equation}
\end{subequations}
The symmetry under the transformation
(\ref{eq: def interchange upper and lower spins})
follows from the invariance of the bosonic theory
defined by Eq.\
(\ref{eq: Abelian bosonized Hamiltonian for single ladder})
under the transformation
\begin{equation}
\widehat{\theta}\mapsto\widehat{\theta}^{\prime},
\qquad
\widehat{\phi}\mapsto\widehat{\phi}^{\prime},
\qquad
\widehat{\theta}^{\prime}\mapsto\widehat{\theta},
\qquad
\widehat{\phi}^{\prime}\mapsto\widehat{\phi}.
\label{eq: def interchange upper and lowe spins a bis}
\end{equation}
The symmetry under the transformation (\ref{eq: trsf i to i+1 and i' to i'+1})
follows from the invariance of the bosonic theory
defined by Eq.\
(\ref{eq: Abelian bosonized Hamiltonian for single ladder})
under the transformation
\begin{equation}
\widehat{\phi}\mapsto\widehat{\phi}+\sqrt{\frac{\pi}{2}},
\qquad
\widehat{\theta}\mapsto\widehat{\theta}+\sqrt{\frac{\pi}{2}},
\qquad
\widehat{\phi}^{\prime}\mapsto\widehat{\phi}^{\prime}+\sqrt{\frac{\pi}{2}},
\qquad
\widehat{\theta}^{\prime}\mapsto\widehat{\theta}^{\prime}+\sqrt{\frac{\pi}{2}}.
\label{eq: trsf i to i+1 and i' to i'+1 bis}
\end{equation}

\subsection{Majorana representation}
\label{subsec: Majorana representation of the single ladder}

Left- and right-moving Majorana fields are defined by
\begin{subequations}
\label{eq: bosonized Majorana fields}
\begin{align}
&
\widehat{\chi}^{1}_{\mathrm{L}}\:=
\frac{1}{\sqrt{\pi\mathfrak{a}}}\,
\cos
\left(
\sqrt{4\pi}\,
\widehat{\phi}^{\,}_{+,\mathrm{L}}
\right)
\equiv
\frac{1}{\sqrt{\pi\mathfrak{a}}}\,
\cos
\left(
\sqrt{\pi}
\left(
\widehat{\phi}^{\,}_{+}
+
\widehat{\theta}^{\,}_{+}
\right)
\right)\equiv
\frac{1}{\sqrt{\pi\mathfrak{a}}}\,
\cos
\left(
\sqrt{\frac{\pi}{2}}
\left(
\widehat{\phi}
+
\widehat{\phi}^{\prime}
+
\widehat{\theta}
+
\widehat{\theta}^{\prime}
\right)
\right),
\\
&
\widehat{\chi}^{2}_{\mathrm{L}}\:=
\frac{-1}{\sqrt{\pi\mathfrak{a}}}\,
\sin
\left(
\sqrt{4\pi}\,
\widehat{\phi}^{\,}_{+,\mathrm{L}}
\right)\equiv
\frac{-1}{\sqrt{\pi\mathfrak{a}}}\,
\sin
\left(
\sqrt{\pi}
\left(
\widehat{\phi}^{\,}_{+}
+
\widehat{\theta}^{\,}_{+}
\right)
\right)\equiv
\frac{-1}{\sqrt{\pi\mathfrak{a}}}\,
\sin
\left(
\sqrt{\frac{\pi}{2}}
\left(
\widehat{\phi}
+
\widehat{\phi}^{\prime}
+
\widehat{\theta}
+
\widehat{\theta}^{\prime}
\right)
\right),
\\
&
\widehat{\chi}^{3}_{\mathrm{L}}\:=
\frac{1}{\sqrt{\pi\mathfrak{a}}}\,
\cos
\left(
\sqrt{4\pi}\,
\widehat{\phi}^{\,}_{-,\mathrm{L}}
\right)\equiv
\frac{1}{\sqrt{\pi\mathfrak{a}}}\,
\cos
\left(
\sqrt{\pi}
\left(
\widehat{\phi}^{\,}_{-}
+
\widehat{\theta}^{\,}_{-}
\right)
\right)\equiv
\frac{1}{\sqrt{\pi\mathfrak{a}}}\,
\cos
\left(
\sqrt{\frac{\pi}{2}}
\left(
\widehat{\phi}
-
\widehat{\phi}^{\prime}
+
\widehat{\theta}
-
\widehat{\theta}^{\prime}
\right)
\right),
\\
&
\widehat{\chi}^{0}_{\mathrm{L}}\:=
\frac{-1}{\sqrt{\pi\mathfrak{a}}}\,
\sin
\left(
\sqrt{4\pi}\,
\widehat{\phi}^{\,}_{-,\mathrm{L}}
\right)\equiv
\frac{-1}{\sqrt{\pi\mathfrak{a}}}\,
\sin
\left(
\sqrt{\pi}
\left(
\widehat{\phi}^{\,}_{-}
+
\widehat{\theta}^{\,}_{-}
\right)
\right)\equiv
\frac{-1}{\sqrt{\pi\mathfrak{a}}}\,
\sin
\left(
\sqrt{\frac{\pi}{2}}
\left(
\widehat{\phi}
-
\widehat{\phi}^{\prime}
+
\widehat{\theta}
-
\widehat{\theta}^{\prime}
\right)
\right),
\end{align}
and
\begin{align}
&
\widehat{\chi}^{1}_{\mathrm{R}}\:=
\frac{1}{\sqrt{\pi\mathfrak{a}}}\,
\cos
\left(
\sqrt{4\pi}\,
\widehat{\phi}^{\,}_{+,\mathrm{R}}
\right)
\equiv
\frac{1}{\sqrt{\pi\mathfrak{a}}}\,
\cos
\left(
\sqrt{\pi}
\left(
\widehat{\phi}^{\,}_{+}
-
\widehat{\theta}^{\,}_{+}
\right)
\right)\equiv
\frac{1}{\sqrt{\pi\mathfrak{a}}}\,
\cos
\left(
\sqrt{\frac{\pi}{2}}
\left(
\widehat{\phi}
+
\widehat{\phi}^{\prime}
-
\widehat{\theta}
-
\widehat{\theta}^{\prime}
\right)
\right),
\\
&
\widehat{\chi}^{2}_{\mathrm{R}}\:=
\frac{1}{\sqrt{\pi\mathfrak{a}}}\,
\sin
\left(
\sqrt{4\pi}\,
\widehat{\phi}^{\,}_{+,\mathrm{R}}
\right)\equiv
\frac{1}{\sqrt{\pi\mathfrak{a}}}\,
\sin
\left(
\sqrt{\pi}
\left(
\widehat{\phi}^{\,}_{+}
-
\widehat{\theta}^{\,}_{+}
\right)
\right)\equiv
\frac{1}{\sqrt{\pi\mathfrak{a}}}\,
\sin
\left(
\sqrt{\frac{\pi}{2}}
\left(
\widehat{\phi}
+
\widehat{\phi}^{\prime}
-
\widehat{\theta}
-
\widehat{\theta}^{\prime}
\right)
\right),
\\
&
\widehat{\chi}^{3}_{\mathrm{R}}\:=
\frac{1}{\sqrt{\pi\mathfrak{a}}}\,
\cos
\left(
\sqrt{4\pi}\,
\widehat{\phi}^{\,}_{-,\mathrm{R}}
\right)\equiv
\frac{1}{\sqrt{\pi\mathfrak{a}}}\,
\cos
\left(
\sqrt{\pi}
\left(
\widehat{\phi}^{\,}_{-}
-
\widehat{\theta}^{\,}_{-}
\right)
\right)\equiv
\frac{1}{\sqrt{\pi\mathfrak{a}}}\,
\cos
\left(
\sqrt{\frac{\pi}{2}}
\left(
\widehat{\phi}
-
\widehat{\phi}^{\prime}
-
\widehat{\theta}
+
\widehat{\theta}^{\prime}
\right)
\right),
\\
&
\widehat{\chi}^{0}_{\mathrm{R}}\:=
\frac{1}{\sqrt{\pi\mathfrak{a}}}\,
\sin
\left(
\sqrt{4\pi}\,
\widehat{\phi}^{\,}_{-,\mathrm{R}}
\right)\equiv
\frac{1}{\sqrt{\pi\mathfrak{a}}}\,
\sin
\left(
\sqrt{\pi}
\left(
\widehat{\phi}^{\,}_{-}
-
\widehat{\theta}^{\,}_{-}
\right)
\right)\equiv
\frac{1}{\sqrt{\pi\mathfrak{a}}}\,
\sin
\left(
\sqrt{\frac{\pi}{2}}
\left(
\widehat{\phi}
-
\widehat{\phi}^{\prime}
-
\widehat{\theta}
+
\widehat{\theta}^{\prime}
\right)
\right),
\end{align} 
\end{subequations}
respectively.
\end{widetext}

After some algebra, we arrive at the Majorana representation
of the $\widehat{su}(2)^{\,}_{1}\oplus\widehat{su}(2)^{\,}_{1}$ 
WZNW model with the conserved currents
$\widehat{\bm{J}}^{\,}_{\mathrm{L}}$,
$\widehat{\bm{J}}^{\,}_{\mathrm{R}}$,
$\widehat{\bm{J}}^{\prime}_{\mathrm{L}}$,
and
$\widehat{\bm{J}}^{\prime}_{\mathrm{R}}$
perturbed by the intra-ladder interaction density
(\ref{eq: fine tuned intra-ladder continuum limit})
that is given by
\begin{subequations}
\label{eq: intra-ladder after fermionization}
\begin{equation}
\widehat{\mathcal{H}}^{\mathrm{tuned}}_{\mathrm{ladder}}\:=
\sum_{\mu=0,1,2,3}
\widehat{\mathcal{H}}^{\mathrm{tuned}}_{\mathrm{ladder},\mu}
\label{eq: intra-ladder after fermionization a}
\end{equation}
with
\begin{equation}
\widehat{\mathcal{H}}^{\mathrm{tuned}}_{\mathrm{ladder},\mu}\:=
\frac{\mathrm{i}}{2}
v
\left(
\widehat{\chi}^{\mu}_{\mathrm{L}}\partial^{\,}_{x}\widehat{\chi}^{\mu}_{\mathrm{L}}
-
\widehat{\chi}^{\mu}_{\mathrm{R}}\partial^{\,}_{x}\widehat{\chi}^{\mu}_{\mathrm{R}}
\right)
+
\mathrm{i}\,m^{\mathrm{tuned}}_{\mu}\,
\widehat{\chi}^{\mu}_{\mathrm{L}}\,\widehat{\chi}^{\mu}_{\mathrm{R}},
\label{eq: intra-ladder after fermionization b}
\end{equation}
where
\begin{equation}
v\propto J^{\,}_{1}\,\mathfrak{a}
\end{equation}
and
\begin{equation}
m^{\mathrm{tuned}}_{\mu}=
\begin{cases}
m^{\mathrm{tuned}}_{\mathrm{s}},& \mu=0,
\\
m^{\mathrm{tuned}}_{\mathrm{t}},& \mu=1,2,3.
\end{cases}
\label{eq: intra-ladder after fermionization c}
\end{equation}
\end{subequations}
The singlet mass $m^{\mathrm{tuned}}_{\mathrm{s}}$
and the triplet mass $m^{\mathrm{tuned}}_{\mathrm{t}}$ 
are here given by
\begin{subequations}
\label{eq: triplet and singlet mass tuned form}
\begin{align}
m^{\mathrm{tuned}}_{\mathrm{s}}\:=&\,
\frac{-1}{2\pi\mathfrak{a}}
\left(	
3g^{\mathrm{tuned}}_{nn}
+
g^{\,}_{\varepsilon\varepsilon}
\right)
\nonumber
\\
=\,&
\frac{-1}{2\pi}
\left(
12\,
J^{\,}_{\perp}
+
J^{\,}_{U}
\right),
\label{eq: triplet and singlet mass tuned form a}
\\
m^{\mathrm{tuned}}_{\mathrm{t}}\:=&\,
\frac{1}{2\pi\mathfrak{a}}
\left(	
g^{\mathrm{tuned}}_{nn}
-
g^{\,}_{\varepsilon\varepsilon}
\right)
\nonumber
\\
=\,&
\frac{1}{2\pi}
\left(
4\,
J^{\,}_{\perp}
-
J^{\,}_{U}
\right),
\label{eq: triplet and singlet mass tuned form b}
\end{align}
\end{subequations}
respectively. Upon tuning the ratio of $J^{\,}_{U}/J^{\,}_{\perp}$
such that the singlet (triplet) mass $m^{\mathrm{tuned}}_{\mathrm{s}}$
($m^{\mathrm{tuned}}_{\mathrm{t}}$ ) vanish, we achieve the critical
point with central charge $1/2$ ($3/2$) in a single ladder
(\ref{eq: def H ladder}).

\begin{figure*}[t]
\begin{center}
(a)
\includegraphics[width=0.45\textwidth]{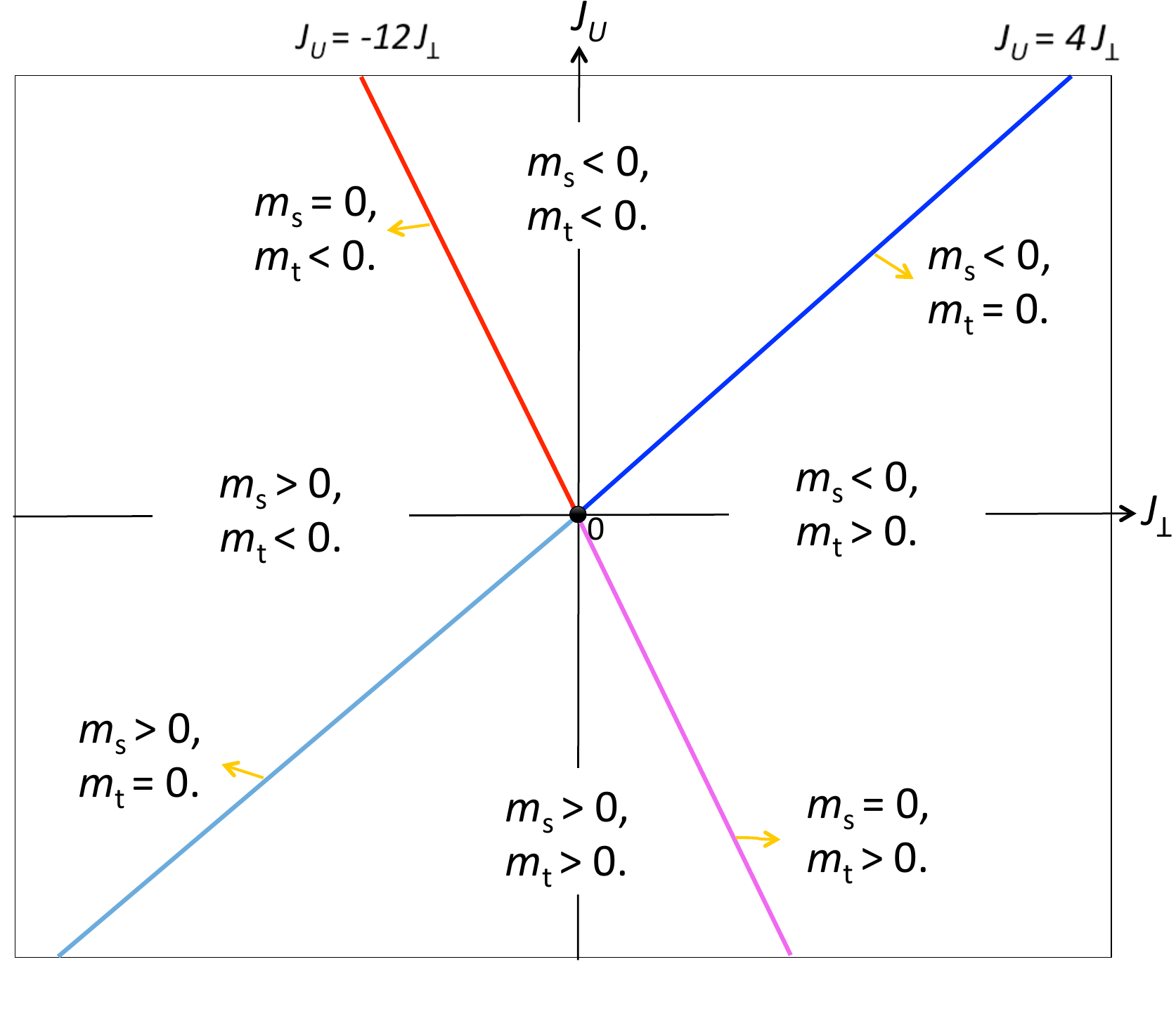}
(b)
\includegraphics[width=0.45\textwidth]{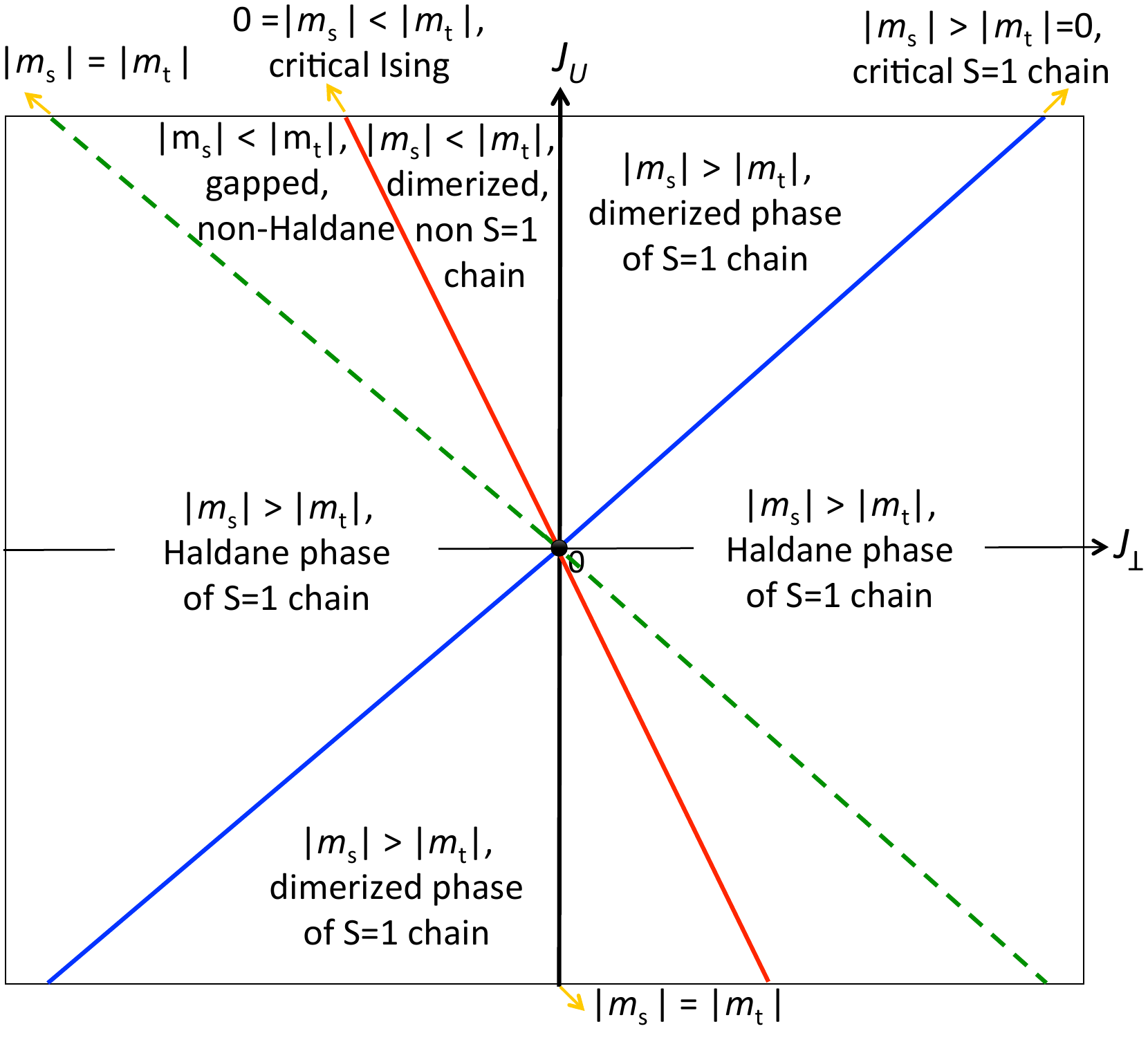}
\caption{(Color online)
(a)
Phase diagram of the fine-tuned quantum spin-1/2 ladder
(\ref{eq: condition for marginal term disappear}) 
based on the sign of the singlet mass and the triplet mass (\ref{eq: triplet and singlet mass tuned form}).
(b)
A refined version of the phase diagram in
Fig.\ \ref{Fig: ladder phase diagram}(a)
is obtained  by considering the difference of the magnitude
of the singlet mass and the triplet mass.
As long as $|m^{\,}_{\mathrm{s}}|>|m^{\,}_{\mathrm{t}}|$,
namely the triplet branch of the spectrum remains the lowest,
the phase is related to the phase of the bilinear and biquadratic spin-1 chain.
\label{Fig: ladder phase diagram}
         }
\end{center}
\end{figure*}

The symmetry under the transformation
(\ref{eq: def interchange upper and lower spins})
is represented by the invariance of the Majorana theory
defined by Eq.\
(\ref{eq: intra-ladder after fermionization})
under the transformation
\begin{subequations}
\label{eq: def interchange upper and lowe spins a bis bis}
\begin{align}
&
\widehat{\chi}^{1}_{\mathrm{M}}\mapsto+\widehat{\chi}^{1}_{\mathrm{M}},
\\
&
\widehat{\chi}^{2}_{\mathrm{M}}\mapsto+\widehat{\chi}^{2}_{\mathrm{M}},
\\
&
\widehat{\chi}^{3}_{\mathrm{M}}\mapsto+\widehat{\chi}^{3}_{\mathrm{M}},
\\
&
\widehat{\chi}^{0}_{\mathrm{M}}\mapsto-\widehat{\chi}^{0}_{\mathrm{M}},
\end{align}
\end{subequations}
for any $\mathrm{M}=\mathrm{L},\mathrm{R}$.

The symmetry under the transformation
(\ref{eq: trsf i to i+1 and i' to i'+1})
is represented in a trivial way for the Majorana theory
defined by Eq.\
(\ref{eq: intra-ladder after fermionization}),
for the transformation (\ref{eq: trsf i to i+1 and i' to i'+1})
is represented by the identity
\begin{equation}
\widehat{\chi}^{\mu}_{\mathrm{M}}\mapsto\widehat{\chi}^{\mu}_{\mathrm{M}}
\label{eq: trsf i to i+1 and i' to i'+1 bis bis}
\end{equation}
for any $\mu=0,1,2,3$ and $\mathrm{M}=\mathrm{L},\mathrm{R}$
according to Eq.\ (\ref{eq: bosonized Majorana fields}).

We follow Ref.~\onlinecite{Nersesyan_97} to discuss the nature of the
phase transition.  According to the sign of the singlet mass and the
triplet mass at the fine-tuned point
(\ref{eq: triplet and singlet mass tuned form}),
we sketch the phase diagram in Fig.\ \ref{Fig: ladder phase diagram}(a).
There are 9 pairs of the signature of
$m^{\,}_{\mathrm{s}}$ and $m^{\,}_{\mathrm{t}}.$ These 9 pairs label 4
phases, 4 critical line and 1 trivial point at the origin.  In
Fig.~\ref{Fig: ladder phase diagram}(b) a refined version of the phase
diagram in Fig.~\ref{Fig: ladder phase diagram}(a) is obtained by
considering the difference of the magnitude of the singlet mass and
the triplet mass.  As long as
$|m^{\,}_{\mathrm{s}}|>|m^{\,}_{\mathrm{t}}|$, namely the triplet
branch of the spectrum remain the lowest, the phase is related to the
phase of the bilinear and biquadratic spin-1 chain.  We note that the
dashed green line is the mirror image of the blue line around the
$J^{\,}_{U}$ axis.  We also remark that the dashed green line is not
present in Fig.~\ref{Fig: ladder phase diagram}(a).

\section{Coupled two-leg ladders}
\label{sec: Coupled two-leg ladders}

\subsection{Microscopic lattice model and its continuum limit}
\label{subsec: The microscopic realization and its continuum limit}

We consider the following inter-ladder interaction
\begin{subequations}
\label{eq: def couplings between consecutive ladders}
\begin{equation}
\widehat{H}^{\,}_{\text{inter-ladder}}\:=
\widehat{H}^{\,}_{\triangle}+\widehat{H}^{\prime}_{\triangle}
+
\widehat{H}^{\,}_{\square}+\widehat{H}^{\prime}_{\square},
\end{equation}  
where
\begin{equation}
\begin{split}
\widehat{H}^{\,}_{\triangle}\:=&\,
\frac{J^{\,}_{\chi}}{2}\,
\sum_{i=1}^{N}
\sum_{\mathtt{m}=1}^{n-1}
\Big[
\widehat{\bm{S}}^{\,}_{i,\mathtt{m}+1}
\cdot
\left(
\widehat{\bm{S}}^{\,}_{i+1,\mathtt{m}}
\wedge
\widehat{\bm{S}}^{\,}_{i,\mathtt{m}}
\right)
\\
&\,
+
\widehat{\bm{S}}^{\,}_{i+1,\mathtt{m}}
\cdot
\left(
\widehat{\bm{S}}^{\,}_{i,\mathtt{m}+1}
\wedge
\widehat{\bm{S}}^{\,}_{i+1,\mathtt{m}+1}
\right)
\Big]
\end{split}
\label{eq: def H triangle}
\end{equation}
and
\begin{equation}
\begin{split}
\widehat{H}^{\,}_{\square}\:=&\,
J^{\,}_{\vee}\,
\sum_{i=1}^{N}
\sum_{\mathtt{m}=1}^{n-1}
\Big(
\widehat{\bm{S}}^{\,}_{i,\mathtt{m}}
\cdot
\widehat{\bm{S}}^{\,}_{i,\mathtt{m}+1}
\\
&\,
+
\kappa^{\,}_{\backslash}\,
\widehat{\bm{S}}^{\,}_{i,\mathtt{m}+1}
\cdot
\widehat{\bm{S}}^{\,}_{i+1,\mathtt{m}}
+
\kappa^{\,}_{/}\,
\widehat{\bm{S}}^{\,}_{i,\mathtt{m}}
\cdot
\widehat{\bm{S}}^{\,}_{i+1,\mathtt{m}+1}
\Big),
\end{split}
\end{equation}
\end{subequations}
with $\widehat{H}^{\prime}_{\triangle}$ and
$\widehat{H}^{\prime}_{\square}$ deduced from
$\widehat{H}^{\,}_{\triangle}$ and $\widehat{H}^{\,}_{\square}$ by the
substitution
$\widehat{\bm{S}}^{\,}_{i,\mathtt{m}}\to\widehat{\bm{S}}^{\prime}_{i,\mathtt{m}}$.
The couplings $\kappa^{\,}_{\backslash}$
and $\kappa^{\,}_{/}$ are dimensionless.
(The choice $\kappa^{\,}_{\backslash}=\kappa^{\,}_{/}=1/2$
is shown in Fig.\ \ref{Fig: 2Dlattice}.)

The inter-ladder Hamiltonian
(\ref{eq: def couplings between consecutive ladders})
has a global $SU(2)\times SU(2)$ symmetry
that reflects the fact that there is no coupling between
the quantum spin $\widehat{\boldsymbol{S}}^{\,}_{i,\mathtt{m}}$
and the quantum spin
$\widehat{\boldsymbol{S}}^{\prime}_{i',\mathtt{m}+1}$
for all $i,i'=1,\cdots,N$.

For the same reason,
the inter-ladder Hamiltonian
(\ref{eq: def couplings between consecutive ladders})
has a global $\mathbb{Z}^{\,}_{2}$ symmetry
under the transformation
[recall Eq.\ (\ref{eq: def interchange upper and lower spins})]
\begin{equation}
\widehat{\bm{S}}^{\,}_{i,\mathtt{m}}
\mapsto
\widehat{\bm{S}}^{\prime}_{i,\mathtt{m}},
\qquad
\widehat{\bm{S}}^{\prime}_{i,\mathtt{m}}
\mapsto
\widehat{\bm{S}}^{\,}_{i,\mathtt{m}},
\label{eq: def interchange upper and lower spins with m}
\end{equation}
for
$i=1,\cdots,N$ and $\mathtt{m}=1,\cdots,n$.

If PBC are imposed on the indices $i$,
the inter-ladder Hamiltonian
(\ref{eq: def couplings between consecutive ladders})
is then invariant under all lattice translations generated by
the transformation
[recall Eq.\ (\ref{eq: trsf i to i+1 and i' to i'+1})]
\begin{equation}
\widehat{\bm{S}}^{\,}_{i,\mathtt{m}}
\mapsto
\widehat{\bm{S}}^{\,}_{i+1,\mathtt{m}},
\qquad
\widehat{\bm{S}}^{\prime}_{i,\mathtt{m}}
\mapsto
\widehat{\bm{S}}^{\prime}_{i+1,\mathtt{m}},
\label{eq: trsf i to i+1 and i' to i'+1 m}
\end{equation}
$i=1,\cdots,N$ and $\mathtt{m}=1,\cdots,n$.

Finally, the inter-ladder Hamiltonian
(\ref{eq: def couplings between consecutive ladders})
is invariant under reversal of time, a global anti-unitary transformation
under which
\begin{subequations}
\label{eq: def global Ising symmetry bis}
\begin{equation}
\widehat{\bm{S}}^{\,}_{i,\mathtt{m}}
\to
-\widehat{\bm{S}}^{\,}_{i,\mathtt{m}},
\qquad
\widehat{\bm{S}}^{\prime}_{i^{\prime},\mathtt{m}'}
\to
-
\widehat{\bm{S}}^{\prime}_{i^{\prime},\mathtt{m}'},
\label{eq: def global Ising symmetry bis a}
\end{equation}
for all
$i,i^{\prime}=1,\cdots,N$ and $\mathtt{m},\mathtt{m}^{\prime}=1,\cdots,n$,
combined with the transformation
\begin{equation}
J^{\,}_{\chi}\mapsto
-J^{\,}_{\chi}.
\label{eq: def global Ising symmetry bis b}
\end{equation}
\end{subequations}
Any fixed non-vanishing $J^{\,}_{\chi}$ breaks time-reversal symmetry.

The naive continuum limit of $\widehat{H}^{\,}_{\text{inter-ladder}}$ 
defined by Eq.~(\ref{eq: def couplings between consecutive ladders})
was derived in Ref.\ \onlinecite{Huang17} (see also Ref.\
\onlinecite{Gorohovsky15}).
All the bare values of the coupling constants
entering Eq.~(\ref{eq: def couplings between consecutive ladders})
that are relevant from the point of view of a
one-loop renormalization group analysis at the WZNW critical point
vanish at the fined-tuned point
\begin{equation}
\kappa^{\,}_{\backslash}=\kappa^{\,}_{/}=1/2,
\label{eq: fine-tuned point relevant inter ladder}
\end{equation}
and the leading-order contribution is simply the current-current interaction
\begin{subequations}
\label{eq: continuum limit of the coupled ladders}
\begin{widetext}
\begin{equation}
\begin{split}
\widehat{\mathcal{H}}^{\,}_{\triangle,\square}(x)\:=
\sum^{n-1}_{\mathtt{m}=1}
\sum^{3}_{a=1}
\left\{
\lambda\,
\left[
\widehat{{J}}^{a}_{\mathrm{L},\mathtt{m}}(x)\,
\widehat{{J}}^{a}_{\mathrm{R},\mathtt{m}+1}(x)
+
\widehat{{J}}^{\prime a}_{\mathrm{L},\mathtt{m}}(x)\,
\widehat{{J}}^{\prime a}_{\mathrm{R},\mathtt{m}+1}(x)
\right]
+
\tilde{\lambda}\,
\left[
\widehat{{J}}^{a}_{\mathrm{R},\mathtt{m}}(x)\,
\widehat{{J}}^{a}_{\mathrm{L},\mathtt{m}+1}(x)
+
\widehat{{J}}^{\prime a}_{\mathrm{R},\mathtt{m}}(x)\,
\widehat{{J}}^{\prime a}_{\mathrm{L},\mathtt{m}+1}(x)
\right]
\right\}
\end{split}
\label{eq: continuum limit of the coupled ladders a}
\end{equation}
with
\begin{equation}
\lambda=2\mathfrak{a}\left[(J^{\,}_{\chi}/\pi)+2J^{\,}_{\vee}\right],
\qquad
\tilde{\lambda}=2\mathfrak{a}\left[-(J^{\,}_{\chi}/\pi)+2J^{\,}_{\vee}\right].
\label{eq: continuum limit of the coupled ladders b}
\end{equation}
\end{widetext}
\end{subequations} 
Here, $\widehat{{J}}^{a}_{\mathrm{M},\mathtt{m}}(x)\in\widehat{su}(2)^{\,}_{k=1}$
and
$\widehat{{J}}^{\prime a}_{\mathrm{M},\mathtt{m}}(x)\in\widehat{su}(2)^{\,}_{k'=1}$
with $\mathrm{M}=\mathrm{L},\mathrm{R}$, i.e., their equal-time commutators
are those of the affine Lie algebras
$\widehat{su}(2)^{\,}_{k=1}$
and
$\widehat{su}(2)^{\,}_{k'=1}$,
respectively.

The global $SU(2)\times SU(2)$ symmetry
of the inter-ladder Hamiltonian
(\ref{eq: def couplings between consecutive ladders})
is manifest in that there is no coupling
between the currents
$\widehat{{J}}^{a}_{\mathrm{M},\mathtt{m}}(x)$
and
$\widehat{{J}}^{\prime a}_{\mathrm{M},\mathtt{m}+1}(x)$.
They are related to the original quantum spin 1/2
by adding the label $\mathtt{m}$ on both sides of
Eqs.\
(\ref{appeq: Ansatz for continuum limit a}),
(\ref{appeq: Ansatz for continuum limit b}),
and (\ref{appeq: Ansatz for continuum limit c})
for the unprimed fields, say.

The global $\mathbb{Z}^{\,}_{2}$ symmetry under
\begin{equation}
\widehat{{J}}^{a}_{\mathrm{M},\mathtt{m}}(x)\mapsto
\widehat{{J}}^{\prime a}_{\mathrm{M},\mathtt{m}}(x),
\qquad
\widehat{{J}}^{\prime a}_{\mathrm{M},\mathtt{m}}(x)\mapsto
\widehat{{J}}^{a}_{\mathrm{M},\mathtt{m}}(x)
\end{equation}
is also manifest.

If PBC are imposed with respect to $x$,
the symmetry of the inter-ladder Hamiltonian
(\ref{eq: def couplings between consecutive ladders})
under all lattice translations generated by
the transformation 
(\ref{eq: trsf i to i+1 and i' to i'+1 m})
is then also manifest in the continuum Hamiltonian density
(\ref{eq: continuum limit of the coupled ladders}),
since the currents
$\widehat{{J}}^{a}_{\mathrm{M},\mathtt{m}}(x)$
and
$\widehat{{J}}^{\prime a}_{\mathrm{M},\mathtt{m}}(x)$
are unchanged by the
transformation
(\ref{eq: trsf i to i+1 and i' to i'+1 m}),
unlike the staggered fields on the right-hand sides of
Eqs.\
(\ref{appeq: Ansatz for continuum limit a}),
(\ref{appeq: Ansatz for continuum limit b}),
and (\ref{appeq: Ansatz for continuum limit c})
for the unprimed fields, say.

Remarkably, the continuum Hamiltonian density
(\ref{eq: continuum limit of the coupled ladders})
has acquired an emergent symmetry, namely it is invariant
under the $\mathtt{m}$-resolved transformations
\begin{subequations}
\begin{align}
&
\widehat{\bm{n}}^{\,}_{\mathtt{m}}\mapsto
\sigma^{\,}_{\mathtt{m}}\,\widehat{\bm{n}}^{\,}_{\mathtt{m}},
\qquad
&
\widehat{\bm{n}}^{\prime}_{\mathtt{m}}\mapsto
\sigma^{\,}_{\mathtt{m}}\,\widehat{\bm{n}}^{\prime}_{\mathtt{m}},
\\
&
\widehat{\varepsilon}^{\,}_{\mathtt{m}}\mapsto
\sigma^{\,}_{\mathtt{m}}\,\widehat{\varepsilon}^{\,}_{\mathtt{m}},
\qquad
&
\widehat{\varepsilon}^{\prime}_{\mathtt{m}}\mapsto
\sigma^{\,}_{\mathtt{m}}\,\widehat{\varepsilon}^{\prime}_{\mathtt{m}},
\end{align}
\end{subequations}
where $\sigma^{\,}_{\mathtt{m}}=\pm1$ for $\mathtt{m}=1,\cdots,n$,
unlike the microscopic inter-ladder Hamiltonian
(\ref{eq: def couplings between consecutive ladders})
for which the lattice translation
(\ref{eq: trsf i to i+1 and i' to i'+1 m})
must act simultaneously on all ladder for it to leave
the microscopic inter-ladder Hamiltonian
(\ref{eq: def couplings between consecutive ladders})
invariant.

Reversal of time is explicitly broken by any non-vanishing
$\lambda\neq\tilde{\lambda}$.

Observe that the bare value of $\tilde{\lambda}$ vanishes if 
\begin{equation}
\frac{J^{\,}_{\chi}}{\pi}=2J^{\,}_{\vee}.
\label{eq: fine-tuned point inter ladder break TRS}
\end{equation}	
Upon the fine tuning
(\ref{eq: fine-tuned point inter ladder break TRS}),
the current-current interaction
(\ref{eq: continuum limit of the coupled ladders a}) simplifies to
\begin{equation}
\begin{split}
\widehat{\mathcal{H}}^{\,}_{\mathrm{inter}-\mathrm{ladder}}(x)\:=
\sum^{n-1}_{\mathtt{m}=1}
\sum^{3}_{a=1}
\lambda\,
\Big[&\,
\widehat{{J}}^{a}_{\mathrm{L},\mathtt{m}}(x)
\widehat{{J}}^{a}_{\mathrm{R},\mathtt{m}+1}(x)
\\
&\,
+
\widehat{{J}}^{\prime a}_{\mathrm{L},\mathtt{m}}(x)
\widehat{{J}}^{\prime a}_{\mathrm{R},\mathtt{m}+1}(x)
\Big].
\end{split}
\label{eq: continuum limit of the coupled ladders fine-tuned}
\end{equation}
The interaction $\widehat{\mathcal{H}}^{\,}_{\mathrm{inter}-\mathrm{ladder}}(x)$
is represented by the directed arcs in Fig.\ \ref{Fig: bundle}(a).
The arrow on the arcs indicates that this choice of
current-current interaction completely breaks time-reversal symmetry.

To summarize,
we are considering a set of $n$ ladders labeled by the index
$\mathtt{m}=1,\cdots,n$.
We are assigning the coordinate $x\in\mathbb{R}$
along the direction of the leg to each ladder.
The ladders are all parallel and equally spaced along
a direction $y$ perpendicular to the $x$ axis.
The Hamiltonian for this set of ladders is approximated by
\begin{subequations}
\label{eq: continuum theory of the coupled ladders}
\begin{equation}
\begin{split}
\widehat{H}\:=
\int\limits_{0}^{L^{\,}_{x}}\mathrm{d}x\,
&\,
\left[
\sum^{n}_{\mathtt{m}=1}
\widehat{\mathcal{H}}^{\,}_{\mathrm{WZNW},\mathtt{m}}(x)
+
\sum^{n}_{\mathtt{m}=1}
\widehat{\mathcal{H}}^{\,}_{\mathrm{intra}-\mathrm{ladder},\mathtt{m}}(x)
\right.
\\
&\,
+
\left.
\sum^{n-1}_{\mathtt{m}=1}
\widehat{\mathcal{H}}^{\,}_{\mathrm{inter}-\mathrm{ladder},\mathtt{m}}(x)
\right].
\end{split}
\label{eq: continuum theory of the coupled ladders a}
\end{equation}
The Hamiltonian density
$\widehat{\mathcal{H}}^{\,}_{\mathrm{WZNW},\mathtt{m}}(x)$
encodes the conformal-field theory in two-dimensional space time
with the affine Lie algebra
$\widehat{su}(2)^{\,}_{1}\oplus\widehat{su}(2)^{\,}_{1}$.
It describes a ladder at
a quantum critical point with central charge $c^{\,}_{\mathtt{m}}=2$
where $\mathtt{m}=1,\cdots,n$.
The intra-ladder interaction is
[c.f. Eq.\ (\ref{eq: fine tuned intra-ladder continuum limit})]
\begin{equation}
\begin{split}
\widehat{\mathcal{H}}^{\,}_{\mathrm{intra}-\mathrm{ladder},\mathtt{m}}(x)\:=&\,
g^{\mathrm{tuned}}_{nn}\,
\widehat{\bm{n}}^{\,}_{\mathtt{m}}(x)\cdot\widehat{\bm{n}}^{\prime}_{\mathtt{m}}(x)
\\
&\,
+
g^{\,}_{\varepsilon\varepsilon}\,
\widehat{\varepsilon}^{\,}_{\mathtt{m}}(x)\,
\widehat{\varepsilon}^{\,\prime}_{\mathtt{m}}(x).
\end{split}
\label{eq: continuum theory of the coupled ladders b}
\end{equation}
The couplings
$g^{\mathrm{tuned}}_{nn}$
and
$g^{\,}_{\varepsilon\varepsilon}$
are dimensionless.
They are related to the microscopic data
of the spin-1/2 ladder depicted in Fig.\
\ref{Fig: 2Dlattice}
by Eq.\
(\ref{eq: fine tuned intra-ladder continuum limit b}).
The inter-ladder interaction is
[c.f. Eq.\ (\ref{eq: continuum limit of the coupled ladders fine-tuned})]
\begin{equation}
\begin{split}
\widehat{\mathcal{H}}^{\,}_{\mathrm{inter}-\mathrm{ladder},\mathtt{m}}(x)\:=&\,
\sum^{3}_{a=1}
\lambda\,
\big[
\widehat{{J}}^{a}_{\mathrm{L},\mathtt{m}}(x)
\widehat{{J}}^{a}_{\mathrm{R},\mathtt{m}+1}(x)
\\
&\,\hphantom{\sum^{3}_{a=1}\lambda\,\big[}
+
\widehat{{J}}^{\prime a}_{\mathrm{L},\mathtt{m}}(x)
\widehat{{J}}^{\prime a}_{\mathrm{R},\mathtt{m}+1}(x)
\big],
\end{split}
\label{eq: continuum theory of the coupled ladders c}
\end{equation}
\end{subequations} 
where $\widehat{{J}}^{a}_{\mathrm{M},\mathtt{m}}(x)\in\widehat{su}(2)^{\,}_{k=1}$
and
$\widehat{{J}}^{\prime a}_{\mathrm{M},\mathtt{m}}(x)\in\widehat{su}(2)^{\,}_{k'=1}$
with $\mathrm{M}=\mathrm{L},\mathrm{R}$, i.e., their equal-time commutators
are those of the affine Lie algebras
$\widehat{su}(2)^{\,}_{k=1}$
and
$\widehat{su}(2)^{\,}_{k'=1}$,
respectively. The coupling $\lambda$ is dimensionless.
It is related to the microscopic data
of the spin-1/2 ladder depicted in Fig.\
\ref{Fig: 2Dlattice}
by Eq.\
(\ref{eq: continuum limit of the coupled ladders b}).
The symmetries of Hamiltonian
(\ref{eq: continuum theory of the coupled ladders})
are the following.

The local symmetry with the affine Lie algebra
$\widehat{su}(2)^{\,}_{1}\oplus\widehat{su}(2)^{\,}_{1}$
associated to $\widehat{\mathcal{H}}^{\,}_{\mathrm{WZNW}}(x)$
is reduced to the global $SU(2)\times SU(2)$ symmetry
by the inter-ladder interaction densities,
owing to its invariance under the interchange of unprimed and primed fields.

There is a global $\mathbb{Z}^{\,}_{2}$ symmetry under the
interchange of unprimed and primed fields.

If PBC are imposed,
there is an emergent $\mathtt{m}$-resolved $\mathbb{Z}^{\,}_{2}$ symmetry
under the transformation
\begin{subequations}
\label{eq: continuum implementation of i to i+1 and i' to i'+1}
\begin{align}
&
\widehat{\bm{n}}^{\,}_{\mathtt{m}}\mapsto
\sigma^{\,}_{\mathtt{m}}\,\widehat{\bm{n}}^{\,}_{\mathtt{m}},
\qquad
&
\widehat{\bm{n}}^{\prime}_{\mathtt{m}}\mapsto
\sigma^{\,}_{\mathtt{m}}\,\widehat{\bm{n}}^{\prime}_{\mathtt{m}},
\\
&
\widehat{\varepsilon}^{\,}_{\mathtt{m}}\mapsto
\sigma^{\,}_{\mathtt{m}}\,\widehat{\varepsilon}^{\,}_{\mathtt{m}},
\qquad
&
\widehat{\varepsilon}^{\prime}_{\mathtt{m}}\mapsto
\sigma^{\,}_{\mathtt{m}}\,\widehat{\varepsilon}^{\prime}_{\mathtt{m}},
\\
&
\widehat{\bm{J}}^{\,}_{\mathrm{M},\mathtt{m}}\mapsto
\widehat{\bm{J}}^{\,}_{\mathrm{M},\mathtt{m}},
\qquad
&
\widehat{\bm{J}}^{\prime}_{\mathrm{M},\mathtt{m}}\mapsto
\widehat{\bm{J}}^{\prime}_{\mathrm{M},\mathtt{m}},
\end{align}
\end{subequations}
where $\sigma^{\,}_{\mathtt{m}}=\pm1$ for $\mathtt{m}=1,\cdots,n$
and $\mathrm{M}=\mathrm{L},\mathrm{R}$.

\subsection{Majorana representation}
\label{subsec: Majorana representation of the coupled two-leg ladders}

Since
$\widehat{so}(4)^{\,}_{1}=\widehat{su}(2)^{\,}_{1}\oplus\widehat{su}(2)^{\,}_{1}$,
we can employ four Majorana fields 
$\widehat{\chi}^{\mu}_{\mathrm{M},\mathtt{m}}(x)$ ($\mu=0,1,2,3$) with
$\mathrm{M}=\mathrm{L},\mathrm{R}$ obeying the
equal-time anti-commutators
\begin{equation}
\left\{
\widehat{\chi}^{\mu}_{\mathrm{M},\mathtt{m}}(x),
\widehat{\chi}^{\mu'}_{\mathrm{M}',\mathtt{m}'}(x')
\right\}=
\delta^{\,}_{\mathrm{M}\mathrm{M}'}\,
\delta^{\,}_{\mathtt{m}\mathtt{m}'}\,	
\delta^{\,}_{\mu\mu'}\,
\delta(x-x')
\end{equation}
to describe the
$\widehat{su}(2)^{\,}_{1}\oplus\widehat{su}(2)^{\,}_{1}$ WZNW model for
two decoupled chains making up a single ladder through the Hamiltonian density
\begin{equation}
\widehat{\mathcal{H}}^{\,}_{\mathrm{WZNW},\mathtt{m}}=
\sum^{3}_{\mu=0}
\frac{\mathrm{i}}{2}
v\,
\left(
\widehat{\chi}^{\mu}_{\mathrm{L},\mathtt{m}}
\partial^{\,}_{x}
\widehat{\chi}^{\mu}_{\mathrm{L},\mathtt{m}}
-
\widehat{\chi}^{\mu}_{\mathrm{R},\mathtt{m}}
\partial^{\,}_{x}
\widehat{\chi}^{\mu}_{\mathrm{R},\mathtt{m}}
\right).
\end{equation}
Here, $v$ is the Fermi velocity.
Furthermore, the $\widehat{su}(2)^{\,}_{k=1}$ currents
$\widehat{{J}}^{a}_{\mathrm{M},\mathtt{m}}(x)$ and the
$\widehat{su}(2)^{\,}_{k'=1}$ currents
$\widehat{{J}}^{\prime a}_{\mathrm{M},\mathtt{m}}(x)$
with
$\mathrm{M}=\mathrm{L},\mathrm{R}$, $a=1,2,3$,
and
$\mathtt{m}=1,\cdots,n$ can be represented 
by the bonding linear combination 
\begin{subequations}
\label{eq: currents in terms of Majoranas}
\begin{equation}
\widehat{\mathcal{K}}^{a}_{\mathrm{M},\mathtt{m}}(x)=
\widehat{{J}}^{a}_{\mathrm{M},\mathtt{m}}(x)
+
\widehat{{J}}^{\prime a}_{\mathrm{M},\mathtt{m}}(x)=
-\,
\frac{\mathrm{i}}{2}\epsilon^{abc}
\widehat{\chi}^{b}_{\mathrm{M},\mathtt{m}}(x)\widehat{\chi}^{c}_{\mathrm{M},\mathtt{m}}(x),
\label{eq: currents in terms of Majoranas a}
\end{equation}
and the anti-bonding linear combination
\begin{equation}
\widehat{\mathcal{I}}^{a}_{\mathrm{M},\mathtt{m}}(x)=
\widehat{{J}}^{a}_{\mathrm{M},\mathtt{m}}(x)
-
\widehat{{J}}^{\prime a}_{\mathrm{M},\mathtt{m}}(x)=
-\,
\mathrm{i}
\widehat{\chi}^{0}_{\mathrm{M},\mathtt{m}}(x)\widehat{\chi}^{a}_{\mathrm{M},\mathtt{m}}(x),
\label{eq: currents in terms of Majoranas b}
\end{equation}
\end{subequations}
respectively.
One verifies that $\widehat{\mathcal{K}}^{a}_{\mathrm{M},\mathtt{m}}(x)$ 
and $\widehat{\mathcal{I}}^{a}_{\mathrm{M},\mathtt{m}}(x)$ 
generate a closed $\widehat{su}(2)^{\,}_{1}\oplus\widehat{su}(2)^{\,}_{1}$ algebra.
For later use, we invert Eq.\
(\ref{eq: currents in terms of Majoranas}) to obtain
\vskip 20 true pt
\begin{widetext}
\begin{subequations}
\label{eq: currents in terms of Majoranas invert}
\begin{align}
\widehat{{J}}^{a}_{\mathrm{M},\mathtt{m}}(x)=
\frac{1}{2}
\left(
\widehat{\mathcal{K}}^{a}_{\mathrm{M},\mathtt{m}}(x)
+
\widehat{\mathcal{I}}^{a}_{\mathrm{M},\mathtt{m}}(x)
\right)
=
\frac{-1}{2}
\left(
\frac{\mathrm{i}}{2}\epsilon^{abc}\,
\widehat{\chi}^{b}_{\mathrm{M},\mathtt{m}}(x)\,
\widehat{\chi}^{c}_{\mathrm{M},\mathtt{m}}(x)
+
\mathrm{i}
\widehat{\chi}^{0}_{\mathrm{M},\mathtt{m}}(x)\,
\widehat{\chi}^{a}_{\mathrm{M},\mathtt{m}}(x)
\right),
\label{eq: currents in terms of Majoranas invert a}
\\
\widehat{{J}}^{\prime a}_{\mathrm{M},\mathtt{m}}(x)=
\frac{1}{2}
\left(
\widehat{\mathcal{K}}^{a}_{\mathrm{M},\mathtt{m}}(x)
-
\widehat{\mathcal{I}}^{a}_{\mathrm{M},\mathtt{m}}(x)
\right)
=
\frac{-1}{2}
\left(
\frac{\mathrm{i}}{2}\epsilon^{abc}\,
\widehat{\chi}^{b}_{\mathrm{M},\mathtt{m}}(x)\,
\widehat{\chi}^{c}_{\mathrm{M},\mathtt{m}}(x)
-
\mathrm{i}
\widehat{\chi}^{0}_{\mathrm{M},\mathtt{m}}(x)\,
\widehat{\chi}^{a}_{\mathrm{M},\mathtt{m}}(x)
\right).
\label{eq: currents in terms of Majoranas invert b}
\end{align}
\end{subequations}
There follows several important consequences from
Eq.\ (\ref{eq: currents in terms of Majoranas invert}).
\end{widetext}

First, $\widehat{\chi}^{0}_{\mathrm{M},\mathtt{m}}(x)$
transforms under the global diagonal $SU(2)$ symmetry of the
WZNW Hamiltonian $\widehat{\mathcal{H}}^{\,}_{\mathrm{WZNW}}(x)$
as the singlet (trivial) representation,
while the triplet $\widehat{\chi}^{a}_{\mathrm{M},\mathtt{m}}(x)$
with $a=1,2,3$
transforms under the same $SU(2)$ as the adjoint representation.

Second, reversal of time that is defined by exchanging
left- and right-moving labels together with sign reversal of
$\widehat{{J}}^{a}_{\mathrm{M},\mathtt{m}}(x)$
is represented by complex conjugation in the Fock space spanned
by the Majorana fields together with exchanging
left- and right-moving labels.

Third, the symmetry under
\begin{equation}
\widehat{{J}}^{a}_{\mathrm{M},\mathtt{m}}(x)\mapsto
\widehat{{J}}^{\prime a}_{\mathrm{M},\mathtt{m}}(x)
\qquad
\widehat{{J}}^{\prime a}_{\mathrm{M},\mathtt{m}}(x)\mapsto
\widehat{{J}}^{a}_{\mathrm{M},\mathtt{m}}(x)
\end{equation}
of the WZNW Hamiltonian $\widehat{\mathcal{H}}^{\,}_{\mathrm{WZNW}}(x)$
is represented by
\begin{equation}
\widehat{\chi}^{0}_{\mathrm{M},\mathtt{m}}(x)\mapsto
-\widehat{\chi}^{0}_{\mathrm{M},\mathtt{m}}(x)
\qquad
\widehat{\chi}^{a}_{\mathrm{M},\mathtt{m}}(x)\mapsto
+\widehat{\chi}^{a}_{\mathrm{M},\mathtt{m}}(x)
\end{equation}
in the Majorana representation.

Fourth, the relation between the currents
and the  Majorana fields is one to many since the local gauge
transformation
\begin{equation}
\widehat{\chi}^{\mu}_{\mathrm{M},\mathtt{m}}(x)\mapsto
\sigma^{\,}_{\mathrm{M},\mathtt{m}}(x)\,
\widehat{\chi}^{\mu}_{\mathrm{M},\mathtt{m}}(x)
\end{equation}
where $\sigma^{\,}_{\mathrm{M},\mathtt{m}}(x)=\pm1$
leaves the right-hand side of Eq.\
(\ref{eq: currents in terms of Majoranas invert})
unchanged.

Given the Majorana representation of the 
$\widehat{su}(2)^{\,}_{1}\oplus\widehat{su}(2)^{\,}_{1}$ currents 
entering the inter-ladder interaction
(\ref{eq: currents in terms of Majoranas}), 
we can rewrite
the inter-ladder current-current interactions (\ref{eq: continuum theory of the coupled ladders c}) 
in terms of $4n$ Majorana fields.
More specifically, we calculate the inter-ladder interactions
(\ref{eq: continuum theory of the coupled ladders c})
by making use of Eq.\ (\ref{eq: currents in terms of Majoranas invert}).
For any $\mathtt{m}=1,\cdots,n-1$ and for any $a=1,2,3,$
we start from the inter-ladder interaction
(\ref{eq: continuum theory of the coupled ladders c}),
\begin{widetext}
\begin{align}
\left(
\widehat{{J}}^{a}_{\mathrm{L},\mathtt{m}}\,
\widehat{{J}}^{a}_{\mathrm{R},\mathtt{m}+1}
+
\widehat{{J}}^{\prime a}_{\mathrm{L},\mathtt{m}}\,
\widehat{{J}}^{\prime a}_{\mathrm{R},\mathtt{m}+1}
\right)=
\frac{1}{2}
\left(
\widehat{\mathcal{K}}^{a}_{\mathrm{L},\mathtt{m}}\,
\widehat{\mathcal{K}}^{a}_{\mathrm{R},\mathtt{m}+1}
+
\widehat{\mathcal{I}}^{a}_{\mathrm{L},\mathtt{m}}\,
\widehat{\mathcal{I}}^{a}_{\mathrm{R},\mathtt{m}+1}
\right).
\end{align}
This bilinear form in the currents can be rewritten as a quartic form
in terms of the Majorana fields.
Thus, the inter-ladder interactions
(\ref{eq: continuum theory of the coupled ladders c})
can be written as
\begin{subequations}
\begin{align}
\widehat{\mathcal{H}}^{\,}_{\mathrm{inter}-\mathrm{ladder},\mathtt{m}}=&\, 
\frac{\lambda}{4}
\left[
\left(
\sum^{3}_{a=1}
\widehat{\chi}^{a}_{\mathrm{L},\mathtt{m}}\,
\widehat{\chi}^{a}_{\mathrm{R},\mathtt{m}+1}
\right)^{2}
+
2
\left(
\widehat{\chi}^{0}_{\mathrm{L},\mathtt{m}}\,
\widehat{\chi}^{0}_{\mathrm{R},\mathtt{m}+1}
\right)
\left(
\sum^{3}_{a=1}
\widehat{\chi}^{a}_{\mathrm{L},\mathtt{m}}\,
\widehat{\chi}^{a}_{\mathrm{R},\mathtt{m}+1}
\right)
+
\mathrm{const}
\right]
\label{eq: inter-ladder current current interaction Majorana}
\\
=&\,
\frac{\lambda}{4}
\left(
\sum^{3}_{\mu=0}
\widehat{\chi}^{\mu}_{\mathrm{L},\mathtt{m}}\,
\widehat{\chi}^{\mu}_{\mathrm{R},\mathtt{m}+1}
\right)^{2}
+
\mathrm{const}'.
\label{eq: inter-ladder current current interaction Majorana b O(4) symmetric}
\end{align}
\end{subequations}
We make three observations. 
\end{widetext}

First,
Eq.\ (\ref{eq: inter-ladder current current interaction Majorana})
does not follow if we assume that the coupling $\lambda$
breaks the $SU(2)$ symmetry through a dependence on the index $a=1,2,3$.

Second,
Eq.\
(\ref{eq: inter-ladder current current interaction Majorana b O(4) symmetric})
displays an explicit global
\begin{equation}
O(4)=\mathbb{Z}^{\,}_{2}\times SO(4)=\mathbb{Z}^{\,}_{2}\times SO(3)\times SO(3)
\end{equation}
symmetry. This symmetry is broken down to the diagonal subgroup
\begin{equation}
SO(3)\subset SO(4)=SO(3)\times SO(3)
\end{equation}
if Heisenberg interactions between the quantum spin
$\widehat{\boldsymbol{S}}^{\,}_{i,\mathtt{m}}$
and the quantum spin
$\widehat{\boldsymbol{S}}^{\prime}_{i',\mathtt{m}+1}$
are added to the interaction
(\ref{eq: continuum theory of the coupled ladders}).
Indeed, one verifies that
such microscopic perturbations generate
$SO(3,1)$-symmetric
perturbations of the form
\begin{equation}
\left(
\sum^{3}_{a=1}
\chi^{a}_{\mathrm{L},\mathtt{m}}\,
\chi^{a}_{\mathrm{R},\mathtt{m}+1}
-
\chi^{0}_{\mathrm{L},\mathtt{m}}\,
\chi^{0}_{\mathrm{R},\mathtt{m}+1}
\right)^{2}
\end{equation}
and $SO(3)$-symmetric
perturbations of the form
\begin{equation}
\sum^{3}_{a,b,c=1}
\epsilon^{abc}
\chi^{a}_{\mathrm{L},\mathtt{m}}
\chi^{b}_{\mathrm{R},\mathtt{m}+1}\,
\left(
\chi^{c}_{\mathrm{L},\mathtt{m}}\,
\chi^{0}_{\mathrm{R},\mathtt{m}+1}
+
\chi^{0}_{\mathrm{L},\mathtt{m}}\,
\chi^{c}_{\mathrm{R},\mathtt{m}+1}
\right)
\end{equation}
to the Gross-Neveu-like interaction
(\ref{eq: inter-ladder current current interaction Majorana b O(4) symmetric}).

Third,
the inter-ladder interactions
(\ref{eq: inter-ladder current current interaction Majorana})
resembles the interactions considered in the paper of
Fidkowski and Kitaev~\cite{Fidkowski10}
(see also Ref.\ \onlinecite{Yao13})
in the context of the stability of the 
topological classification of free fermions when perturbed by interactions.

On the other hand,
the intra-ladder interaction
(\ref{eq: continuum theory of the coupled ladders b})
is a mere quadratic form when expressed in terms of the Majoranas
[c.f. Eq.\ (\ref{eq: intra-ladder after fermionization})],
\begin{equation}
\widehat{\mathcal{H}}^{\,}_{\mathrm{intra}-\mathrm{ladder},\mathtt{m}}=
\mathrm{i}\,m^{\,}_{\mathrm{s}}\,
\widehat{\chi}^{0}_{\mathrm{L},\mathtt{m}}\,
\widehat{\chi}^{0}_{\mathrm{R},\mathtt{m}}
+
\sum^{3}_{a=1}
\mathrm{i}\,m^{\,}_{\mathrm{t}}\,
\widehat{\chi}^{a}_{\mathrm{L},\mathtt{m}}\,
\widehat{\chi}^{a}_{\mathrm{R},\mathtt{m}}.
\label{eq: intra-ladder mass Majorana}
\end{equation}
Here, $m^{\,}_{\mathrm{s}},m^{\,}_{\mathrm{t}}\in\mathbb{R}$
are the bare masses of the Majorana fields.

In short, the lattice model presented in Fig.\ \ref{Fig: 2Dlattice}
provides a microscopic realization of the Majorana field theory
(\ref{eq: desired Majorana theory})
with $v^{\,}_{\mu}\equiv v$ for $\mu=0,1,2,3$,
$m^{\,}_{0}\equiv m^{\,}_{\mathrm{s}}$,
and
$m^{\,}_{a}\equiv m^{\,}_{\mathrm{t}}$ for $a=1,2,3$.
Upon the fine tuning
(\ref{eq: condition for marginal term disappear}),
(\ref{eq: fine-tuned point relevant inter ladder}),
(\ref{eq: fine-tuned point inter ladder break TRS}),
and $m^{\,}_{\mathrm{s}}=0$
from Eq.\ (\ref{eq: triplet and singlet mass tuned form a}),
there only remains three independent couplings
out of the seven couplings from the microscopic lattice model.
We choose these three independent microscopic couplings to be
$J^{\,}_{1}$,
$J^{\,}_{\vee}$,
and
$J^{\,}_{\perp}$.  
They condition the values of the velocity $v$,
triplet mass $m^{\,}_{\mathrm{t}}$,
and the coupling constant $\lambda$
for the current-current interaction through
\begin{equation}
v\propto
\mathfrak{a}\,J^{\,}_{1},
\qquad
m^{\,}_{\mathrm{t}}\propto
J^{\,}_{\perp},
\qquad
\lambda\propto
\mathfrak{a}\,J^{\,}_{\vee}.
\end{equation}
To realize a topologically ordered phase, 
we need to choose $J^{\,}_{\vee}>0,$ 
while the signature of $J^{\,}_{\perp}$ is arbitrary
(see Fig.\ \ref{Fig: phase diagram topological order a}).

To summarize, the Majorana representation of
Hamiltonian (\ref{eq: continuum theory of the coupled ladders})
obeying periodic boundary conditions with respect to the coordinates
$x\in[0,L^{\,}_{x}]$ and $\mathtt{m}=1,\cdots,n$
is given by
\begin{subequations}
\label{eq: summary for Majoran rep of coupled ladders}
\begin{align}
&
\widehat{\mathcal{H}}=
\sum_{\mathtt{m}=1}^{n}
\left(
\widehat{\mathcal{H}}^{\,}_{\mathrm{WZNW},\mathtt{m}}
+
\widehat{\mathcal{H}}^{\,}_{\mathrm{intra-ladder},\mathtt{m}}
+
\widehat{\mathcal{H}}^{\,}_{\mathrm{inter-ladder},\mathtt{m}}
\right),
\label{eq: summary for Majoran rep of coupled ladders a}
\\
&
\widehat{\mathcal{H}}^{\,}_{\mathrm{WZNW},\mathtt{m}}=
\sum^{3}_{\mu=0}
\frac{\mathrm{i}}{2}v\,
\left(
\widehat{\chi}^{\mu}_{\mathrm{L},\mathtt{m}}
\partial^{\,}_{x}
\widehat{\chi}^{\mu}_{\mathrm{L},\mathtt{m}}
-
\widehat{\chi}^{\mu}_{\mathrm{R},\mathtt{m}}
\partial^{\,}_{x}
\widehat{\chi}^{\mu}_{\mathrm{R},\mathtt{m}}
\right),
\label{eq: summary for Majoran rep of coupled ladders b}
\\
&
\widehat{\mathcal{H}}^{\,}_{\mathrm{intra}-\mathrm{ladder},\mathtt{m}}= 
\mathrm{i}\,m^{\,}_{\mathrm{s}}\,
\widehat{\chi}^{0}_{\mathrm{L},\mathtt{m}}\,
\widehat{\chi}^{0}_{\mathrm{R},\mathtt{m}}
+
\sum^{3}_{a=1}
\mathrm{i}\,m^{\,}_{\mathrm{t}}\,
\widehat{\chi}^{a}_{\mathrm{L},\mathtt{m}}\,
\widehat{\chi}^{a}_{\mathrm{R},\mathtt{m}},
\label{eq: summary for Majoran rep of coupled ladders c}
\\
&
\widehat{\mathcal{H}}^{\,}_{\mathrm{inter}-\mathrm{ladder},\mathtt{m}}= 
\frac{\lambda}{4}
\left(
\sum^{3}_{\mu=0}
\widehat{\chi}^{\mu}_{\mathrm{L},\mathtt{m}}\,
\widehat{\chi}^{\mu}_{\mathrm{R},\mathtt{m}+1}
\right)^{2}
+
\mathrm{const}'.
\label{eq: summary for Majoran rep of coupled ladders d}
\end{align}
\end{subequations}

\subsection{Abelian bosonization}
\label{subsec: Abelian bosonization of the  coupled two-leg ladder}

It is instructive to use Abelian bosonization to trade the Majorana
representation in Eq.\ (\ref{eq: summary for Majoran rep of coupled ladders})
for a bosonic one. With the help of the conventions from
Secs.\
\ref{subsec: Abelian bosonization of the single ladder}
and
\ref{subsec: Majorana representation of the single ladder}
together with some trigonometric identities, one finds
\begin{widetext}
\begin{subequations}
\label{eq: Abelian bosonized Hamiltonian for coupled ladders}
\begin{align}
&
\widehat{\mathcal{H}}=
\sum_{\mathtt{m}=1}^{n}
\left(
\widehat{\mathcal{H}}^{\mathrm{upper}}_{\mathrm{leg},\mathtt{m}}
+
\widehat{\mathcal{H}}^{\mathrm{lower}}_{\mathrm{leg},\mathtt{m}}
+
\widehat{\mathcal{H}}^{\,}_{\mathrm{intra-ladder},\mathtt{m}}
+
\widehat{\mathcal{H}}^{\,}_{\mathrm{inter-ladder},\mathtt{m}}
\right),
\label{eq: Abelian bosonized Hamiltonian for coupled ladders a}
\\
&
\widehat{\mathcal{H}}^{\mathrm{upper}}_{\mathrm{leg},\mathtt{m}}=
\frac{v}{2}
\left[
\widehat{\Pi}^{2}_{\mathtt{m}}
+
\left(\partial^{\,}_{x}\widehat{\phi}^{\,}_{\mathtt{m}}\right)^{2}
\right],
\qquad
\widehat{\mathcal{H}}^{\mathrm{lower}}_{\mathrm{leg},\mathtt{m}}=
\frac{v}{2}
\left[
\widehat{\Pi}^{\prime 2}_{\mathtt{m}}
+
\left(\partial^{\,}_{x}\widehat{\phi}^{\prime}_{\mathtt{m}}\right)^{2}
\right],
\label{eq: Abelian bosonized Hamiltonian for coupled ladders b}
\\
&
\widehat{\mathcal{H}}^{\,}_{\mathrm{intra-ladder},\mathtt{m}}=
-
\frac{m^{\,}_{\mathrm{t}}}{\pi\mathfrak{a}}\,
\cos
\left(
\sqrt{2\pi}
\left(
\widehat{\phi}^{\,}_{\mathtt{m}}
+
\widehat{\phi}^{\prime}_{\mathtt{m}}
\right)
\right)
-
\frac{
m^{\,}_{\mathrm{t}}
+
m^{\,}_{\mathrm{s}}
     }
     {
2\pi\mathfrak{a}
     }\,
\cos
\left(
\sqrt{2\pi}
\left(
\widehat{\phi}^{\,}_{\mathtt{m}}
-
\widehat{\phi}^{\prime}_{\mathtt{m}}
\right)
\right)
+
\frac{
m^{\,}_{\mathrm{t}}
-
m^{\,}_{\mathrm{s}}
     }
     {
2\pi\mathfrak{a}
     }
\cos
\left(
\sqrt{2\pi}
\left(
\widehat{\theta}^{\,}_{\mathtt{m}}
-
\widehat{\theta}^{\prime}_{\mathtt{m}}
\right)
\right),
\label{eq: Abelian bosonized Hamiltonian for coupled ladders c}
\\
&
\widehat{\mathcal{H}}^{\,}_{\mathrm{inter-ladder},\mathtt{m}}=
-
\frac{\lambda}{4}
\left(
\frac{1}{\pi\mathfrak{a}}\,
\right)^{2}
\left[
\sum_{\pm}
\cos
\left(
\sqrt{\frac{\pi}{2}}
\left[
\widehat{\phi}^{\,}_{\mathtt{m}}
+
\widehat{\theta}^{\,}_{\mathtt{m}}
+
\widehat{\phi}^{\,}_{\mathtt{m}+1}
-
\widehat{\theta}^{\,}_{\mathtt{m}+1}
\pm
\left(
\mathrm{unprimed}
\to
\mathrm{primed}
\right)
\right]
\right)
\right]^{2},
\label{eq: Abelian bosonized Hamiltonian for coupled ladders d}
\end{align}
\end{subequations}
\end{widetext}
where it is understood that the trigonometric functions of the
bosonic fields must be normal ordered.

The symmetry under the transformation
(\ref{eq: def interchange upper and lower spins with m})
becomes the invariance of the bosonic theory
defined by Eq.\
(\ref{eq: Abelian bosonized Hamiltonian for coupled ladders})
under the global transformation
\begin{equation}
\widehat{\theta}^{\,}_{\mathtt{m}}\mapsto
\widehat{\theta}^{\prime}_{\mathtt{m}},
\qquad
\widehat{\phi}^{\,}_{\mathtt{m}}\mapsto
\widehat{\phi}^{\prime}_{\mathtt{m}},
\qquad
\widehat{\theta}^{\prime}_{\mathtt{m}}\mapsto
\widehat{\theta}^{\,}_{\mathtt{m}},
\qquad
\widehat{\phi}^{\prime}_{\mathtt{m}}\mapsto\widehat{\phi}^{\,}_{\mathtt{m}},
\label{eq: def interchange upper and lower spins a bis bis}
\end{equation}
for $\mathtt{m}=1,\cdots,n$.

The symmetry under the transformation
(\ref{eq: trsf i to i+1 and i' to i'+1 m})
becomes the invariance of the bosonic theory
defined by Eq.\
(\ref{eq: Abelian bosonized Hamiltonian for coupled ladders})
under the emergent $\mathtt{m}$-resolved transformation
\begin{equation}
\begin{split}
&
\widehat{\phi}^{\,}_{\mathtt{m}}\mapsto
\widehat{\phi}^{\,}_{\mathtt{m}}
+
\sigma^{\mathrm{stag}}_{\mathtt{m}}\,
\sqrt{\frac{\pi}{2}},
\qquad
\widehat{\theta}^{\,}_{\mathtt{m}}\mapsto
\widehat{\theta}^{\,}_{\mathtt{m}}
+
\sigma^{\mathrm{stag}}_{\mathtt{m}}\,
\sqrt{\frac{\pi}{2}},
\\
&
\widehat{\phi}^{\prime}_{\mathtt{m}}\mapsto
\widehat{\phi}^{\prime}_{\mathtt{m}}
+
\sigma^{\mathrm{stag}}_{\mathtt{m}}\,
\sqrt{\frac{\pi}{2}},
\qquad
\widehat{\theta}^{\prime}_{\mathtt{m}}\mapsto
\widehat{\theta}^{\prime}_{\mathtt{m}}
+
\sigma^{\mathrm{stag}}_{\mathtt{m}}\,
\sqrt{\frac{\pi}{2}},
\end{split}
\label{eq: trsf i to i+1 and i' to i'+1 bis bis bosonic}
\end{equation}
where $\sigma^{\mathrm{stag}}_{\mathtt{m}}\,=0,1$
for $\mathtt{m}=1,\cdots,n$,
for the arguments of the two cosines
on the right-hand side of Eq.\
(\ref{eq: Abelian bosonized Hamiltonian for coupled ladders d})
change at most by $2\pi$ under any one of these transformations.

Evidently,
$\widehat{\mathcal{H}}^{\,}_{\mathrm{intra-ladder},\mathtt{m}}$
and
$\widehat{\mathcal{H}}^{\,}_{\mathrm{intra-ladder},\mathtt{m}+1}$
do not commute with
$\widehat{\mathcal{H}}^{\,}_{\mathrm{inter-ladder},\mathtt{m}}$.
Moreover, it is far from obvious that the limit $\lambda=0$
is nothing but a noninteracting theory of Majorana fields.

However, the bosonic representation
(\ref{eq: Abelian bosonized Hamiltonian for coupled ladders})
becomes advantageous in the limit $m^{\,}_{\mathrm{s}}=m^{\,}_{\mathrm{t}}=0$
for which the intra-ladder interaction vanish,
as we now explain. In this limit,
we are left with the inter-ladder interaction only.
The inter-ladder interaction density consists of squaring the sum over two
cosines that are given by
\begin{subequations}
\label{eq: sum two cosines}
\begin{equation}
\cos
\left(
\sqrt{2\pi}\,
\left(
\widehat{\Xi}^{\,}_{\mathtt{m},\mathtt{m}+1}
+
\widehat{\Xi}^{\prime}_{\mathtt{m},\mathtt{m}+1}
\right)
\right)
\label{eq: sum two cosines a}
\end{equation}
and
\begin{equation}
\cos
\left(
\sqrt{2\pi}\,
\left(
\widehat{\Xi}^{\,}_{\mathtt{m},\mathtt{m}+1}
-
\widehat{\Xi}^{\prime}_{\mathtt{m},\mathtt{m}+1}
\right)
\right),
\label{eq: sum two cosines b}
\end{equation}
respectively, where
\begin{equation}
\widehat{\Xi}^{\,}_{\mathtt{m},\mathtt{m}+1}\:=
\frac{1}{\sqrt{4}}
\left(
\widehat{\phi}^{\,}_{\mathtt{m}}
+
\widehat{\theta}^{\,}_{\mathtt{m}}
+
\widehat{\phi}^{\,}_{\mathtt{m}+1}
-
\widehat{\theta}^{\,}_{\mathtt{m}+1}
\right),
\label{eq: sum two cosines c}
\end{equation}
and
\begin{equation}
\widehat{\Xi}^{\prime}_{\mathtt{m},\mathtt{m}+1}\:=
\frac{1}{\sqrt{4}}
\left(
\widehat{\phi}^{\prime}_{\mathtt{m}}
+
\widehat{\theta}^{\prime}_{\mathtt{m}}
+
\widehat{\phi}^{\prime}_{\mathtt{m}+1}
-
\widehat{\theta}^{\prime}_{\mathtt{m}+1}
\right).
\label{eq: sum two cosines d}
\end{equation}
\end{subequations}
Now, the linear combination
[recall Eq.\ (\ref{eq: bosonized Majorana fields})]
\begin{equation}
\widehat{\phi}^{\,}_{\mathrm{L},\mathtt{m}}\:=
\widehat{\phi}^{\,}_{\mathtt{m}}
+
\widehat{\theta}^{\,}_{\mathtt{m}}
\qquad
\left(
\widehat{\phi}^{\prime}_{\mathrm{L},\mathtt{m}}\:=
\widehat{\phi}^{\prime}_{\mathtt{m}}
+
\widehat{\theta}^{\prime}_{\mathtt{m}}
\right)
\end{equation}
defines a left-moving bosonic field on the upper (lower) leg of ladder
$\mathtt{m}$, while the linear combination
[recall Eq.\ (\ref{eq: bosonized Majorana fields})]
\begin{equation}
\widehat{\phi}^{\,}_{\mathrm{R},\mathtt{m}+1}\:=
\widehat{\phi}^{\,}_{\mathtt{m}+1}
-
\widehat{\theta}^{\,}_{\mathtt{m}+1}
\qquad
\left(
\widehat{\phi}^{\prime}_{\mathrm{L},\mathtt{m}+1}\:=
\widehat{\phi}^{\prime}_{\mathtt{m}+1}
-
\widehat{\theta}^{\prime}_{\mathtt{m}+1}
\right)
\end{equation}
defines a right-moving bosonic field on the upper (lower) leg of ladder
$\mathtt{m}+1$. It follows that, at equal times,
$\widehat{\phi}^{\,}_{\mathrm{L},\mathtt{m}}$
must commute with
$\widehat{\phi}^{\,}_{\mathrm{R},\mathtt{m}+1}$,
$\widehat{\phi}^{\prime}_{\mathrm{L},\mathtt{m}}$
must commute with
$\widehat{\phi}^{\prime}_{\mathrm{R},\mathtt{m}+1}$,
$\widehat{\Xi}^{\,}_{\mathtt{m},\mathtt{m}+1}$
must commute with
$\widehat{\Xi}^{\,}_{\mathtt{m}+1,\mathtt{m}+2}$,
$\widehat{\Xi}^{\prime}_{\mathtt{m},\mathtt{m}+1}$
must commute with
$\widehat{\Xi}^{\prime}_{\mathtt{m}+1,\mathtt{m}+2}$,
the cosine
(\ref{eq: sum two cosines a})
must commute with the cosine
(\ref{eq: sum two cosines b}),
and
\begin{widetext}
\begin{equation}
\widehat{\mathcal{H}}^{\,}_{\mathrm{inter-ladder},\mathtt{m}}=
-
\frac{\lambda}{4}
\left(
\frac{1}{\pi\mathfrak{a}}\,
\right)^{2}
\left[
\cos
\left(
\sqrt{2\pi}
\left(
\widehat{\Xi}^{\,}_{\mathtt{m},\mathtt{m}+1}
+
\widehat{\Xi}^{\prime}_{\mathtt{m},\mathtt{m}+1}
\right)
\right)
+
\cos
\left(
\sqrt{2\pi}
\left(
\widehat{\Xi}^{\,}_{\mathtt{m},\mathtt{m}+1}
-
\widehat{\Xi}^{\prime}_{\mathtt{m},\mathtt{m}+1}
\right)
\right)
\right]^{2}
\label{eq: inter ladder interaction after Abelian bosonization}
\end{equation}
\end{widetext}
must commute with
$\widehat{\mathcal{H}}^{\,}_{\mathrm{inter-ladder},\mathtt{m}'}$
for all $\mathtt{m},\mathtt{m}'=1,\cdots,n$.
Hence, the set of operators
$\{\widehat{\mathcal{H}}^{\,}_{\mathrm{inter-ladder},\mathtt{m}}\}$
labeled by $\mathtt{m}=1,\cdots,n$
can be simultaneously diagonalized by choosing the eigenfields
\begin{subequations}
\label{eq: diagonalization inter ladder interaction}
\begin{equation}
\Xi^{\,}_{\mathtt{m},\mathtt{m}+1}(x)
\pm
\Xi^{\prime}_{\mathtt{m},\mathtt{m}+1}(x)
\label{eq: diagonalization inter ladder interaction a}
\end{equation}
of
\begin{equation}
\widehat{\Xi}^{\,}_{\mathtt{m},\mathtt{m}+1}(x)
\pm
\widehat{\Xi}^{\prime}_{\mathtt{m},\mathtt{m}+1}(x)
\label{eq: diagonalization inter ladder interaction b}
\end{equation}
to be either
\begin{equation}
\Xi^{\,}_{\mathtt{m},\mathtt{m}+1}(x)
\pm
\Xi^{\prime}_{\mathtt{m},\mathtt{m}+1}(x)=
0	
+
\sqrt{2\pi}\,n^{\pm}_{\mathtt{m},\mathtt{m}+1},
\
n^{\pm}_{\mathtt{m},\mathtt{m}+1}\in\mathbb{Z},
\label{eq: diagonalization inter ladder interaction c}
\end{equation}
or
\begin{equation}
\Xi^{\,}_{\mathtt{m},\mathtt{m}+1}(x)
\pm
\Xi^{\prime}_{\mathtt{m},\mathtt{m}+1}(x)=
\sqrt{\frac{\pi}{2}}
+
\sqrt{2\pi}\,n^{\pm}_{\mathtt{m},\mathtt{m}+1},
\
n^{\pm}_{\mathtt{m},\mathtt{m}+1}\in\mathbb{Z}.
\label{eq: diagonalization inter ladder interaction d}
\end{equation}
\end{subequations}
Any eigenvalue from the family
(\ref{eq: diagonalization inter ladder interaction c})
is to be interpreted as the positive expectation value
\begin{subequations}
\begin{equation}
\left\langle\mathrm{GS};+\left|  
\sum^{3}_{\mu=0}
\mathrm{i}
\widehat{\chi}^{\mu}_{\mathrm{L},\mathtt{m}}\,
\widehat{\chi}^{\mu}_{\mathrm{R},\mathtt{m}+1}
\right|\mathrm{GS};+\right\rangle\equiv+C>0
\end{equation}
in the ground state $|\mathrm{GS};+\rangle$.
Any eigenvalue from the family
(\ref{eq: diagonalization inter ladder interaction d})
is to be interpreted as the negative expectation value
\begin{equation}
\left\langle\mathrm{GS};-\left|  
\sum^{3}_{\mu=0}
\mathrm{i}
\widehat{\chi}^{\mu}_{\mathrm{L},\mathtt{m}}\,
\widehat{\chi}^{\mu}_{\mathrm{R},\mathtt{m}+1}
\right|\mathrm{GS};-\right\rangle\equiv-C<0
\end{equation}
\end{subequations}
in the ground state $|\mathrm{GS};-\rangle$.
Any non-vanishing value of $C>0$ breaks spontaneously
the $\mathrm{M}$- and $\mathtt{m}$-resolved symmetry under the
transformation
(\ref{eq: symmetries c chiral})
of Hamiltonian (\ref{eq: summary for Majoran rep of coupled ladders})
in the limit $m^{\,}_{\mathrm{s}}=m^{\,}_{\mathrm{t}}=0$
[any non-vanishing value of $C>0$ also breaks spontaneously
the $\mathtt{m}$-resolved symmetry under the
transformation
(\ref{eq: symmetries c})
of Hamiltonian (\ref{eq: summary for Majoran rep of coupled ladders})
for any one of
$m^{\,}_{\mathrm{s}}$ or $m^{\,}_{\mathrm{t}}$ non-vanishing].

Classical static $\mathtt{m}$-resolved solitons are time-independent eigenfields
(\ref{eq: diagonalization inter ladder interaction a})
that (i) interpolate between any pair from the classical minima enumerated
in Eqs.\
(\ref{eq: diagonalization inter ladder interaction c})
and
(\ref{eq: diagonalization inter ladder interaction d})
as $x$ interpolates from $x=-\infty$ to $x=+\infty$
(ii) and whose energy density is of compact support
with respect to $x\in\mathbb{R}$.

Following Refs.\ \onlinecite{Witten78} and \onlinecite{Shankar78},
we identify among all such solitons
four types of $\mathtt{m}$-resolved elementary solitons.
A type-I $\mathtt{m}$-resolved soliton corresponds to both 
$\Xi^{\,}_{\mathtt{m},\mathtt{m}+1}
+
\Xi^{\prime}_{\mathtt{m},\mathtt{m}+1}$ 
and
$\Xi^{\,}_{\mathtt{m},\mathtt{m}+1}
-
\Xi^{\prime}_{\mathtt{m},\mathtt{m}+1}$ 
increasing monotonically in their values
by the amount $\sqrt{\pi/2}$
between $x=-\infty$ to $x=+\infty$.
A type-II $\mathtt{m}$-resolved soliton corresponds to both 
$\Xi^{\,}_{\mathtt{m},\mathtt{m}+1}
+
\Xi^{\prime}_{\mathtt{m},\mathtt{m}+1}$ 
and
$\Xi^{\,}_{\mathtt{m},\mathtt{m}+1}
-
\Xi^{\prime}_{\mathtt{m},\mathtt{m}+1}$ 
decreasing monotonically in their values
by the amount $\sqrt{\pi/2}$
between $x=-\infty$ to $x=+\infty$.
A type-II $\mathtt{m}$-resolved soliton
can be thought of as an $\mathtt{m}$-resolved anti-soliton of type I.
A type-III $\mathtt{m}$-resolved soliton corresponds to
$\Xi^{\,}_{\mathtt{m},\mathtt{m}+1}
+
\Xi^{\prime}_{\mathtt{m},\mathtt{m}+1}$ 
($\Xi^{\,}_{\mathtt{m},\mathtt{m}+1}
-
\Xi^{\prime}_{\mathtt{m},\mathtt{m}+1}$)
increasing (decreasing) monotonically in value
by the amount $\sqrt{\pi/2}$ between $x=-\infty$ to $x=+\infty$.
A type-IV soliton corresponds to
$\Xi^{\,}_{\mathtt{m},\mathtt{m}+1}
+
\Xi^{\prime}_{\mathtt{m},\mathtt{m}+1}$ 
($\Xi^{\,}_{\mathtt{m},\mathtt{m}+1}
-
\Xi^{\prime}_{\mathtt{m},\mathtt{m}+1}$)
decreasing (increasing) monotonically in values
by the amount $\sqrt{\pi/2}$
between $x=-\infty$ to $x=+\infty$.
A type-IV $\mathtt{m}$-resolved soliton
can be thought of as and $\mathtt{m}$-resolved anti-soliton of type III.
Upon quantization, Witten has shown in
Ref.\ \onlinecite{Witten78}
that we may associate these four types
of elementary solitons to point-like many-body excitations that form
a four-dimensional irreducible representation of a Clifford algebra
with four generators.

Solitons of type I, II, III, and IV
interpolate between any pair with one classical minima
from the family
(\ref{eq: diagonalization inter ladder interaction c})
and the other classical minima from the family
(\ref{eq: diagonalization inter ladder interaction d})
as $x$ interpolates from $x=-\infty$ to $x=+\infty$.
Such solitons should be distinguished from those solitons
that interpolate between any pair with both classical minima
from either one of the two families
(\ref{eq: diagonalization inter ladder interaction c})
and
(\ref{eq: diagonalization inter ladder interaction d})
as $x$ interpolates from $x=-\infty$ to $x=+\infty$.
The former solitons are associated with the spontaneous breaking
of the chiral symmetry. The solitons
associated with two classical minima of either one
of the families 
(\ref{eq: diagonalization inter ladder interaction c})
or
(\ref{eq: diagonalization inter ladder interaction d})
differing by
$\delta n^{+}_{\mathtt{m},\mathtt{m}+1}=\pm1$
while
$\delta n^{-}_{\mathtt{m},\mathtt{m}+1}=0$
are associated with the spontaneous breaking
of the symmetry (\ref{eq: trsf i to i+1 and i' to i'+1 bis bis bosonic}).

We close this discussion by deriving the sine-Gordon
representation (\ref{eq: def double SG theory})
of Hamiltonian
(\ref{eq: Abelian bosonized Hamiltonian for coupled ladders})
in the limit of vanishing intra-ladder interaction.
The sine-Gordon Hamiltonian (\ref{eq: def double SG theory})
follows from expanding the squared bracket on the right-hand side of  
Eq.\ (\ref{eq: inter ladder interaction after Abelian bosonization}).
There are four products of normal-ordered cosine interactions
in this expansion. Two of them involve squaring the same
normal-ordered cosine operator. This exercise requires
combining point splitting with the operator product expansion
and results in a renormalization of the kinetic contributions
(\ref{eq: Abelian bosonized Hamiltonian for coupled ladders b}).
The remaining two products of normal-ordered cosine interactions
involve distinct commuting operators for which we can use
the decomposition rule for the multiplication of two cosines
into the addition of two cosines.
The sine-Gordon Hamiltonian (\ref{eq: def double SG theory})
follows with the identifications
\begin{subequations}
\begin{align}
&
\widehat{\varphi}^{\,}_{\mathrm{s}}\:=
\frac{\sqrt{8\pi}}{\beta}\,
\widehat{\Xi}^{\,}_{\mathtt{m},\mathtt{m}+1},  
\qquad
\widehat{\varphi}^{\,}_{\mathrm{c}}\:=
\frac{\sqrt{8\pi}}{\beta}\,
\widehat{\Xi}^{\prime}_{\mathtt{m},\mathtt{m}+1}.
\end{align}
\end{subequations}

\subsection{Spontaneous symmetry breaking}

We have shown that Hamiltonian
(\ref{eq: Abelian bosonized Hamiltonian for coupled ladders})
commutes with any one of the 2n+1 transformations
\begin{widetext}
\begin{subequations}
\label{eq: def gauge, layers, lattice}
\begin{align}
&
\widehat{\phi}^{\,}_{\mathtt{m}}\mapsto
\widehat{\phi}^{\,}_{\mathtt{m}}
+
2\sigma^{\,}_{\mathtt{m}}\,\sqrt{\frac{\pi}{2}},
\qquad
\widehat{\theta}^{\,}_{\mathtt{m}}\mapsto
\widehat{\theta}^{\,}_{\mathtt{m}},
\qquad
\widehat{\phi}^{\prime}_{\mathtt{m}}\mapsto
\widehat{\phi}^{\prime}_{\mathtt{m}},
\qquad
\widehat{\theta}^{\prime}_{\mathtt{m}}\mapsto
\widehat{\theta}^{\prime}_{\mathtt{m}},
&
\hbox{Majorana redundancy$\qquad\qquad\qquad$}
\label{eq: def gauge, layers, lattice a}
\\
&
\widehat{\phi}^{\,}_{\mathtt{m}}\mapsto
\widehat{\phi}^{\prime}_{\mathtt{m}}
\qquad
\widehat{\theta}^{\,}_{\mathtt{m}}\mapsto
\widehat{\theta}^{\prime}_{\mathtt{m}},
\qquad
\widehat{\phi}^{\prime}_{\mathtt{m}}\mapsto
\widehat{\phi}^{\,}_{\mathtt{m}},
\qquad
\widehat{\theta}^{\prime}_{\mathtt{m}}\mapsto
\widehat{\theta}^{\,}_{\mathtt{m}},
&
\hbox{
$
\widehat{\bm{S}}^{\,}_{i,\mathrm{m}}\leftrightarrow
\widehat{\bm{S}}^{\prime}_{i,\mathrm{m}}\qquad\qquad\qquad
$}
\label{eq: def gauge, layers, lattice b}
\\
&
\widehat{\phi}^{\,}_{\mathtt{m}}\mapsto
\widehat{\phi}^{\,}_{\mathtt{m}}
+
\sigma^{\,}_{\mathtt{m}}\,\sqrt{\frac{\pi}{2}},
\qquad
\widehat{\theta}^{\,}_{\mathtt{m}}\mapsto
\widehat{\theta}^{\,}_{\mathtt{m}}
+
\sigma^{\,}_{\mathtt{m}}\,\sqrt{\frac{\pi}{2}},
\qquad
\widehat{\phi}^{\prime}_{\mathtt{m}}\mapsto
\widehat{\phi}^{\prime}_{\mathtt{m}}
+
\sigma^{\,}_{\mathtt{m}}\,\sqrt{\frac{\pi}{2}},
\qquad
\widehat{\theta}^{\prime}_{\mathtt{m}}\mapsto
\widehat{\theta}^{\prime}_{\mathtt{m}}
+
\sigma^{\,}_{\mathtt{m}}\,\sqrt{\frac{\pi}{2}},
&
\hbox{
$
\widehat{\bm{S}}^{(\prime)}_{i,\mathrm{m}}\mapsto
\widehat{\bm{S}}^{(\prime)}_{i+1,\mathrm{m}}\qquad\qquad\qquad
$}
\label{eq: def gauge, layers, lattice c}
\end{align}
\end{subequations}
with $\sigma^{\,}_{\mathtt{m}}=0,1$ for $\mathtt{m}=1,\cdots,n$.
\end{widetext}
All these transformations commute pairwise and their action on
the trigonometric functions entering Hamiltonian
(\ref{eq: Abelian bosonized Hamiltonian for coupled ladders})
is involutive. Consequently, all many-body energy eigenstates
of Hamiltonian
(\ref{eq: Abelian bosonized Hamiltonian for coupled ladders})
are $2^{2n}$-fold degenerate with the decomposition
\begin{equation}
2^{2n}=2^{n-1}\times2\times2^{n},
\end{equation}
whereby the factor $2^{n-1}$ arises from the Majorana redundancy
as was deduced after Eq.\ (\ref{eq: prod_of_sigmas}),
the factor $2$ arises from the global symmetry under the exchange
of unprimed and primed fields in all ladders, and
the factor $2^{n}$ arises from the reversal in sign of all
the staggered fields in an arbitrarily chosen ladder.
The degeneracy $2^{n-1}$ was shown to be broken spontaneously
at and only at zero temperature through the breaking of the chiral symmetry
encoded by the order parameter (\ref{eq: definition of bond order parameter})
in the Majorana representation. The remaining degeneracy
$2\times2^{n}$ remains unbroken all the way to and at zero temperature,
as is evident from the fact that the chiral order parameter for the Majorana
fermions is invariant under the transformations
(\ref{eq: def gauge, layers, lattice b})
and
(\ref{eq: def gauge, layers, lattice c}).
In particular, the degeneracy $2^{n}$ is invisible to the
Majorana fields.
Hence, the topological degeneracy in the phase diagram
from Fig.\ 
\ref{Fig: phase diagram topological order a}
coexists with the $2^{2n}$ degeneracy associated with
the symmetries
(\ref{eq: def gauge, layers, lattice a}),
(\ref{eq: def gauge, layers, lattice b}),
and
(\ref{eq: def gauge, layers, lattice c}).
Whereas the degeneracy $2^{n-1}$ associated to
the symmetry (\ref{eq: def gauge, layers, lattice a})
originates from the redundancy of the $\mathtt{m}$-resolved
Majorana representation of the conserved chiral currents
(\ref{eq: currents in terms of Majoranas})
and, as such, is intrinsic to the Majorana representation
and invisible to any probe from the Fock space generated by the spin-1/2,
the degeneracy $2\times2^{n}$ associated with the symmetries
(\ref{eq: def gauge, layers, lattice b})
and
(\ref{eq: def gauge, layers, lattice c})
is specific to any microscopic Hamiltonian
with the global symmetry
(\ref{eq: def interchange upper and lower spins with m})
and a ladder-resolved extension of the global symmetry
(\ref{eq: trsf i to i+1 and i' to i'+1 m}).
In other words, the degeneracy $2^{n}$ resulting from the
$\mathtt{m}$-resolved symmetry under the transformation
(\ref{eq: def gauge, layers, lattice c})
is not intrinsic to the microscopic inter-ladder interaction
(\ref{eq: def couplings between consecutive ladders}),
but emerges from neglecting perturbations
to the inter-ladder Hamiltonian
(\ref{eq: continuum theory of the coupled ladders c})
[see also
Eqs.\ (\ref{eq: summary for Majoran rep of coupled ladders d})
or
(\ref{eq: Abelian bosonized Hamiltonian for coupled ladders d})]
that we now discuss.

\subsection{Competing instabilities}

The ATO and NATO phases of the effective low-energy theory
encoded by Hamiltonian (\ref{eq: desired Majorana theory})
compete with the ordered phases that are stabilized
by the longer-range interaction densities of the forms
\begin{subequations}
\label{eq: competing interactions}
\begin{align}
&
\left[
\widehat{\bm{n}}^{\,}_{\mathtt{m}}(x)
+
\sigma\,
\widehat{\bm{n}}^{\prime}_{\mathtt{m}}(x)
\right]
\cdot
\left[
\widehat{\bm{n}}^{\,}_{\mathtt{m}+\mathtt{r}}(x)
+
\sigma'\,
\widehat{\bm{n}}^{\prime}_{\mathtt{m}+\mathtt{r}}(x)
\right],
\label{rel1}
\\
&
\left[
\widehat{\varepsilon}^{\,}_{\mathtt{m}}(x)
+
\sigma\,
\widehat{\varepsilon}^{\prime}_{\mathtt{m}}(x)
\right]
\left[
\widehat{\varepsilon}^{\,}_{\mathtt{m}+\mathtt{r}}(x)
+
\sigma'\,
\widehat{\varepsilon}^{\prime}_{\mathtt{m}+\mathtt{r}}(x)
\right],
\label{rel2}
\\
&
\left[
\widehat{\bm{n}}^{\,}_{\mathtt{m}}(x)
+
\sigma\,
\widehat{\bm{n}}^{\prime}_{\mathtt{m}}(x)
\right]
\cdot
\partial^{\,}_{x}
\left[
\widehat{\bm{n}}^{\,}_{\mathtt{m}+\mathtt{r}}(x)
+
\sigma'\,
\widehat{\bm{n}}^{\prime}_{\mathtt{m}+\mathtt{r}}(x)
\right],
\label{twist1}
\\
&
\left[
\widehat{\varepsilon}^{\,}_{\mathtt{m}}(x)
+
\sigma\,
\widehat{\varepsilon}^{\prime}_{\mathtt{m}}(x)
\right]
\cdot
\partial^{\,}_{x}
\left[
\widehat{\varepsilon}^{\,}_{\mathtt{m}+\mathtt{r}}(x)
+
\sigma'\,
\widehat{\varepsilon}^{\prime}_{\mathtt{m}+\mathtt{r}}(x)
\right],
\label{twist2}
\end{align}
\end{subequations}
where $\sigma,\sigma'=\pm1$ and $|\mathtt{r}|>1$. (Here we recall
that, by design, all the bare couplings vanish for $\mathtt{r}=1$.)

One consequence of these interaction densities is that they
can remove the emergent $\mathtt{m}$-resolved symmetry under the
transformation
(\ref{eq: continuum implementation of i to i+1 and i' to i'+1})
[(\ref{eq: def gauge, layers, lattice c})].
The fate (confinement versus deconfinement) of the solitons associated with
the spontaneous breaking of the symmetry
(\ref{eq: continuum implementation of i to i+1 and i' to i'+1})
[(\ref{eq: def gauge, layers, lattice c})]
in the presence of such symmetry-breaking interactions
is left for future work.

The perturbative renormalization group allows to
assess the potency of the competing interaction densities
(\ref{eq: competing interactions})
relative to the inter-ladder interaction density
(\ref{eq: continuum theory of the coupled ladders c})
[see also
Eqs.\ (\ref{eq: summary for Majoran rep of coupled ladders d})
or
(\ref{eq: Abelian bosonized Hamiltonian for coupled ladders d})].

On the one hand, the interactions densities
(\ref{twist1}) and (\ref{twist2}) are marginal
perturbations of the critical theory
(\ref{eq: summary for Majoran rep of coupled ladders b}).
As such, they can be safely ignored provided their
coupling constants are smaller than the couplings of the leading
current-current interactions.

On the other hand, the interaction densities 
(\ref{rel1}) and (\ref{rel2})
are relevant perturbations of the critical theory
(\ref{eq: summary for Majoran rep of coupled ladders b}).
As such, they present certain difficulties.
For the spin-1/2 lattice Hamiltonian that we have chosen,
their coupling constants are of the order of
$\sim J^{2}_{\vee}/J^{\,}_{1}$,
but cannot be reliably determined otherwise. The problem is
that at small microscopic bare couplings $\chi\ll J^{\,}_{1}$ 
and $J^{\,}_{\vee}\ll J^{\,}_{1}$, the spectral gap $M$
of the effectively one-dimensional Hamiltonian
(\ref{eq: desired Majorana theory mt=0})
that originates from the marginal current-current interaction is
exponentially small in the effective coupling $\lambda$
of Hamiltonian (\ref{eq: desired Majorana theory mt=0}).
Hence, to make $M$ larger than $\sim J^{2}_{\vee}/J^{\,}_{1}$,
we have to go in the region of intermediate to strong bare couplings
$\lambda/J^{\,}_{1}$, with $\lambda$ defined in
Eq.\ (\ref{eq: continuum limit of the coupled ladders b}).
Our methods do not allow to establish whether the required parameter regime 
exists or whether it can be reached for some modification of
the lattice Hamiltonian. 
We take heart from the fact that the required regime of large energy gap
$M\sim J^{\,}_{1}$
exists in  the frustrated quantum spin 1/2 zig-zag ladder with a
three-spin interaction that
was studied by Frahm and R\"odenbeck in Ref.\ \onlinecite{Frahm97}.
They demonstrated that,
under certain conditions corresponding to our $\lambda=0$
in Eq.\ (\ref{eq: continuum theory of the coupled ladders c}),
their ladder is integrable with a spectrum that is partially gapped
with a gap of order of the leading exchange interaction,
$J^{\,}_{1}$ in our setting. In this case,
the mass $M$ exceeds the energy scale for the characteristic energy
scales associated to the relevant interactions from Eq.\
(\ref{eq: competing interactions}), in which case their
neglect in the effective Hamiltonian (\ref{eq: desired Majorana theory})
would be a posteriori consistent.
Ultimately, however, only numerical calculations can establish that
the Majorana field theory (\ref{eq: desired Majorana theory})
captures the low energy physics of the microscopic spin
1/2 lattice model.

\section{Summary}
\label{sec: Summary}

In this paper, we put an emphasis on finding a field theory which
would lead to non-Abelian topological order.  We used the
methods of quantum-field theory to solve a fermionic model with
a topologically nontrivial ground state.
We also presented a candidate lattice spin model 
whose low energy sector is described by this fermionic theory, 
namely, a model of quantum spin $S=1/2$ ladders coupled by a
three-spin interaction.
We found that the bulk spectrum of the fermionic model 
is gapped and that there are robust chiral gapless modes on the
boundaries. There are two topologically nontrivial phases. In one of
them the boundary modes are described by a single species of gapless
Majorana fermions. This phase realizes a Nonabelian Topological Order (NATO).
In the other phase, the boundary modes are four gapless modes. This phase
realizes an Abelian Topological Order (ATO).
In both cases the bulk excitations are gapped. We found that their
spectrum consists of two types of particles. Particles of one type are
Majoranas. They propagate in two-dimensional space. Particles of the
other type are fractionalized solitons which remain confined to
individual ladders. Since the Majoranas are bound states of these
particles, their spectrum is situated below the soliton-antisoliton
continuum.

Since spin operators are bosonic, single Majoranas cannot be
observed by measuring spin-spin correlation functions. However, their
presence can be ascertained by measurements of thermal transport which
are sensitive to their statistics. 

The presence of the solitons is a sign of the extensive
ground-state degeneracy in the fermionic sector of the
theory. This degeneracy is associated with the existence of an order parameter
that is not local in the original spin variables. Hence, any
operator that is local in the spins has no access to this degeneracy.

\section*{Acknowledgments}

J.-H. C. was supported by the Swiss national Science Foundation (SNSF)
under Grant No.\ 2000021 153648. 
C. C. was supported by the U.S. Department of Energy (DOE),
Division of Condensed Matter Physics and Materials
Science, under Contract No.\ DE-FG02-06ER46316.
A. M. T. was supported by the U.S. Department of Energy (DOE),
Division of Condensed Matter Physics and Materials
Science, under Contract No.\ DE-AC02-98CH10886.
We acknowledge the Condensed Matter Theory Visitors Program at
Boston University for support.
A. M. T. is grateful to Alexander Abanov for valueable discussions. We thank Yohei Fuji for constructive criticisms.

\bibliography{microscopics-for-non-abelian-wires}

\begin{thebibliography}{33}%
\makeatletter
\providecommand \@ifxundefined [1]{%
 \@ifx{#1\undefined}
}%
\providecommand \@ifnum [1]{%
 \ifnum #1\expandafter \@firstoftwo
 \else \expandafter \@secondoftwo
 \fi
}%
\providecommand \@ifx [1]{%
 \ifx #1\expandafter \@firstoftwo
 \else \expandafter \@secondoftwo
 \fi
}%
\providecommand \natexlab [1]{#1}%
\providecommand \enquote  [1]{``#1''}%
\providecommand \bibnamefont  [1]{#1}%
\providecommand \bibfnamefont [1]{#1}%
\providecommand \citenamefont [1]{#1}%
\providecommand \href@noop [0]{\@secondoftwo}%
\providecommand \href [0]{\begingroup \@sanitize@url \@href}%
\providecommand \@href[1]{\@@startlink{#1}\@@href}%
\providecommand \@@href[1]{\endgroup#1\@@endlink}%
\providecommand \@sanitize@url [0]{\catcode `\\12\catcode `\$12\catcode
  `\&12\catcode `\#12\catcode `\^12\catcode `\_12\catcode `\%12\relax}%
\providecommand \@@startlink[1]{}%
\providecommand \@@endlink[0]{}%
\providecommand \url  [0]{\begingroup\@sanitize@url \@url }%
\providecommand \@url [1]{\endgroup\@href {#1}{\urlprefix }}%
\providecommand \urlprefix  [0]{URL }%
\providecommand \Eprint [0]{\href }%
\providecommand \doibase [0]{http://dx.doi.org/}%
\providecommand \selectlanguage [0]{\@gobble}%
\providecommand \bibinfo  [0]{\@secondoftwo}%
\providecommand \bibfield  [0]{\@secondoftwo}%
\providecommand \translation [1]{[#1]}%
\providecommand \BibitemOpen [0]{}%
\providecommand \bibitemStop [0]{}%
\providecommand \bibitemNoStop [0]{.\EOS\space}%
\providecommand \EOS [0]{\spacefactor3000\relax}%
\providecommand \BibitemShut  [1]{\csname bibitem#1\endcsname}%
\let\auto@bib@innerbib\@empty
\bibitem [{\citenamefont {Wen}(1991)}]{Wen91a}%
  \BibitemOpen
  \bibfield  {author} {\bibinfo {author} {\bibfnamefont {X.-G.}\ \bibnamefont
  {Wen}},\ }\href {\doibase 10.1142/S0217979291001541} {\bibfield  {journal}
  {\bibinfo  {journal} {Int. J. Mod. Phys. B}\ }\textbf {\bibinfo {volume}
  {05}},\ \bibinfo {pages} {1641} (\bibinfo {year} {1991})}\BibitemShut
  {NoStop}%
\bibitem [{\citenamefont {Kitaev}\ and\ \citenamefont
  {Preskill}(2006)}]{Kitaev06b}%
  \BibitemOpen
  \bibfield  {author} {\bibinfo {author} {\bibfnamefont {A.}~\bibnamefont
  {Kitaev}}\ and\ \bibinfo {author} {\bibfnamefont {J.}~\bibnamefont
  {Preskill}},\ }\href {\doibase 10.1103/PhysRevLett.96.110404} {\bibfield
  {journal} {\bibinfo  {journal} {Phys. Rev. Lett.}\ }\textbf {\bibinfo
  {volume} {96}},\ \bibinfo {pages} {110404} (\bibinfo {year}
  {2006})}\BibitemShut {NoStop}%
\bibitem [{\citenamefont {Levin}\ and\ \citenamefont {Wen}(2006)}]{Levin06}%
  \BibitemOpen
  \bibfield  {author} {\bibinfo {author} {\bibfnamefont {M.}~\bibnamefont
  {Levin}}\ and\ \bibinfo {author} {\bibfnamefont {X.-G.}\ \bibnamefont
  {Wen}},\ }\href {\doibase 10.1103/PhysRevLett.96.110405} {\bibfield
  {journal} {\bibinfo  {journal} {Phys. Rev. Lett.}\ }\textbf {\bibinfo
  {volume} {96}},\ \bibinfo {pages} {110405} (\bibinfo {year}
  {2006})}\BibitemShut {NoStop}%
\bibitem [{\citenamefont {Chen}\ \emph {et~al.}(2010)\citenamefont {Chen},
  \citenamefont {Gu},\ and\ \citenamefont {Wen}}]{Chen10}%
  \BibitemOpen
  \bibfield  {author} {\bibinfo {author} {\bibfnamefont {X.}~\bibnamefont
  {Chen}}, \bibinfo {author} {\bibfnamefont {Z.-C.}\ \bibnamefont {Gu}}, \ and\
  \bibinfo {author} {\bibfnamefont {X.-G.}\ \bibnamefont {Wen}},\ }\href
  {\doibase 10.1103/PhysRevB.82.155138} {\bibfield  {journal} {\bibinfo
  {journal} {Phys. Rev. B}\ }\textbf {\bibinfo {volume} {82}},\ \bibinfo
  {pages} {155138} (\bibinfo {year} {2010})}\BibitemShut {NoStop}%
\bibitem [{\citenamefont {Wen}(2016)}]{Wen16}%
  \BibitemOpen
  \bibfield  {author} {\bibinfo {author} {\bibfnamefont {X.-G.}\ \bibnamefont
  {Wen}},\ }\href {\doibase 10.1093/nsr/nwv077} {\bibfield  {journal} {\bibinfo
   {journal} {Nat. Sci. Rev.}\ }\textbf {\bibinfo {volume} {3}},\ \bibinfo
  {pages} {68} (\bibinfo {year} {2016})}\BibitemShut {NoStop}%
\bibitem [{\citenamefont {Lan}\ \emph {et~al.}(2016)\citenamefont {Lan},
  \citenamefont {Kong},\ and\ \citenamefont {Wen}}]{Lan16a}%
  \BibitemOpen
  \bibfield  {author} {\bibinfo {author} {\bibfnamefont {T.}~\bibnamefont
  {Lan}}, \bibinfo {author} {\bibfnamefont {L.}~\bibnamefont {Kong}}, \ and\
  \bibinfo {author} {\bibfnamefont {X.-G.}\ \bibnamefont {Wen}},\ }\href
  {\doibase 10.1103/PhysRevB.94.155113} {\bibfield  {journal} {\bibinfo
  {journal} {Phys. Rev. B}\ }\textbf {\bibinfo {volume} {94}},\ \bibinfo
  {pages} {155113} (\bibinfo {year} {2016})}\BibitemShut {NoStop}%
\bibitem [{\citenamefont {Moessner}\ and\ \citenamefont
  {Sondhi}(2001)}]{Moessner01}%
  \BibitemOpen
  \bibfield  {author} {\bibinfo {author} {\bibfnamefont {R.}~\bibnamefont
  {Moessner}}\ and\ \bibinfo {author} {\bibfnamefont {S.~L.}\ \bibnamefont
  {Sondhi}},\ }\href {\doibase 10.1103/PhysRevLett.86.1881} {\bibfield
  {journal} {\bibinfo  {journal} {Phys. Rev. Lett.}\ }\textbf {\bibinfo
  {volume} {86}},\ \bibinfo {pages} {1881} (\bibinfo {year}
  {2001})}\BibitemShut {NoStop}%
\bibitem [{\citenamefont {Kitaev}(2006)}]{Kitaev06a}%
  \BibitemOpen
  \bibfield  {author} {\bibinfo {author} {\bibfnamefont {A.}~\bibnamefont
  {Kitaev}},\ }\href {\doibase http://dx.doi.org/10.1016/j.aop.2005.10.005}
  {\bibfield  {journal} {\bibinfo  {journal} {Annals of Physics}\ }\textbf
  {\bibinfo {volume} {321}},\ \bibinfo {pages} {2 } (\bibinfo {year}
  {2006})}\BibitemShut {NoStop}%
\bibitem [{\citenamefont {Mukhopadhyay}\ \emph {et~al.}(2001)\citenamefont
  {Mukhopadhyay}, \citenamefont {Kane},\ and\ \citenamefont
  {Lubensky}}]{Mukhopadhyay01}%
  \BibitemOpen
  \bibfield  {author} {\bibinfo {author} {\bibfnamefont {R.}~\bibnamefont
  {Mukhopadhyay}}, \bibinfo {author} {\bibfnamefont {C.~L.}\ \bibnamefont
  {Kane}}, \ and\ \bibinfo {author} {\bibfnamefont {T.~C.}\ \bibnamefont
  {Lubensky}},\ }\href {\doibase 10.1103/PhysRevB.63.081103} {\bibfield
  {journal} {\bibinfo  {journal} {Phys. Rev. B}\ }\textbf {\bibinfo {volume}
  {63}},\ \bibinfo {pages} {081103} (\bibinfo {year} {2001})}\BibitemShut
  {NoStop}%
\bibitem [{\citenamefont {Kane}\ \emph {et~al.}(2002)\citenamefont {Kane},
  \citenamefont {Mukhopadhyay},\ and\ \citenamefont {Lubensky}}]{Kane02}%
  \BibitemOpen
  \bibfield  {author} {\bibinfo {author} {\bibfnamefont {C.~L.}\ \bibnamefont
  {Kane}}, \bibinfo {author} {\bibfnamefont {R.}~\bibnamefont {Mukhopadhyay}},
  \ and\ \bibinfo {author} {\bibfnamefont {T.~C.}\ \bibnamefont {Lubensky}},\
  }\href {\doibase 10.1103/PhysRevLett.88.036401} {\bibfield  {journal}
  {\bibinfo  {journal} {Phys. Rev. Lett.}\ }\textbf {\bibinfo {volume} {88}},\
  \bibinfo {pages} {036401} (\bibinfo {year} {2002})}\BibitemShut {NoStop}%
\bibitem [{\citenamefont {Teo}\ and\ \citenamefont {Kane}(2014)}]{Teo14}%
  \BibitemOpen
  \bibfield  {author} {\bibinfo {author} {\bibfnamefont {J.~C.~Y.}\
  \bibnamefont {Teo}}\ and\ \bibinfo {author} {\bibfnamefont {C.~L.}\
  \bibnamefont {Kane}},\ }\href {\doibase 10.1103/PhysRevB.89.085101}
  {\bibfield  {journal} {\bibinfo  {journal} {Phys. Rev. B}\ }\textbf {\bibinfo
  {volume} {89}},\ \bibinfo {pages} {085101} (\bibinfo {year}
  {2014})}\BibitemShut {NoStop}%
\bibitem [{\citenamefont {Kane}\ \emph {et~al.}(2017)\citenamefont {Kane},
  \citenamefont {Stern},\ and\ \citenamefont {Halperin}}]{Kane17}%
  \BibitemOpen
  \bibfield  {author} {\bibinfo {author} {\bibfnamefont {C.~L.}\ \bibnamefont
  {Kane}}, \bibinfo {author} {\bibfnamefont {A.}~\bibnamefont {Stern}}, \ and\
  \bibinfo {author} {\bibfnamefont {B.~I.}\ \bibnamefont {Halperin}},\ }\href
  {\doibase 10.1103/PhysRevX.7.031009} {\bibfield  {journal} {\bibinfo
  {journal} {Phys. Rev. X}\ }\textbf {\bibinfo {volume} {7}},\ \bibinfo {pages}
  {031009} (\bibinfo {year} {2017})}\BibitemShut {NoStop}%
\bibitem [{\citenamefont {Chakravarty}(1996)}]{Chakravarty96}%
  \BibitemOpen
  \bibfield  {author} {\bibinfo {author} {\bibfnamefont {S.}~\bibnamefont
  {Chakravarty}},\ }\href {\doibase 10.1103/PhysRevLett.77.4446} {\bibfield
  {journal} {\bibinfo  {journal} {Phys. Rev. Lett.}\ }\textbf {\bibinfo
  {volume} {77}},\ \bibinfo {pages} {4446} (\bibinfo {year}
  {1996})}\BibitemShut {NoStop}%
\bibitem [{\citenamefont {Sierra}(1996)}]{Sierra96}%
  \BibitemOpen
  \bibfield  {author} {\bibinfo {author} {\bibfnamefont {G.}~\bibnamefont
  {Sierra}},\ }\href {http://stacks.iop.org/0305-4470/29/i=12/a=032} {\bibfield
   {journal} {\bibinfo  {journal} {J. Phys. A: Math. Gen.}\ }\textbf {\bibinfo
  {volume} {29}},\ \bibinfo {pages} {3299} (\bibinfo {year}
  {1996})}\BibitemShut {NoStop}%
\bibitem [{\citenamefont {Sierra}(1997)}]{Sierra97}%
  \BibitemOpen
  \bibfield  {author} {\bibinfo {author} {\bibfnamefont {G.}~\bibnamefont
  {Sierra}},\ }\enquote {\bibinfo {title} {On the application of the non-linear
  sigma model to spin chains and spin ladders},}\ in\ \href {\doibase
  10.1007/BFb0104637} {\emph {\bibinfo {booktitle} {Strongly Correlated
  Magnetic and Superconducting Systems: Proceedings of the El Escorial Summer
  School Held in Madrid, Spain, 15--19 July 1996}}},\ \bibinfo {editor} {edited
  by\ \bibinfo {editor} {\bibfnamefont {G.}~\bibnamefont {Sierra}}\ and\
  \bibinfo {editor} {\bibfnamefont {M.~A.}\ \bibnamefont
  {Mart{\'i}n-Delgado}}}\ (\bibinfo  {publisher} {Springer Berlin Heidelberg},\
  \bibinfo {address} {Berlin, Heidelberg},\ \bibinfo {year} {1997})\ pp.\
  \bibinfo {pages} {137--166}\BibitemShut {NoStop}%
\bibitem [{\citenamefont {Sylju\aa{}sen}\ \emph {et~al.}(1997)\citenamefont
  {Sylju\aa{}sen}, \citenamefont {Chakravarty},\ and\ \citenamefont
  {Greven}}]{Syljuasen97}%
  \BibitemOpen
  \bibfield  {author} {\bibinfo {author} {\bibfnamefont {O.~F.}\ \bibnamefont
  {Sylju\aa{}sen}}, \bibinfo {author} {\bibfnamefont {S.}~\bibnamefont
  {Chakravarty}}, \ and\ \bibinfo {author} {\bibfnamefont {M.}~\bibnamefont
  {Greven}},\ }\href {\doibase 10.1103/PhysRevLett.78.4115} {\bibfield
  {journal} {\bibinfo  {journal} {Phys. Rev. Lett.}\ }\textbf {\bibinfo
  {volume} {78}},\ \bibinfo {pages} {4115} (\bibinfo {year}
  {1997})}\BibitemShut {NoStop}%
\bibitem [{\citenamefont {Neupert}\ \emph {et~al.}(2014)\citenamefont
  {Neupert}, \citenamefont {Chamon}, \citenamefont {Mudry},\ and\ \citenamefont
  {Thomale}}]{Neupert14}%
  \BibitemOpen
  \bibfield  {author} {\bibinfo {author} {\bibfnamefont {T.}~\bibnamefont
  {Neupert}}, \bibinfo {author} {\bibfnamefont {C.}~\bibnamefont {Chamon}},
  \bibinfo {author} {\bibfnamefont {C.}~\bibnamefont {Mudry}}, \ and\ \bibinfo
  {author} {\bibfnamefont {R.}~\bibnamefont {Thomale}},\ }\href {\doibase
  10.1103/PhysRevB.90.205101} {\bibfield  {journal} {\bibinfo  {journal} {Phys.
  Rev. B}\ }\textbf {\bibinfo {volume} {90}},\ \bibinfo {pages} {205101}
  (\bibinfo {year} {2014})}\BibitemShut {NoStop}%
\bibitem [{\citenamefont {Huang}\ \emph {et~al.}(2016)\citenamefont {Huang},
  \citenamefont {Chen}, \citenamefont {Gomes}, \citenamefont {Neupert},
  \citenamefont {Chamon},\ and\ \citenamefont {Mudry}}]{Huang16a}%
  \BibitemOpen
  \bibfield  {author} {\bibinfo {author} {\bibfnamefont {P.-H.}\ \bibnamefont
  {Huang}}, \bibinfo {author} {\bibfnamefont {J.-H.}\ \bibnamefont {Chen}},
  \bibinfo {author} {\bibfnamefont {P.~R.~S.}\ \bibnamefont {Gomes}}, \bibinfo
  {author} {\bibfnamefont {T.}~\bibnamefont {Neupert}}, \bibinfo {author}
  {\bibfnamefont {C.}~\bibnamefont {Chamon}}, \ and\ \bibinfo {author}
  {\bibfnamefont {C.}~\bibnamefont {Mudry}},\ }\href {\doibase
  10.1103/PhysRevB.93.205123} {\bibfield  {journal} {\bibinfo  {journal} {Phys.
  Rev. B}\ }\textbf {\bibinfo {volume} {93}},\ \bibinfo {pages} {205123}
  (\bibinfo {year} {2016})}\BibitemShut {NoStop}%
\bibitem [{\citenamefont {Huang}\ \emph {et~al.}(2017)\citenamefont {Huang},
  \citenamefont {Chen}, \citenamefont {Feiguin}, \citenamefont {Chamon},\ and\
  \citenamefont {Mudry}}]{Huang17}%
  \BibitemOpen
  \bibfield  {author} {\bibinfo {author} {\bibfnamefont {P.-H.}\ \bibnamefont
  {Huang}}, \bibinfo {author} {\bibfnamefont {J.-H.}\ \bibnamefont {Chen}},
  \bibinfo {author} {\bibfnamefont {A.~E.}\ \bibnamefont {Feiguin}}, \bibinfo
  {author} {\bibfnamefont {C.}~\bibnamefont {Chamon}}, \ and\ \bibinfo {author}
  {\bibfnamefont {C.}~\bibnamefont {Mudry}},\ }\href {\doibase
  10.1103/PhysRevB.95.144413} {\bibfield  {journal} {\bibinfo  {journal} {Phys.
  Rev. B}\ }\textbf {\bibinfo {volume} {95}},\ \bibinfo {pages} {144413}
  (\bibinfo {year} {2017})}\BibitemShut {NoStop}%
\bibitem [{\citenamefont {Witten}(1978)}]{Witten78}%
  \BibitemOpen
  \bibfield  {author} {\bibinfo {author} {\bibfnamefont {E.}~\bibnamefont
  {Witten}},\ }\href {\doibase http://dx.doi.org/10.1016/0550-3213(78)90204-3}
  {\bibfield  {journal} {\bibinfo  {journal} {Nuclear Physics B}\ }\textbf
  {\bibinfo {volume} {142}},\ \bibinfo {pages} {285 } (\bibinfo {year}
  {1978})}\BibitemShut {NoStop}%
\bibitem [{\citenamefont {Lukyanov}\ and\ \citenamefont
  {Zamolodchikov}(2001)}]{Lukyanov01}%
  \BibitemOpen
  \bibfield  {author} {\bibinfo {author} {\bibfnamefont {S.}~\bibnamefont
  {Lukyanov}}\ and\ \bibinfo {author} {\bibfnamefont {A.}~\bibnamefont
  {Zamolodchikov}},\ }\href {\doibase
  https://doi.org/10.1016/S0550-3213(01)00262-0} {\bibfield  {journal}
  {\bibinfo  {journal} {Nuclear Physics B}\ }\textbf {\bibinfo {volume}
  {607}},\ \bibinfo {pages} {437 } (\bibinfo {year} {2001})}\BibitemShut
  {NoStop}%
\bibitem [{\citenamefont {Essler}\ and\ \citenamefont
  {Tsvelik}(2002)}]{Essler01}%
  \BibitemOpen
  \bibfield  {author} {\bibinfo {author} {\bibfnamefont {F.~H.~L.}\
  \bibnamefont {Essler}}\ and\ \bibinfo {author} {\bibfnamefont {A.~M.}\
  \bibnamefont {Tsvelik}},\ }\href {\doibase 10.1103/PhysRevB.65.115117}
  {\bibfield  {journal} {\bibinfo  {journal} {Phys. Rev. B}\ }\textbf {\bibinfo
  {volume} {65}},\ \bibinfo {pages} {115117} (\bibinfo {year}
  {2002})}\BibitemShut {NoStop}%
\bibitem [{\citenamefont {Essler}\ and\ \citenamefont
  {Tsvelik}(2005)}]{Essler05a}%
  \BibitemOpen
  \bibfield  {author} {\bibinfo {author} {\bibfnamefont {F.~H.~L.}\
  \bibnamefont {Essler}}\ and\ \bibinfo {author} {\bibfnamefont {A.~M.}\
  \bibnamefont {Tsvelik}},\ }\href {\doibase 10.1103/PhysRevB.71.195116}
  {\bibfield  {journal} {\bibinfo  {journal} {Phys. Rev. B}\ }\textbf {\bibinfo
  {volume} {71}},\ \bibinfo {pages} {195116} (\bibinfo {year}
  {2005})}\BibitemShut {NoStop}%
\bibitem [{\citenamefont {Shelton}\ \emph {et~al.}(1996)\citenamefont
  {Shelton}, \citenamefont {Nersesyan},\ and\ \citenamefont
  {Tsvelik}}]{Shelton96}%
  \BibitemOpen
  \bibfield  {author} {\bibinfo {author} {\bibfnamefont {D.~G.}\ \bibnamefont
  {Shelton}}, \bibinfo {author} {\bibfnamefont {A.~A.}\ \bibnamefont
  {Nersesyan}}, \ and\ \bibinfo {author} {\bibfnamefont {A.~M.}\ \bibnamefont
  {Tsvelik}},\ }\href {\doibase 10.1103/PhysRevB.53.8521} {\bibfield  {journal}
  {\bibinfo  {journal} {Phys. Rev. B}\ }\textbf {\bibinfo {volume} {53}},\
  \bibinfo {pages} {8521} (\bibinfo {year} {1996})}\BibitemShut {NoStop}%
\bibitem [{\citenamefont {Nersesyan}\ and\ \citenamefont
  {Tsvelik}(1997)}]{Nersesyan_97}%
  \BibitemOpen
  \bibfield  {author} {\bibinfo {author} {\bibfnamefont {A.~A.}\ \bibnamefont
  {Nersesyan}}\ and\ \bibinfo {author} {\bibfnamefont {A.~M.}\ \bibnamefont
  {Tsvelik}},\ }\href {\doibase 10.1103/PhysRevLett.78.3939} {\bibfield
  {journal} {\bibinfo  {journal} {Phys. Rev. Lett.}\ }\textbf {\bibinfo
  {volume} {78}},\ \bibinfo {pages} {3939} (\bibinfo {year}
  {1997})}\BibitemShut {NoStop}%
\bibitem [{\citenamefont {Gogolin}\ \emph {et~al.}(2004)\citenamefont
  {Gogolin}, \citenamefont {Nersesyan},\ and\ \citenamefont
  {Tsvelik}}]{Gogolin04}%
  \BibitemOpen
  \bibfield  {author} {\bibinfo {author} {\bibfnamefont {A.~O.}\ \bibnamefont
  {Gogolin}}, \bibinfo {author} {\bibfnamefont {A.~A.}\ \bibnamefont
  {Nersesyan}}, \ and\ \bibinfo {author} {\bibfnamefont {A.~M.}\ \bibnamefont
  {Tsvelik}},\ }\href@noop {} {\emph {\bibinfo {title} {Bosonization and
  Strongly Correlated Systems}}}\ (\bibinfo  {publisher} {Cambridge University
  Press},\ \bibinfo {year} {2004})\BibitemShut {NoStop}%
\bibitem [{\citenamefont {Tsvelik}(2003)}]{Tsvelik03}%
  \BibitemOpen
  \bibfield  {author} {\bibinfo {author} {\bibfnamefont {A.~M.}\ \bibnamefont
  {Tsvelik}},\ }\href@noop {} {\emph {\bibinfo {title} {Quantum Field Theory in
  Condensed Matter Physics}}}\ (\bibinfo  {publisher} {Cambridge University
  Press},\ \bibinfo {year} {2003})\BibitemShut {NoStop}%
\bibitem [{\citenamefont {Allen}\ \emph {et~al.}(2000)\citenamefont {Allen},
  \citenamefont {Essler},\ and\ \citenamefont {Nersesyan}}]{Allen2000}%
  \BibitemOpen
  \bibfield  {author} {\bibinfo {author} {\bibfnamefont {D.}~\bibnamefont
  {Allen}}, \bibinfo {author} {\bibfnamefont {F.~H.~L.}\ \bibnamefont
  {Essler}}, \ and\ \bibinfo {author} {\bibfnamefont {A.~A.}\ \bibnamefont
  {Nersesyan}},\ }\href {\doibase 10.1103/PhysRevB.61.8871} {\bibfield
  {journal} {\bibinfo  {journal} {Phys. Rev. B}\ }\textbf {\bibinfo {volume}
  {61}},\ \bibinfo {pages} {8871} (\bibinfo {year} {2000})}\BibitemShut
  {NoStop}%
\bibitem [{\citenamefont {Gorohovsky}\ \emph {et~al.}(2015)\citenamefont
  {Gorohovsky}, \citenamefont {Pereira},\ and\ \citenamefont
  {Sela}}]{Gorohovsky15}%
  \BibitemOpen
  \bibfield  {author} {\bibinfo {author} {\bibfnamefont {G.}~\bibnamefont
  {Gorohovsky}}, \bibinfo {author} {\bibfnamefont {R.~G.}\ \bibnamefont
  {Pereira}}, \ and\ \bibinfo {author} {\bibfnamefont {E.}~\bibnamefont
  {Sela}},\ }\href {\doibase 10.1103/PhysRevB.91.245139} {\bibfield  {journal}
  {\bibinfo  {journal} {Phys. Rev. B}\ }\textbf {\bibinfo {volume} {91}},\
  \bibinfo {pages} {245139} (\bibinfo {year} {2015})}\BibitemShut {NoStop}%
\bibitem [{\citenamefont {Fidkowski}\ and\ \citenamefont
  {Kitaev}(2010)}]{Fidkowski10}%
  \BibitemOpen
  \bibfield  {author} {\bibinfo {author} {\bibfnamefont {L.}~\bibnamefont
  {Fidkowski}}\ and\ \bibinfo {author} {\bibfnamefont {A.}~\bibnamefont
  {Kitaev}},\ }\href {\doibase 10.1103/PhysRevB.81.134509} {\bibfield
  {journal} {\bibinfo  {journal} {Phys. Rev. B}\ }\textbf {\bibinfo {volume}
  {81}},\ \bibinfo {pages} {134509} (\bibinfo {year} {2010})}\BibitemShut
  {NoStop}%
\bibitem [{\citenamefont {Yao}\ and\ \citenamefont {Ryu}(2013)}]{Yao13}%
  \BibitemOpen
  \bibfield  {author} {\bibinfo {author} {\bibfnamefont {H.}~\bibnamefont
  {Yao}}\ and\ \bibinfo {author} {\bibfnamefont {S.}~\bibnamefont {Ryu}},\
  }\href {\doibase 10.1103/PhysRevB.88.064507} {\bibfield  {journal} {\bibinfo
  {journal} {Phys. Rev. B}\ }\textbf {\bibinfo {volume} {88}},\ \bibinfo
  {pages} {064507} (\bibinfo {year} {2013})}\BibitemShut {NoStop}%
\bibitem [{\citenamefont {Shankar}\ and\ \citenamefont
  {Witten}(1978)}]{Shankar78}%
  \BibitemOpen
  \bibfield  {author} {\bibinfo {author} {\bibfnamefont {R.}~\bibnamefont
  {Shankar}}\ and\ \bibinfo {author} {\bibfnamefont {E.}~\bibnamefont
  {Witten}},\ }\href {\doibase http://dx.doi.org/10.1016/0550-3213(78)90031-7}
  {\bibfield  {journal} {\bibinfo  {journal} {Nuclear Physics B}\ }\textbf
  {\bibinfo {volume} {141}},\ \bibinfo {pages} {349 } (\bibinfo {year}
  {1978})}\BibitemShut {NoStop}%
\bibitem [{\citenamefont {Frahm}\ and\ \citenamefont
  {Rödenbeck}(1997)}]{Frahm97}%
  \BibitemOpen
  \bibfield  {author} {\bibinfo {author} {\bibfnamefont {H.}~\bibnamefont
  {Frahm}}\ and\ \bibinfo {author} {\bibfnamefont {C.}~\bibnamefont
  {Rödenbeck}},\ }\href {http://stacks.iop.org/0305-4470/30/i=13/a=005}
  {\bibfield  {journal} {\bibinfo  {journal} {Journal of Physics A:
  Mathematical and General}\ }\textbf {\bibinfo {volume} {30}},\ \bibinfo
  {pages} {4467} (\bibinfo {year} {1997})}\BibitemShut {NoStop}%
\end{thebibliography}%

\end{document}